\documentclass[preprint,12pt]{elsarticle}

\usepackage{standalone}
\usepackage{tikz}
\usepackage{multicol}
\usepackage[ruled,vlined,english]{algorithm2e}
\usepackage{url}

\usepackage{amssymb, amsmath}
\usepackage{xcolor}
\usepackage{relsize}
\def\textsubscript#1{\ensuremath{_{\mbox{\textscale{.6}{#1}}}}}
\usepackage{graphicx}
\usepackage{multirow}

\usepackage{lineno}

\journal{Elsevier}

\begin{document}

\newcommand{\blacklinedashed}{\raisebox{2pt}{\tikz{\draw[-,black,dashed,line width = 1pt](0,0) -- (5.1mm,0);}}}

\newcommand{\blacklinesolid}{\raisebox{2pt}{\tikz{\draw[-,black,solid,line width = 1pt](0,0) -- (5mm,0);}}}

\newcommand{\blacklinedotted}{\raisebox{2pt}{\tikz{\draw[-,black,dotted,line width = 1.5pt](0,0) -- (5mm,0);}}}

\newcommand{\blacklinemarker}{\raisebox{1pt}{\tikz{\draw[-,black,solid,line width = 0.7pt](0,0) -- (5mm,0); \filldraw[black] (2.5mm,0) circle (1.5pt);}}}

\definecolor{greySTRONG}{RGB}{127.5,127.5,127.5}
\definecolor{greySOFT}{RGB}{204,204,204}

\newcommand{\greylinesolidstrong}{\raisebox{2pt}{\tikz{\draw[-,greySTRONG,solid,line width = 1.5pt](0,0) -- (5mm,0)}}}

\newcommand{\greylinedashedstrong}{\raisebox{2pt}{\tikz{\draw[-,greySTRONG,dashed,line width = 1.5pt](0,0) -- (5mm,0)}}}

\newcommand{\greylinedottedstrong}{\raisebox{2pt}{\tikz{\draw[-,greySTRONG,dotted,line width = 1.5pt](0,0) -- (5mm,0)}}}

\newcommand{\greylinesolidsoft}{\raisebox{2pt}{\tikz{\draw[-,greySOFT,solid,line width = 1.5pt](0,0) -- (5mm,0)}}}

\newcommand{\greylinedashedsoft}{\raisebox{2pt}{\tikz{\draw[-,greySOFT,dashed,line width = 1.5pt](0,0) -- (5mm,0)}}}

\newcommand{\redline}{\raisebox{2pt}{\tikz{\draw[-,red,solid,line width = 1pt](0,0) -- (5mm,0);}}}

\newcommand{\redlinedashed}{\raisebox{2pt}{\tikz{\draw[-,red,dashed,line width = 1pt](0,0) -- (5mm,0);}}}

\newcommand{\bluelinedasheddotted}{\raisebox{2pt}{\tikz{\draw[-.,blue,dashdotted,line width = 1pt](0,0) -- (5mm,0);}}}

\newcommand{\blueline}{\raisebox{2pt}{\tikz{\draw[-.,blue,solid,line width = 1pt](0,0) -- (5mm,0);}}}

\definecolor{green}{RGB}{0,150,0}
\newcommand{\greenline}{\raisebox{2pt}{\tikz{\draw[-,green,dotted,line width = 1.5pt](0,0) -- (5mm,0);}}}

\definecolor{magenta}{RGB}{255,0,255}
\newcommand{\magentaline}{\raisebox{2pt}{\tikz{\draw[-,magenta,dotted,line width = 0.7pt](0,0) -- (5mm,0);}}}

\begin{frontmatter}



\title{Enhanced State Estimation for turbulent flows combining Ensemble Data Assimilation and Machine Learning}


\author[inst1]{Miguel M. Valero}
\corref{cor1}
\ead{miguel.martinez\_valero@ensam.eu}
\cortext[cor1]{Corresponding author}

\affiliation[inst1]{Univ. Lille, CNRS, ONERA, Arts et Métiers ParisTech, Centrale Lille, UMR 9014- LMFL- Laboratoire de Mécanique des fluides de Lille - Kampé de Feriet, F-59000 Lille, France}

\author[inst1]{Marcello Meldi}

\begin{abstract}
A novel strategy is proposed to improve the accuracy of state estimation and reconstruction from low-fidelity models and sparse data from sensors. This strategy combines ensemble Data Assimilation (DA) and Machine Learning (ML) tools, exploiting their complementary features. ML techniques rely on the data produced by DA methods during analysis phases to train physics-informed corrective algorithms, which are then coupled with the low-fidelity models when data from sensors is unavailable. The methodology is validated via the analysis of the turbulent plane channel flow test case for $Re_\tau \approx 550$. Here, the low-fidelity model consists of coarse-grained simulations coupled with the Immersed Boundary Method (IBM), while observation is sampled by a highly refined body-fitted calculation. The analysis demonstrates the capabilities of the algorithm based on DA and ML to accurately predict the flow features with significantly reduced computational costs. This approach exhibits potential for future synergistic applications of DA and ML, leveraging the robustness and efficiency of ML models alongside the physical interpretability ensured by DA algorithms.
\end{abstract}


\begin{highlights}
\item A methodology combining EnKF with RFR is developed, showcasing the complementary nature of DA with ML to enhance flow prediction.
\item DA is used to generate a massive, physics-informed, accurate dataset that serves as valuable training data for building an efficient ML model.
\item An ML black-box tool is successfully combined with coarse-grained IBM simulations.
\item The ML-augmented model exhibits strong predictive capability in turbulent boundary layers of wall-bounded flows also outside the training range when the grid resolution in the near-wall region is higher than in the simulations used to create the training dataset.
\end{highlights}


\end{frontmatter}


\section{Introduction}
\label{sec:introduction}


Advances in the development of numerical techniques in fluid mechanics are rapidly removing several barriers precluding the analysis of realistic flow configurations of industrial interest. Current applications of data-driven techniques in Computational Fluid Dynamics (CFD), including Data Assimilation (DA) \cite{Asch2016_siam, Evensen2022_springer} and Machine Learning (ML) \cite{Burkov2019, Brunton2020_ARFM}, have provided accurate results when combined with simulation strategies such as Reynolds-Averaged Navier--Stokes (RANS) approaches \cite{Gorle2013_pof, Edeling2014_jcp, Margheri2014_cf, Tracey2015_AIAA, Wu2018_prf, Srinivasan2019_prf, Volpiani2021_prf, Villanueva2024_cof} and Large Eddy Simulation (LES) \cite{Meldi2011_pof, Vollant_jot, Meldi2018_ftc, Chandramouli2020_jcp, Lozano2020_arb, Mons2021_prf, Moldovan2024_cf, Villanueva2023, Plogman2024}. With the progressive increase in computational resources available at computational centres, one can envision the blossoming of futuristic applications such as Digital Twins (DTs) \cite{Molinaro2021_cf, Wagg2020_asme, Yang2022_sensors, Sharma2022_jiii, Schena2024_jcp} for realistic fluid mechanics cases in the coming decades. 

Among the issues that must be challenged to reach such a level of technological maturity, one central aspect deals with the optimisation of the performances of data-driven tools when only sparse data (\textit{observation}) is available. For ML tools, and in particular, for deep learning, a limited amount of data for training and validation can lead to a lack of convergence of the algorithms, severely impacting the accuracy and robustness of the prediction. The techniques developed to handle this problem are usually referred to as \textit{frugal Artificial Intelligence} \cite{Boosari2019_mdpi, White2020_aiaa, Kochkov_am, Fan2024}. Among these, DA is a very efficient complementary tool. In fact, approaches such as the Ensemble Kalman Filter (EnKF) \cite{Asch2016_siam, Evensen2009_ieee} can be used to produce a complete state estimation on the physical domain investigated, combining the prediction of a model and sparse observation. This state estimation can be very rich in terms of data produced, providing complete information to the ML tools for training. This strategy can, therefore, bypass the potential scarcity of observation. Rigorous mathematical frameworks have been proposed in the literature to combine DA and ML \cite{Brajard2020_trs, Farchi2021_qjrms, Arcucci2021_as}, and applications using CFD have been reported \cite{Villiers2024_ftc, WangArXiV, Quattromini2023, Ling2015_aiaa}. However, the prediction of the DA analysis itself can also be severely degraded because of the scarcity of data. This problem is observed, for example, if observation is available only over a limited time window. During the aforementioned window, DA can succeed in obtaining an accurate prediction, thanks to the combination of the state estimation and the resulting parametric optimisation of the system. However, once the observation is no longer available, the DA state update cannot be performed. This can lead to progressive degradation of the results, particularly if the DA optimisation of the numerical model is not entirely satisfying. Usually, this problem happens because of intrinsic limitations in the structural features of the model itself, such as relying on the eddy viscosity hypothesis for RANS closures. Therefore, if the DA state augmentation provides poor results in terms of state estimation, the ML training phase will also be sub-optimal.      
In the present work, we explore the usage of ML tools to augment the predictive capabilities of an EnKF. To this purpose, a Random Forest Regression (RFR) algorithm \cite{Breiman2001_ML} is trained to reproduce the model optimisation as well as the state update obtained in the analysis phase of the EnKF. This algorithm is then integrated as a black box in a numerical simulation solver for the simulation of a permanent flow in order to assess its accuracy. The validation of the algorithm is performed via the analysis of the turbulent plane channel flow for $Re_\tau \approx 550$. This test case exhibits fully developed turbulent structures, and it is, therefore, a complex case to handle for DA and ML algorithms despite the simplicity of the geometry. Here, the RFR tools are used for two different purposes. First, they are trained to mimic the effect of an Immersed Boundary Method (IBM), whose free coefficients are optimised in the DA analysis phase. Second, they are trained to learn the non-linear state update process performed by the EnKF. It is shown that the global ML tool matches the accuracy of the complete DA strategy, but it is not affected by the time window of investigation and by the availability of further data. These findings highlight the efficiency of the methodology for the analysis of cases for which data can be limited in time but for which analysis over significantly long time windows may be required. This is, for instance, the case of several internal flows, such as those in combustion engines or turbomachinery.

The article is structured as follows. In \S\ref{sec:numerical_ingredients}, we introduce the numerical equations and the numerical background, with a particular emphasis on the data-informed IBM developed using the DA algorithm, along with a presentation of the ensemble-learning ML method employed. In \S\ref{sec:EnKF_limitations}, the test case for the analysis is presented, highlighting the limitations of the DA algorithm and how ML techniques can address these challenges. Later, in \S\ref{sec:ML_stateEstimation}, we explore the potential of various ML models as a complete alternative to the DA tool. Finally, in \S\ref{sec:conclusions}, we provide conclusions and suggest directions for future research.

\section{Numerical ingredients}
\label{sec:numerical_ingredients}

The numerical tools used in the present analysis are now briefly introduced. These tools include the strategies for CFD simulations as well as the DA and ML techniques, which are combined.

\subsection{Numerical solver and Immersed Boundary Method}
\label{sec:IBM}

The numerical simulations are performed using a discretisation of the Navier--Stokes equations for incompressible flows and Newtonian fluids:

\begin{eqnarray}
\boldsymbol{\nabla} \cdot \boldsymbol{u} &=& 0 \label{eqn:NavierStokesM} \label{eqn:NavierStokes_Mass}\\
\frac{\partial \boldsymbol{u}}{\partial t} + \boldsymbol{\nabla} \cdot (\boldsymbol{u}\boldsymbol{u}) &=& -\nabla p + \nu \nabla^2 \boldsymbol{u} + \boldsymbol{f}_P
    \label{eqn:NavierStokes_Mom}
\end{eqnarray}
Here, $\boldsymbol{u}$ is the velocity field, $p$ is the pressure normalised over the density $\rho$, $t$ is the time, and $\nu$ is the kinematic viscosity of the fluid. The coordinates $\boldsymbol{x} = (x, \, y, \, z)$ indicate the physical frame of reference. The term $\boldsymbol{f}_P$ represents a volume forcing affecting the momentum equation. In this analysis, volume forcing is modelled via the Immersed Boundary Method to consider the presence of the physical wall in the domain. The source term $\boldsymbol{f}_P$ included in Equation \ref{eqn:NavierStokes_Mom} is a classical penalisation scheme \cite{Angot1999_nm}:

\begin{equation}
     \boldsymbol{f}_P = \left\{
        \begin{array}{ll}
        \boldsymbol{0} & \textrm{if } \boldsymbol{x} \in \Omega_f \\
        -\nu \, \boldsymbol{D} \left(\boldsymbol{u} - \boldsymbol{u_{ib}} \right) & \textrm{if } \boldsymbol{x} \in (\Sigma_b \cup \Omega_b)
        \end{array} \right.
    \label{eqn:forcePenDarcy}
\end{equation}
$\Omega_f$, $\Omega_b$, and $\Sigma_b$ correspond to the fluid, solid, and interface regions, respectively. $\boldsymbol{u_{ib}}$ is here the target velocity, and it is used to obtain a Dirichlet velocity condition at the wall. The coefficients $D_{ij}$ of the tensor $\boldsymbol{D}$ control the intensity of the volume source term. Guidelines in the literature indicate that their value can be adjusted to optimise performance while ensuring the stability of the numerical algorithm \cite{Verzicco2022_arfm, Kou2021_jcp}. However, optimised values are often highly sensitive to the local features of the flow (in particular to attached or separated boundary layers) and therefore, global optimisation can provide unsatisfying results. 

The numerical discretisation is based on a Finite Volume (FV) approach using a PISO scheme \citep{Ferziger1996_springer, Greenshields2022_springer, ISSA198640, Versteeg2007_pearson}. This algorithm performs a recursive resolution of the discretised set of equations obtained from Equations \ref{eqn:NavierStokes_Mass}-\ref{eqn:NavierStokes_Mom} until convergence is obtained. The solver used is the \textit{PisoFoam} algorithm included in the open-source platform OpenFOAM \cite{OpenFOAM}. Space derivatives are discretised using second-order centred schemes. A second-order backward scheme is applied to advance the solution in time. Details about the grid resolution and the time steps are provided in the test case description in \S\ref{sec:test_case}.

\subsection{Data-informed Immersed Boundary Method using Data Assimilation}
\label{sec:IBM_DA}



Discussion in \S\ref{sec:IBM} indicated that a global calibration of the IBM coefficients $D_{ij}$ may lead to inaccurate results, as optimised values are related to the local features of the flow at the level of the specific grid element. To this purpose, Valero \& Meldi \cite{Valero2023} developed a data-informed procedure for the calibration of the IBM coefficients $D_{ij}$. This procedure relies on an ensemble of $N_e$ \textit{model} realisations and high-fidelity \textit{observation}, which were combined using a Data Assimilation (DA) technique based on the Ensemble Kalman Filter (EnKF). 
The two main ingredients used in this data-driven approach are:
\begin{itemize}
\item The discrete dynamic \textit{model} $\mathcal{M}$ corresponds to the unsteady IBM numerical solver presented in \S\ref{sec:IBM}. This model offers a quasi-continuous representation of the flow field $\boldsymbol{u}$, though its accuracy is affected by the grid refinement selected. 
\item The observation $\boldsymbol{Y}$ is obtained by sampling instantaneous data from a high-fidelity, body-fitted Direct Numerical Simulation (DNS). The sensors would provide local values of the physical field (velocity at the wall) as well as global quantities (time evolution of the friction coefficient $C_f$). 
\end{itemize}
Both sources of information are affected by uncertainties, which are here modelled as unbiased Gaussian distributions. The EnKF optimally adjusts the predicted flow field $\boldsymbol{u}$ to balance the dynamics of both sources of information according to their respective uncertainties. The key elements of the procedure are now discussed, while a complete description of the method is provided in \ref{sec:EnKF}. The EnKF scheme is structured in two steps:

\begin{enumerate}
    \item A forecast ($f$) step where the physical state $\boldsymbol{u}^f$ is advanced in time using the numerical model obtained by the discretisation of Equations \ref{eqn:NavierStokes_Mass}--\ref{eqn:NavierStokes_Mom}. Using the DA formalism, this forecast step is expressed as a time advancement of the flow state $\boldsymbol{u}$ from the time step $k-1$ to $k$ using the model $\mathcal{M}$:
    \begin{equation}
    \boldsymbol{u}^f_{i,k} = \mathcal{M}(\theta_{i,k:k-1}) \, \boldsymbol{u}_{i,k-1}
    \end{equation}
    Here, the time advancement is considered for each ensemble member $i$. The array $\theta$ includes free parameters that can be modified to control the model's prediction $\mathcal{M}$. An example can be numerical values in boundary conditions or the local values of the coefficients $D_{ij}$ governing the IBM.
    \item If observation is sampled at the time step $k$, an analysis ($a$) phase is performed where the DA procedure uses the available information to improve the flow prediction and the model's accuracy. To do so, the model prediction at the forecast is combined with observation via the so-called Kalman gain $\boldsymbol{K}$:
    \begin{equation}
    \boldsymbol{u}_{k,i}^a = \boldsymbol{u}_{k,i}^f + \boldsymbol{K}_k \left( \boldsymbol{Y}_{k,i} - \mathcal{H}(\boldsymbol{u}_{k,i}^f)\right)
    \label{eq::KalmanFilterUpdate}
    \end{equation}
    The projection operator $\mathcal{H}$ usually consists of manipulation (interpolation and/or integration) of the physical field predicted by the model to obtain a consistent comparison with the observation sampled by the sensors. Equation \ref{eq::KalmanFilterUpdate} is usually used for two distinct operations. The first one, usually referred to as \textit{state augmentation} or \textit{state estimation}, updates the physical state $\boldsymbol{u}^{f \rightarrow a}$. The second one, referred to as \textit{parametric optimisation}, uses results from the previous operation to optimise some model's free parameters $\theta^{f \rightarrow a}$ to reduce the discrepancy between model and observation. In this work, the two operations are performed simultaneously thanks to the \emph{extended state} approach \cite{Asch2016_siam}.
\end{enumerate}

The analysis phase is performed only when observation $\boldsymbol{Y}$ is available at time step $k$ during unsteady simulations. The accuracy of the Kalman filter depends on the time window between successive analyses as well as on the size of the time step of the model. For the former, the DA procedure may diverge if the sample rate does not respect the Nyquist theorem with respect to the flow turnover time \cite{Meldi2018_ftc}. On the other hand, the size of the time step of the model must be small enough to ensure efficient physical propagation between analysis phases. 
In this analysis, the EnKF is performed online, which means that ensemble simulations are not stopped and restarted between analysis phases, but they are put on standby while the DA algorithm performs the state estimation and the parametric optimisation. This efficient algorithmic structure is obtained via a team-developed application, CONES \emph{(Coupling OpenFOAM with Numerical EnvironmentS)} \cite{Villanueva2024_cof}, available on \path{https://github.com/MiguelMValero/CONES}. This tool facilitates efficient coupling and field exchange between OpenFOAM modules and the DA algorithm code, eliminating the need for simulation interruptions or restarts, which could otherwise incur higher computational costs than the DA procedure itself.

If the underlying model is unbiased and complete (i.e., its prediction is exact once its free coefficients are optimised), state augmentation is expected to become negligible after the EnKF converges, as the physical state predicted by the model aligns with the high-fidelity space. However, if the model is biased and cannot exactly reproduce the true solution even with an optimised parametric description (e.g., turbulence modelling \cite{Wilcox2006_DCW, Xiao2019_pas}), state augmentation plays a crucial role in reducing the discrepancy between predictions and observations, thereby enhancing accuracy. Hence, DA prediction using state augmentation and parametric optimisation is, in principle, more precise or equivalent to a flow solution obtained solely with a DA-optimised model. However, several factors affect the choices made for the DA strategy. One key criterion deals with the computational resources required. Simultaneously updating both the system's state and the model's coefficients is significantly more computationally expensive than optimising the model's free parameters alone. These differences can span several orders of magnitude. The primary reason for this is that the number of degrees of freedom resolved by the model $n_{DF}$ is usually significantly larger than the number of free parameters to be optimised $n_\theta$ ($n_{DF} \gg n_\theta$). In 3D simulations for incompressible flows without dimensionality reduction techniques like localisation \cite{Asch2016_siam}, $n_{DF}$ is three times the number of cells in the domain, which corresponds to the velocity vector in each grid element. On the other hand, $n_\theta$ typically consists of only a few values if a global optimisation of the models is performed. Additionally, the algebraic operations required by the DA algorithm are computationally intensive, compounded by the need to send and receive arrays containing a high number of data points and the intrusive modifications to the solver, both of which further increase the computation time. Besides, here, the elapsed simulation time $t_\textrm{sim}$ and the computation cost $C.C$ can differ significantly since the former only accounts for a single realisation (plus the DA algorithm occasionally), whereas the latter must include the calculation of the $N_e$ realisations, so $C.C. \approx N_e \,t_\textrm{sim}$.

\begin{figure}
    \centering
    \includegraphics[width=\linewidth,trim={1cm, 0, 0, 0}, clip]{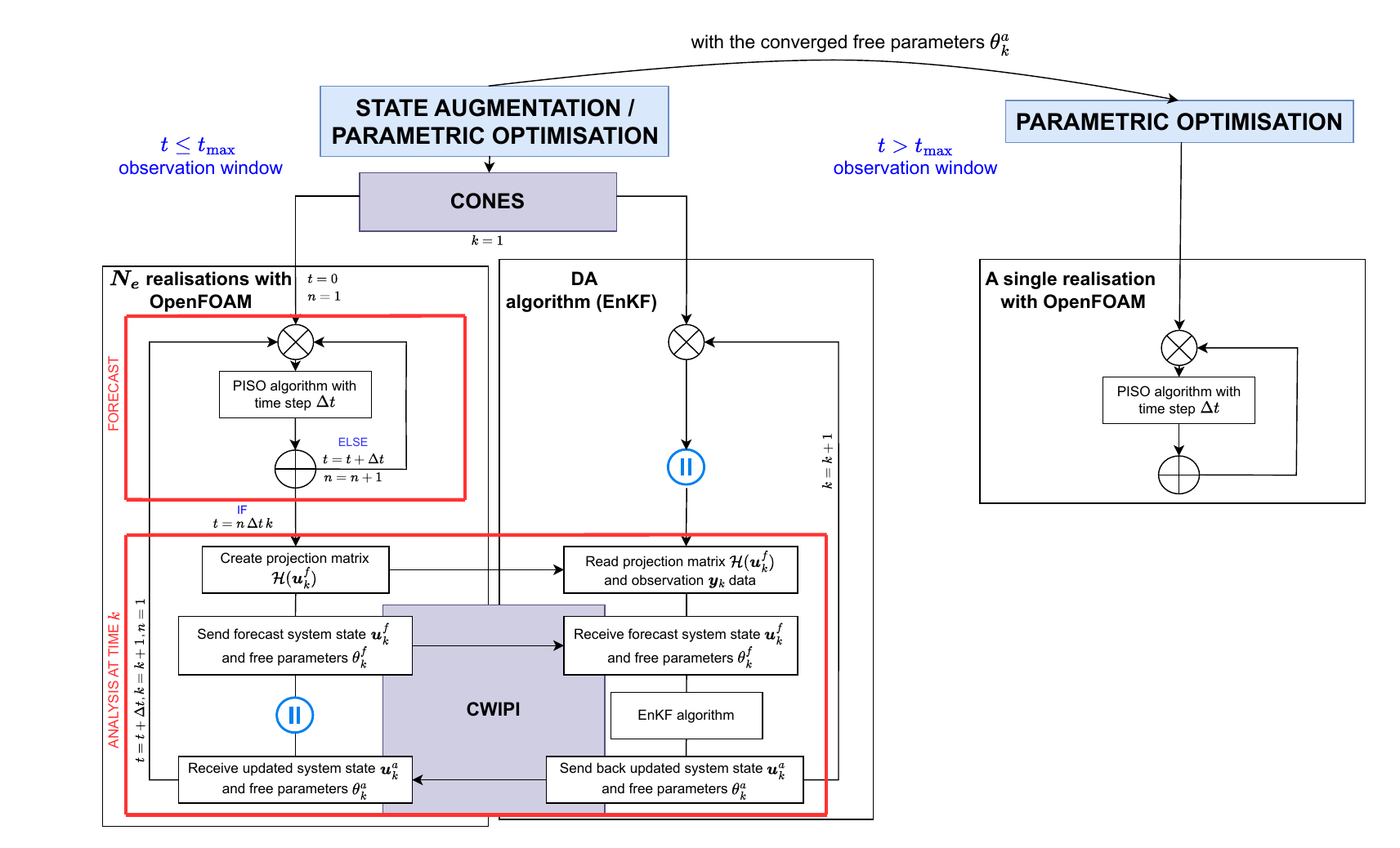}
    \caption{Difference between the state augmentation and parametric optimisation via the EnKF and the pure parametric optimisation once the model's free parameters $\theta_k^a$ are converged.}
    \label{fig:EnKF_se_po}
\end{figure}

A second criterion affecting the accuracy of the DA algorithm is that the synergistic combination of system augmentation and parametric optimisation requires the presence of high-fidelity data within the time interval considered during the simulation. If, for the sake of simplicity, one considers that the first analysis phase is performed for $t=0$, the complete DA scheme with state estimation and parametric optimisation can be performed for the observation window $t_{\boldsymbol{y}} \in [0, t_\textrm{max}]$, where $t_\textrm{max}$ corresponds to the last set of samples of the time series. For predictions beyond the observation window ($t > t_\textrm{max})$, the only way to enhance the flow is by replacing the original model's parameters with the updated free coefficients from the EnKF ($\theta_k^a$) and running a single realisation, as shown in Figure \ref{fig:EnKF_se_po}. 
In the case of the present analysis, a CFD run with the IBM model is first augmented using state estimation and parametric optimisation for $t \leq t_\textrm{max}$, and a CFD with optimised parameters for the local coefficients $D_{ij}$ is run for $t > t_\textrm{max}$. As the IBM model is not complete, one can expect a progressive degradation of the results for $t > t_\textrm{max}$, which is investigated in \S\ref{sec:OptiVSstate}.

\subsection{Machine Learning tools: Random Forest Regression}

The ML tools used in the present work are now described. The Random Forest Regression (RFR) \cite{Breiman2001_ML} is an ensemble learning method that builds multiple decision trees during training and outputs the average of their predictions to enhance accuracy and model robustness. It is widely applied in regression tasks, and it is known for its flexibility and ability to handle both numerical and categorical features. Combining bootstrapping and random feature selection, it aggregates the predictions of multiple decision trees, thus mitigating overfitting and enhancing model stability. Several critical steps are involved in generating decision trees within an RFR model. Initially, one must determine the number $R$ of decision trees to be included in the forest. Increasing the number of trees can improve the accuracy and convergence of the algorithm but also demands increased computational resources.

\begin{figure}[!h]
    \centering
    \includegraphics[width=\linewidth, trim={0, 1cm, 0, 0.5cm}, clip]{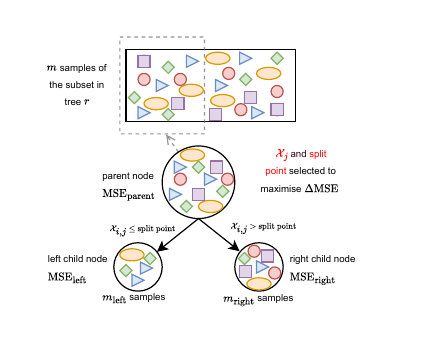}
    \caption{Schematic representation of a parent node and the criteria to split it into two child nodes for a random decision tree $r$.}
    \label{fig:RFR_MSE}
\end{figure}

 Figure \ref{fig:RFR_MSE} provides a schematic overview of the method. Each decision tree $r \in [1,...,R]$ is constructed using a subset of the entire training dataset and selects splitting criteria or nodes to progress in tree depth. Commonly, the splitting criterion aims to minimise the variance of the target variable within resulting subsets, i.e., the split point is chosen to minimise the Mean Square Error (MSE) of the child nodes (MSE\textsubscript{child}) and to maximise the reduction in MSE ($\Delta\textrm{MSE}$) from the parent to the child nodes, which is computed as the difference between the MSE in the parent node and the weighted sum of the MSEs in the resulting child nodes. These quantities are calculated using the following equations:
\begin{eqnarray}
    \textrm{MSE} &=& \frac{1}{m} \sum_{i = 1}^m \left(\mathcal{Y}_i - \hat{\mathcal{Y}_i} \right)^2 \\
    \Delta\textrm{MSE} &=& \mathrm{MSE_{parent}} - \overbrace{\left(\frac{m_{\mathrm{left}}}{m} \,\mathrm{MSE_{left}} + \frac{m_{\mathrm{right}}}{m} \,\mathrm{MSE_{right}} \right)}^{\text{MSE\textsubscript{child}}} 
\end{eqnarray}

Here, $m$ represents the number of samples of the parent node, $\mathcal{Y}_i$ and $\hat{\mathcal{Y}_i}$ indicate the actual and predicted target value for the sample $i$, and $m_{\mathrm{left}}$ and $m_{\mathrm{right}}$ refer to the number of samples in the left and right child nodes. The MSEs for the parent, left child, and right child nodes are $\mathrm{MSE_{parent}}$, $\mathrm{MSE_{left}}$ and $\mathrm{MSE_{right}}$, respectively. 

At each parent node, the dataset is partitioned into two subsets based on a specific feature $\mathcal{X}_j$ and split point, yielding two child nodes. The left child node typically represents data meeting the split criteria  (e.g., feature value $\mathcal{X}_{i,j} \leq$ split point), while the right child node does not (e.g., feature value $\mathcal{X}_{i,j} >$ split point).
Splitting nodes continue recursively for each child node until predefined stopping criteria are met, such as the maximum depth of each decision tree or the minimum number of samples required to split an internal node or form a leaf node. These hyperparameters play a crucial role in preventing overfitting.
By navigating these steps, RFR constructs a robust ensemble model capable of capturing complex relationships within data while maintaining generalisability. Additionally, it provides several advantages over deep learning architectures, such as neural networks (NNs), particularly in terms of interpretability and input preparation. Unlike NNs, it does not require feature normalisation and still delivers strong reliability. Moreover, NNs usually require larger datasets to attain comparable accuracy and are more susceptible to overfitting issues \cite{Roßbach, Wang2017_SIAM}. 

In the context of CFD, RFR has recently been employed in the prediction of 2D scenarios of turbulent flows via the blending of RANS models for a NACA profile \cite{deZordoBanliat2024_jfm} and other test cases like a turbulent flat plate, an axisymmetric subsonic jet, and a wall-mounted hump \cite{Cherroud2023}. Also, enhanced physics-informed RFR have been applied in the prediction of the Reynolds-stress anisotropy tensor for improved turbulence models in RANS simulations \cite{Kaandorp2020_cf, Wang2017_jcp, Heyse2021}.

\section{Performance of the EnKF when using finite observation time windows}
\label{sec:EnKF_limitations}

This section discusses the limitations of the EnKF when observation is available only over a limited time window. As previously mentioned, the availability of limited datasets over time can significantly impact the accuracy of flow predictions, particularly in scenarios involving complex modelling challenges, such as turbulence modelling and IBM. First, the test case of investigation is presented. The DA experiment is then detailed, and the sensitivity of the prediction of the EnKF to the availability of observation is assessed. In particular, the performance of the EnKF within the observation window $t \leq t_\textrm{max}$ (where state augmentation/estimation and parametric optimisation of the model can be performed) is compared with the accuracy obtained after the end of such time window $t> t_\textrm{max}$, where state estimation is not available anymore and the model advances in time using exclusively the DA-optimised free parameters.   

\subsection{Test case: turbulent plane channel flow, $Re_\tau \approx 550$}
\label{sec:test_case}

\begin{figure}
    \begin{tabular}{cc}
    \includegraphics[width=0.58\linewidth]{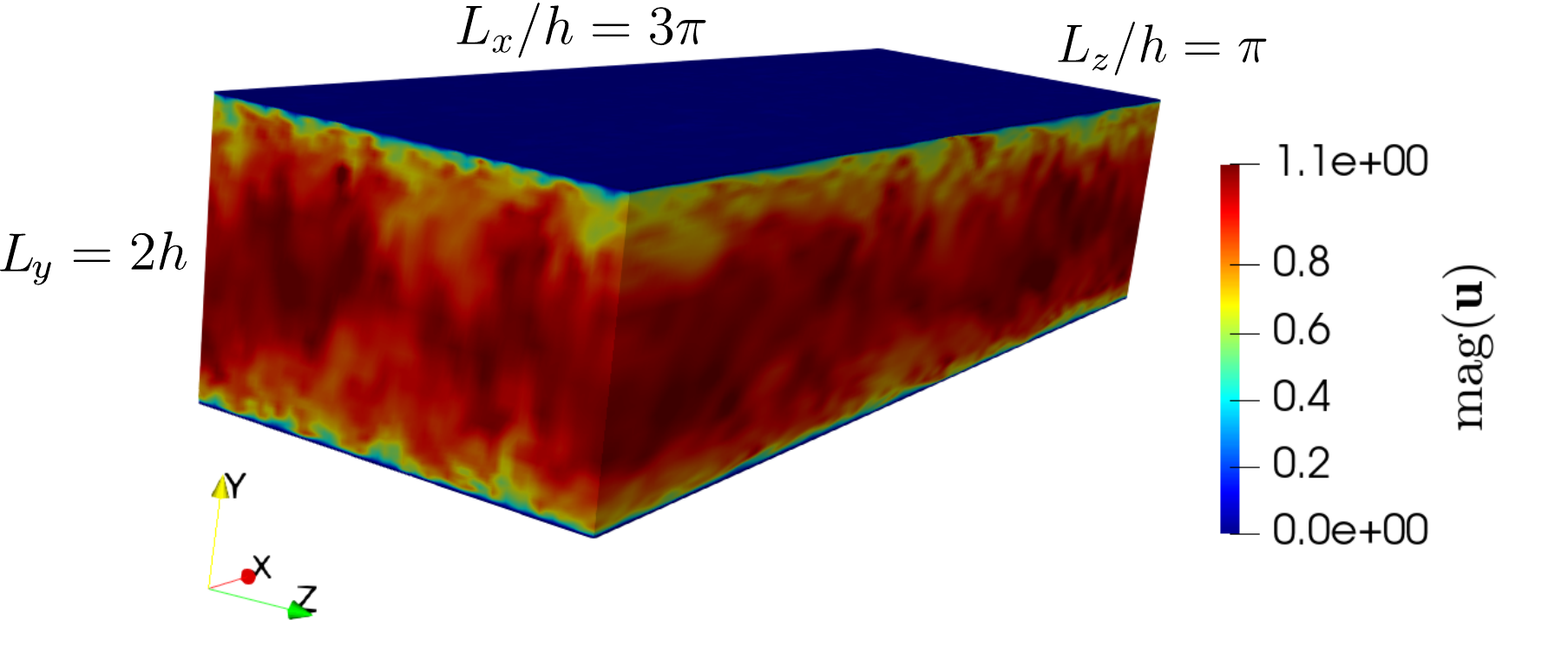} & \includegraphics[width=0.42\linewidth, 
    trim={2cm, 0, 0, 0}, clip]{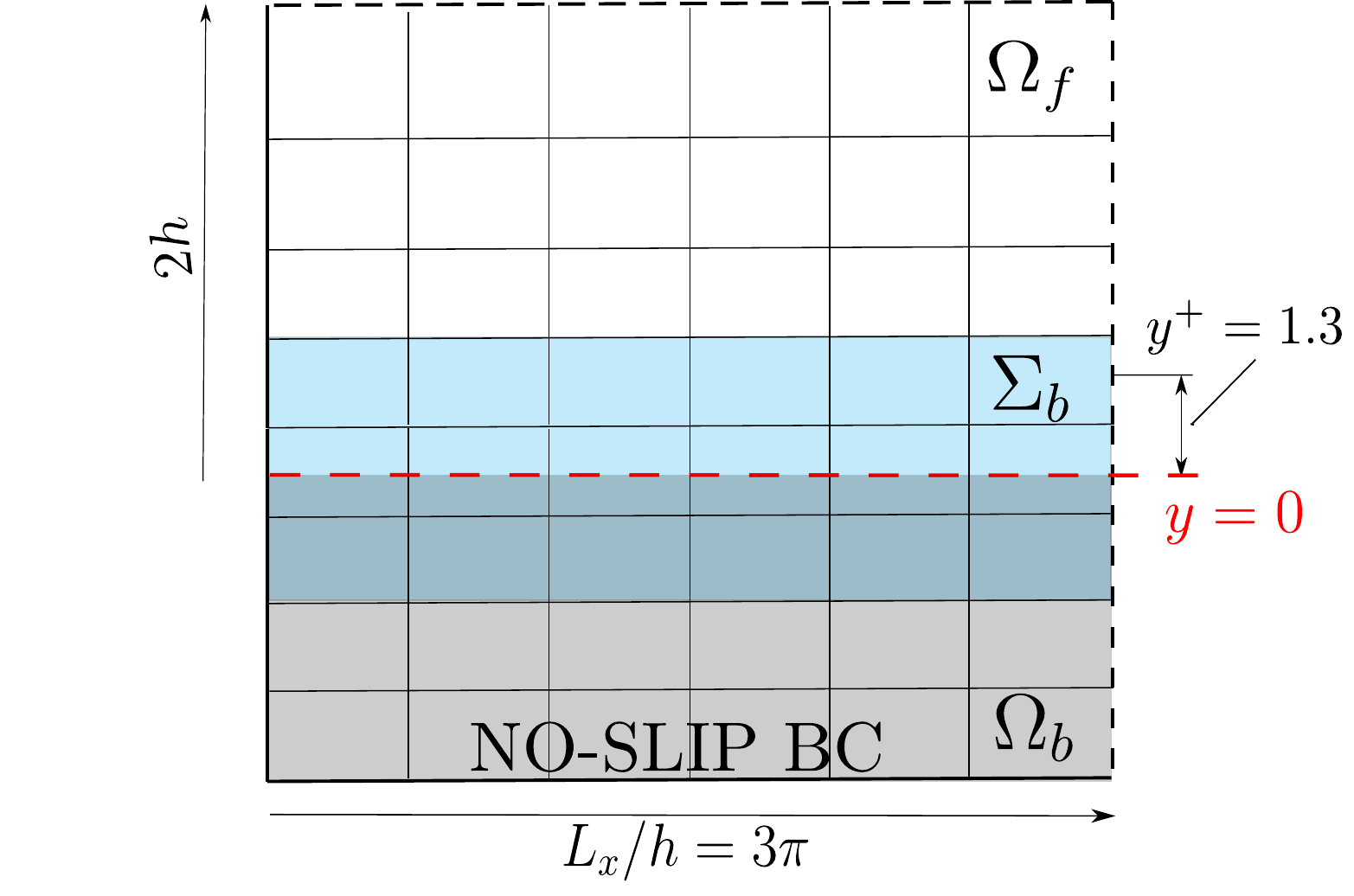} \\
    \textit{(a)} & \textit{(b)} \\
    \end{tabular}
    \caption{\textit{(a)} Scheme of the turbulent plane channel flow, and \textit{(b)} distribution of the cell layers in the wall-normal direction for the solid region $\Omega_b$, the fluid region $\Omega_f$, and the interface regions $\Sigma_b$ regions for the IBM simulations.}
    \label{fig:planeChannel}
\end{figure}

The turbulent plane channel flow is a cornerstone benchmark test case in Computational Fluid Dynamics (CFD), offering readily accessible high-fidelity Direct Numerical Simulations (DNS) open databases online \cite{Kim1987_jfm, DelAlamo2003_pof, Hoyas2008_pof, Bernardini2014_jfm, Cimarelli2015_jfm}. A fully developed turbulent flow oriented in the streamwise direction $x$ is investigated between two parallel static walls positioned at $y = 0$ and $y = 2h$. Here, $y$ denotes the wall-normal direction, $z$ is the spanwise direction, and $h$ is the half-height of the channel. A scheme of the flow is found in Figure \ref{fig:planeChannel}\textit{(a)}.

The friction Reynolds number here investigated is $Re_\tau = u_\tau h /\nu \approx 550$, where $u_\tau = \sqrt{\tau_w / \rho}$ represents the friction velocity derived from the wall shear stress $\tau_w$. A uniform mass flow rate over time is obtained by integrating a source term into the dynamic equations already available in the libraries of OpenFOAM. Periodic boundary conditions are used for the lateral boundaries. A coarse-grained body-fitted DNS, named hereafter DNS-BF, is conducted with high wall resolution, achieving a wall unit value $y^+ \approx 1$. Additionally, a coarse-grained simulation with penalty IBM, called DNS-IBM, is simulated with a grid very similar to the one employed for DNS-BF. Notable differences lie in the treatment of the near-wall region: both simulations incorporate the immersed/physical walls at $y = 0, 2h$, yet for the DNS-IBM run, these positions correspond to the centre of the fourth Eulerian mesh element starting from the outer limit of the computational domain, where a no-slip boundary condition is imposed. As shown in Figure \ref{fig:planeChannel}\textit{(b)}, this configuration results in two cell layers in the wall-normal direction representing the solid $\Omega_b$ region of the domain (grey colour), while three cell layers depict the interface $\Sigma_b$ between the fluid and solid regions (blue colour). These additional layers ensure numerical stability. As expressed in Equation \ref{eqn:forcePenDarcy}, a penalisation term $\boldsymbol{f}_P$ is included in the dynamic equations if the position of the considered grid cell is $\boldsymbol{x} \in \Sigma_b \cup \Omega_b$. The penalisation coefficients $D_{ij}$ included in the tensor $\boldsymbol{D} = D \,\boldsymbol{I}$ are set with a uniform value of $D = 10^5$ in the mesh elements where $\boldsymbol{x} \in \Sigma_b \cup \Omega_b$, and are set to zero elsewhere, following the work by Valero \& Meldi \cite{Valero2023}. To enforce the zero-velocity condition at the immersed walls, $\boldsymbol{u_{ib}} = \boldsymbol{0}$.

In the aforementioned work, a strategy based on the EnKF was proposed to combine high-fidelity data sampled from a well-refined body-fitted DNS (referred to as R-DNS-BF) with the lower-fidelity DNS-IBM model previously presented. It was demonstrated that considering an observation window of $300\,t_A$, where $t_A$ is the advective time defined for this test case as $t_A = h/U_c$, with $U_c$ the time-averaged velocity at the centreline, the DA algorithm successfully achieves accurate flow prediction. This result is accomplished by combining the EnKF state estimation with the local optimisation of the IBM model coefficients $D_{ij}$. In this case, variations of the coefficients $D_{ij}$ in the wall-normal direction $y$ were investigated, which required the optimisation of five coefficients for each component of the volume forcing (see Figure \ref{fig:planeChannel}\textit{(b)}). This last optimisation task reduces the bias of the model, which, therefore, can follow the time evolution of the observed quantities with increased accuracy. However, in cases where the structural limitations of the model can preclude an exact representation of the test case investigated, even with an optimised parametric description, the update obtained by state estimation can also be beneficial in obtaining increased accuracy. The analysis in \S\ref{sec:OptiVSstate} aims to identify the contribution of these two DA concurring aspects.    

\subsection{Accuracy of the parametric optimisation using the EnKF}
\label{sec:OptiVSstate}

The differences in accuracy between the full DA algorithm (which combines state estimation and parametric optimisation) and numerical simulations with optimised DA models are now discussed for the test case of interest. These two DA-augmented IBM runs are referred to as DNS-IBM-DA\textsubscript{s.e.} and DNS-IBM-DA\textsubscript{p.o.}, respectively. 
In the first case, data from the high fidelity simulation R-DNS-BF (in the form of instantaneous values of the friction coefficient $C_f$ and the wall velocity $\boldsymbol{u}=\boldsymbol{0}$ on several wall sensors) are combined with IBM realisations during $2\,500$ analysis phases, which are regularly spaced over an observation window of $300\, t_A$. During these analysis phases, the flow field is updated via state estimation, and the parameters $D_{ij}$ of the IBM are optimised. The complete results of the investigation are provided and discussed in the work by Valero \& Meldi \cite{Valero2023}. For the case DNS-IBM-DA\textsubscript{p.o.}, a single IBM run is performed, using the optimised coefficients $D_{ij}$ determined by the case DNS-IBM-DA\textsubscript{s.e.}. One could see that the case DNS-IBM-DA\textsubscript{p.o.} corresponds to a numerical simulation performed after a DA investigation if, for any reason, the stream of data from the targeted application is interrupted. 

\begin{figure}
    \begin{tabular}{ccc}
    \includegraphics[width=0.32\linewidth]{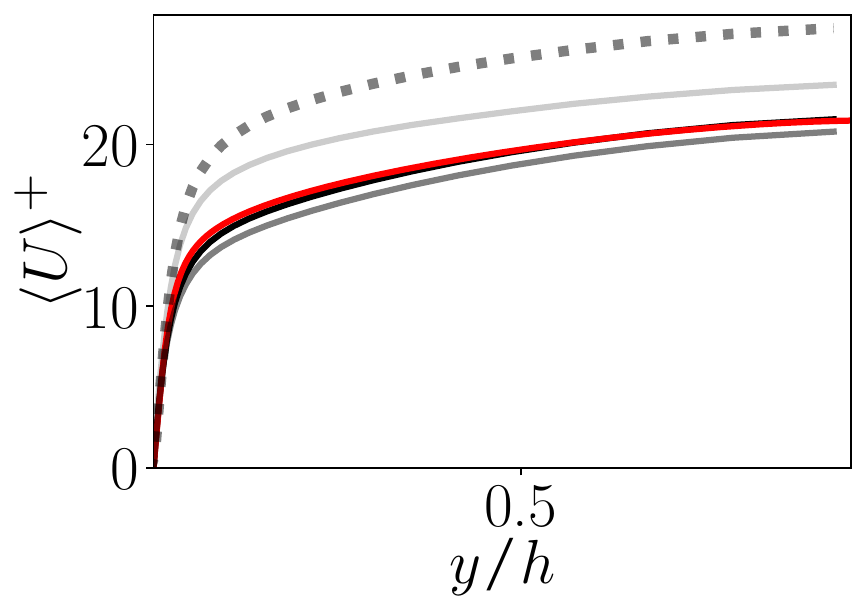} & \includegraphics[width=0.32\linewidth]{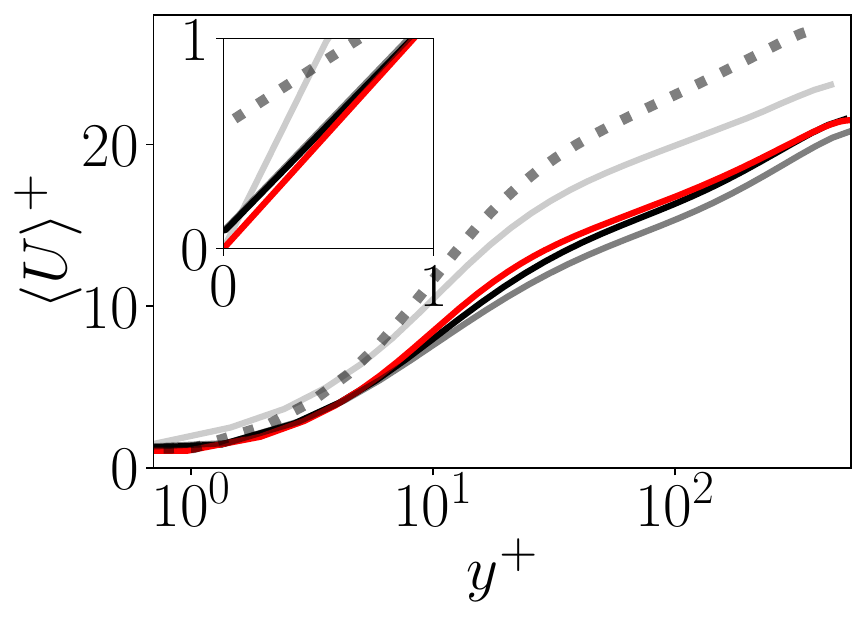} &
    \includegraphics[width=0.32\linewidth]{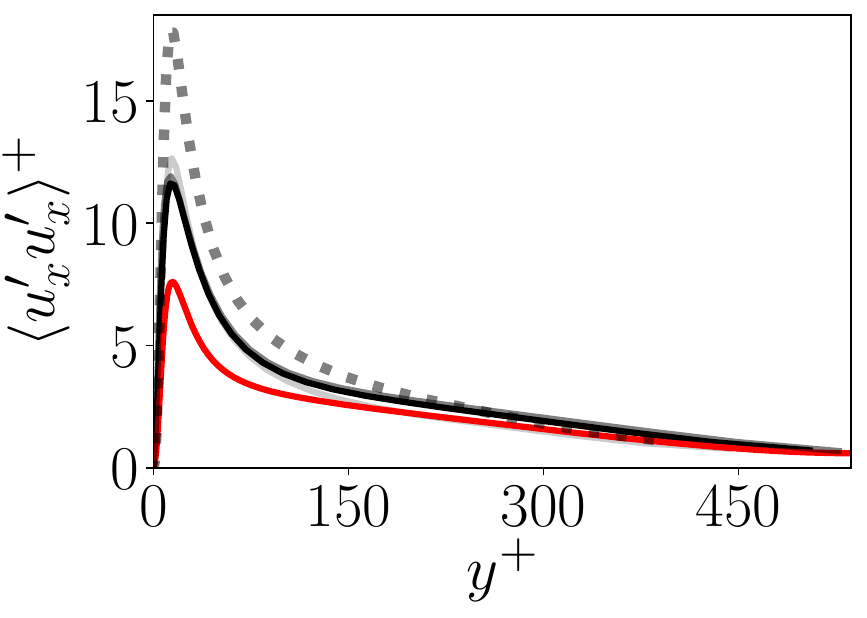} \\
    \textit{(a)} & \textit{(b)} & \textit{(c)} \\
    \includegraphics[width=0.32\linewidth]{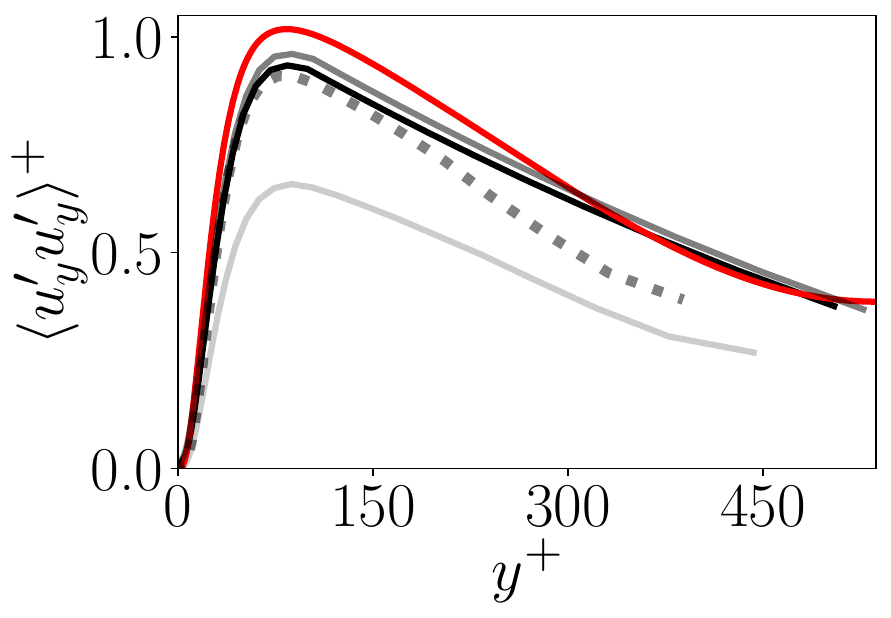} & \includegraphics[width=0.32\linewidth, height=3cm]{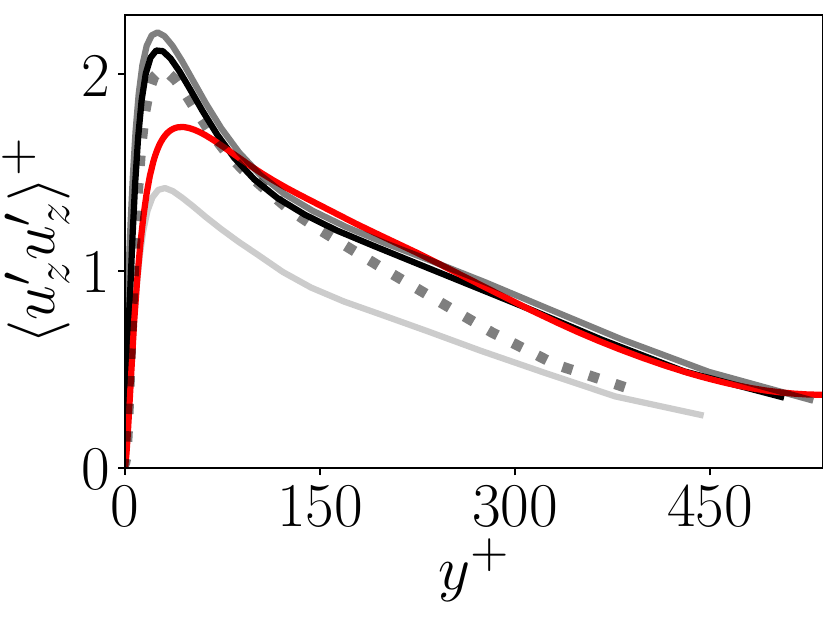} &
    \includegraphics[width=0.32\linewidth]{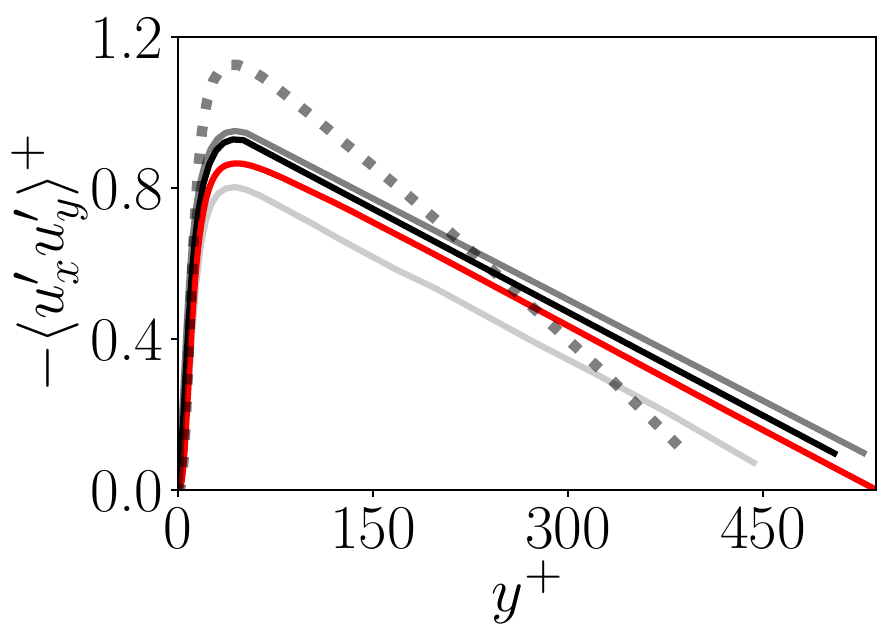} \\
    \textit{(d)} & \textit{(e)} & \textit{(f)}
    \end{tabular}
    \caption{Comparison of the main statistical moments of the velocity field. Results are shown for simulations (\protect\blacklinesolid) DNS-IBM-DA\textsubscript{s.e.}, (\protect\greylinesolidstrong) DNS-IBM-DA\textsubscript{p.o.}, (\protect\greylinesolidsoft) DNS-BF, (\protect\greylinedottedstrong) DNS-IBM, and (\protect\redline) R-DNS-BF.}
    \label{fig:DA_initial}
\end{figure}

The results of this study are illustrated in Figure \ref{fig:DA_initial}, where the data are presented in non-dimensional form $\langle \boldsymbol{\cdot} \rangle^+$ by employing $u_\tau$ and the viscous wall unit $\delta_\nu = \nu / u_\tau$ for normalisation. The $+$ suffix indicates normalisations using the $u_\tau$ calculated by each simulation, while the $\star$ suffix indicates normalisation using the $u_\tau$ obtained by the simulation R-DNS-IBM. Statistical homogeneity is observed across the $x$ and $z$ directions, so the statistical moments are averaged over these two directions. Further details on the test cases can be found in Table \ref{tab:summary1}. The mean velocity profile $\langle U \rangle^+ = \langle u_x \rangle / u_\tau$ and the components of the Reynolds stress tensor $\langle u_i^\prime \, u_j^\prime \rangle^+ = \langle u_i^\prime \, u_j^\prime \rangle / u_\tau^2$ of the simulation R-DNS-BF, which is here used as a reference, were already validated with datasets in the literature \cite{DelAlamo2003_pof, Hoyas2008_pof} in our previous work \cite{Valero2023}. 

One can see in Figure \ref{fig:DA_initial}\textit{(a)}--\textit{(b)} that the mean flow predicted by the data-informed (DNS-IBM-DA\textsubscript{s.e.}) run is extremely close to the reference simulation, unlike the DNS-BF and DNS-IBM simulations. The combined effect of state estimation and parametric optimisation is, therefore, successful in obtaining an accurate prediction. More precisely, DA not only ensures more precise enforcement of the no-slip condition at the immersed wall compared to the pure IBM simulation but also enhances the friction velocity $u_\tau$, with a discrepancy of only $1.46\%$ when compared with the value obtained by the R-DNS-BF run. These enhancements are achieved without incurring the high computational costs $C.C$ that were required for the R-DNS-BF run. In Table \ref{tab:summary1}, $C.C$ is normalised over the computational costs of the coarse-grained body-fitted simulation, indicated as $C.C^\star = C.C / C.C_{\textrm{DNS-BF}}$. As highlighted in \S\ref{sec:IBM_DA}, it is important to observe that state estimation demands significantly higher computational costs compared to parametric optimisation alone---over an order of magnitude difference---due to the additional overhead of the EnKF algorithm and the need to run an ensemble of simulations. Furthermore, when comparing the curves between DNS-IBM-DA\textsubscript{s.e.} and DNS-IBM-DA\textsubscript{p.o.}, a clear degradation is observed in the latter. While its accuracy is still significantly better than the one achieved by the simulations DNS-BF and DNS-IBM, the results show the effect of the cumulative errors produced by the model bias discussed in previous sections, preventing the achievement of the same level of accuracy that was observed for the DA solution with state estimation.

Evaluating the second-order statistics, as illustrated in Figure \ref{fig:DA_initial}\textit{(c)}--\textit{(f)}, presents a nuanced challenge. Overall, subtle improvements are observed in the components of the Reynolds stress tensor with the DA procedure, with minimal distinction between DNS-IBM-DA\textsubscript{s.e.} and DNS-IBM-DA\textsubscript{p.o.}. The limited differences between the two DA strategies arise because the second-order fields are passively updated when local velocity values serve as observations. In both approaches, enhanced accuracy is noted in $\langle u_y^\prime u_y^\prime \rangle^+$, while slight improvements are evident in $\langle u_x^\prime u_x^\prime \rangle^+$ and $\langle u_x^\prime u_y^\prime \rangle^+$. Conversely, minor degradations are observed in $\langle u_z^\prime u_z^\prime \rangle^+$. Upon comparison, the amalgamation of state estimation with parametric optimisation generally yields more precise curves than parametric optimisation alone, except for $\langle u_y^\prime u_y^\prime \rangle^+$.

One can see that the effect of state estimation is, therefore, beneficial for the global accuracy of the solution, in particular when structural deficiencies of the model may preclude a completely satisfying optimisation of the underlying models. This is actually the case for most applications in turbulent flows. In the following, we develop ML techniques that target the emulation of the DA procedures. 

\begin{table}
  \begin{center}
\def~{\hphantom{0}}
\scalebox{0.75}{
  \begin{tabular}{lcccccccccc}
       & $N_x \times N_y \times N_z$ & $\Delta x^\star$ &  $\Delta z^\star$ & $\Delta y^\star_{\text{min}}$ & $\Delta y^\star_{\text{max}} $ & $L_x/h$ & $L_z/h$ & $u_\tau$ & $C.C^\star$ \\
       & & & & & & & & \\
       R-DNS-BF & $1\,024 \times 256 \times 512$ & $9.9$ & $6.6$ & $1$ & $11.2$ & $6\pi$ & $2\pi$ & $0.0480$ & $\approx12k$ \\
       & & & & & & & & \\
       DNS-BF & $128 \times 61 \times 64$ & $39.5$ & $26.3$ & $1.08$ & $52$ & $3\pi$ & $\pi$ & $0.0430$ & $1$ \\
       DNS-IBM & $128 \times 64 \times 64$ & $39.5$ & $26.3$ & $1.3$ & $52$ & $3\pi$ & $\pi$ & $0.0375$ & $1.03$ \\
       & & & & & & & & \\
       DNS-IBM-DA\textsubscript{s.e.} & \multirow{2}{*}{$128 \times 64 \times 64$} & \multirow{2}{*}{$39.5$} & \multirow{2}{*}{$26.3$} & \multirow{2}{*}{$1.3$} & \multirow{2}{*}{$52$} & \multirow{2}{*}{$3\pi$} & \multirow{2}{*}{$\pi$} & $0.0487$ & $37.60^\dag$ \\
       DNS-IBM-DA\textsubscript{p.o.} & & & & & & & & $0.0509$ & $1.03$ \\
       & & & & & & & & \\
       DNS-IBM-ML\textsubscript{p.o.} & \multirow{2}{*}{$128 \times 64 \times 64$} & \multirow{2}{*}{$39.5$} & \multirow{2}{*}{$26.3$} & \multirow{2}{*}{$1.3$} & \multirow{2}{*}{$52$} & \multirow{2}{*}{$3\pi$} & \multirow{2}{*}{$\pi$} & $0.0516$ & $7.03$ \\
       DNS-IBM-ML\textsubscript{s.e.} & & & & & & & & $0.0493$ & $8.78$
  \end{tabular}}
  \caption{Summary of the numerical runs considered for the study. $^\dag$For DNS-IBM-DA\textsubscript{s.e.}, the computational costs are also normalised with respect to the ensemble members $N_e$ to represent the elapsed simulation time. This normalisation is expressed as $C.C^\star / (N_e + 1)$, where $C.C^\star$ represents the computational costs, and an additional CPU core is assumed for performing the analysis phases with the EnKF.}
  \label{tab:summary1}
  \end{center}
\end{table}

\section{Machine Learning for state estimation and parametric optimisation}
\label{sec:ML_stateEstimation}

The usage of DA and ML tools in the framework of scale-resolving streaming data is now presented. The resulting models can be trained and potentially updated online when observation for DA is available, and they can be used as a DA surrogate when observation is not available. Two strategies are proposed here. The first strategy, which is conceptually the simplest, focuses on the DA parametric optimisation phase. Specifically, a black-box IBM model, hereafter referred to as IBM\textsubscript{RFR}, is designed to emulate the DA-augmented IBM tool. The second ML strategy, termed SE\textsubscript{RFR}, captures the nonlinear effects of state estimation, offering the capability to update the flow state during time windows when observations are unavailable.    

\subsection{Strategy one: Machine Learning experiment to obtain a black-box IBM tool}
\label{sec:ML_parameterOpt}

\begin{figure}
    \begin{tabular}{ccc}
        \includegraphics[scale=0.32]{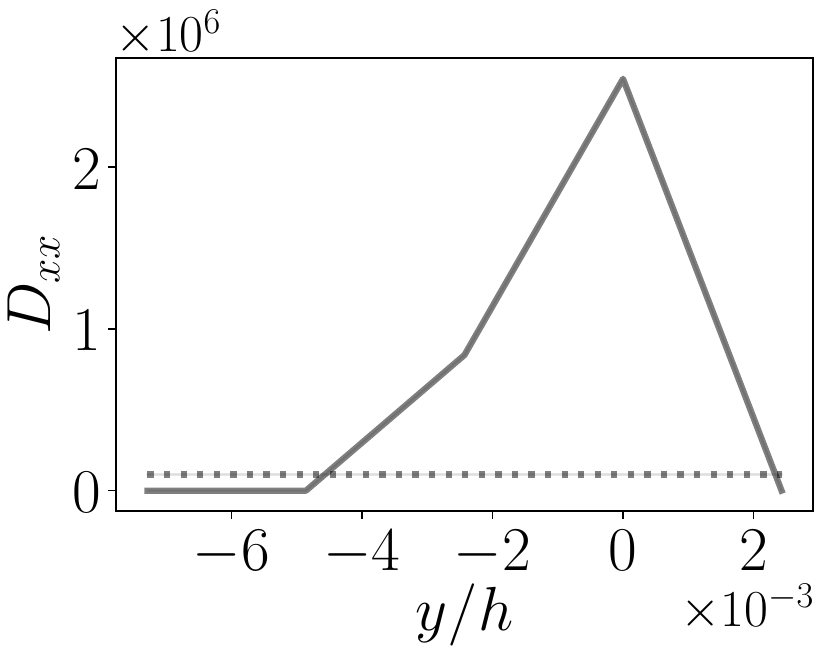}
        \includegraphics[scale=0.32]{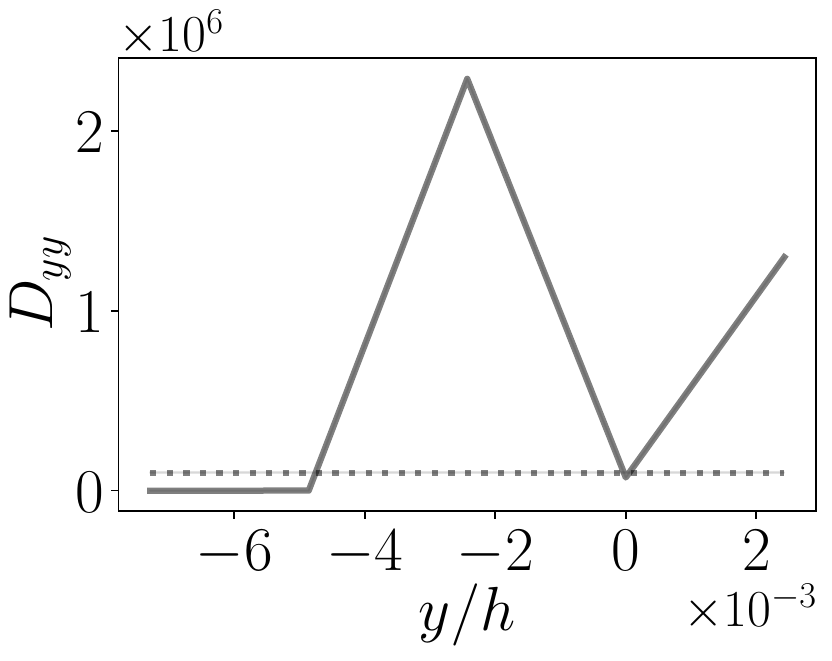}
        \includegraphics[scale=0.32]{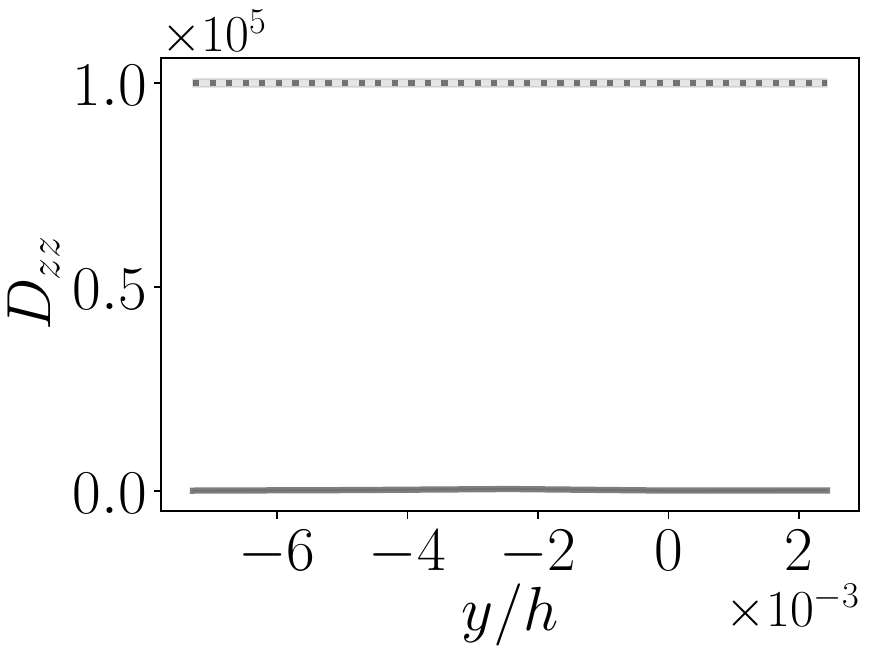}   
    \end{tabular}
    \caption{Penalty coefficients $D_{ij}$ improved by means of (\protect\greylinesolidstrong) DNS-IBM-DA\textsubscript{s.e.}, concerning the ones employed in the classical penalisation (\protect\greylinedottedstrong) DNS-IBM. The standard deviation $\sigma_{ij}$ of the \emph{prior} for the initial Gaussian perturbation of the ensemble is represented in light grey ($\sigma_{ij} = 0.05\, D_{ij}$).}
    \label{fig:coefficientsD}
\end{figure}

The ML strategy proposed here mimics the parametric optimisation procedure performed by the EnKF in order to obtain a black-box tool providing the same performance. DA is an efficient complementary tool for ML, as the instantaneous flow data can be used to train the algorithm. Figure \ref{fig:coefficientsD} illustrates the values of the coefficients $D_{ij}$ employed in the IBM simulations. One can see the differences produced by the DA optimisation process when comparing the prior state of simulation DNS-IBM and the final state used for the run DNS-IBM-DA\textsubscript{p.o.}. By exploiting the rich databases obtained by the DA algorithm, a comprehensive database of local penalty forcing values $\boldsymbol{f}_P$ can be obtained over several time steps. This quantity serves as an output for the black-box ML tool, and therefore it is selected as labelled data or dependent variable $\mathcal{Y}$. The features or independent variables $\mathcal{X}$ selected for this database include the local velocity field values $\boldsymbol{u} = ( u_x, u_y, u_z )$ and the position of the considered cell layer in the wall-normal direction $y/h$. An axial symmetry condition is enforced at half-height of the channel, where mesh elements near the top wall $y_{\textrm{top}}/h$ contribute to the database via the expression $y/h = 2 - y_{\textrm{top}}/h$. Given the abundance of continuous data, leveraging the Random Forest Regression (RFR) ensemble supervised learning method appears to be a promising starting point. This method offers resilience against overfitting and outliers compared to other regression algorithms like linear regression and can effectively handle non-linear relationships in the data. Implementation is simplified through the C++ library \emph{dlib} \cite{King2009_jmlr}, making integration into the CFD solvers of OpenFOAM straightforward. Also, it employs all available CPU cores of the system to serialise the ML models, optimising costs during the learning procedure.

Preliminary tests have shown that, from a few million sampling data, the influence of hyperparameters on training quality is not particularly strong. Consequently, hyperparameter values are selected with a focus on preserving computational efficiency. For the test cases investigated here, the number of decision trees is set to $R = 100$, where each individual decision tree considers a third of the whole dataset for split criteria in the parent nodes. The minimum sampling size to form a leaf node is set to $5$, mitigating the risk of overfitting. 

The training data consists of $76$ consecutive instantaneous snapshots of one of the ensemble members of the run DNS-IBM-DA\textsubscript{s.e.}, which are obtained over a time span of $1.5 \,t_A$, with a time step of $\Delta t = 0.02\, t_A$. The time window for sampling here is set in the range $[298.5 \,t_A, \, 300 \,t_A]$ in order to grant convergence of the optimisation of the coefficients $D_{ij}$. The choice of the specific ensemble member is not relevant as each of them is a different Monte-Carlo realisation of the system uncertainty, and they all provide the same statistical information. The dataset used for training consists of $m_T = 6\,225\,920$ sampling data points. The training process is conducted online, generating three distinct models to predict each of the components of the forcing term $\hat{\boldsymbol{f}}_P = (\hat{f}_{P_x}, \hat{f}_{P_y}, \hat{f}_{P_z})$.

\begin{figure}
    \begin{tabular}{ccc}
        \includegraphics[scale=0.31]{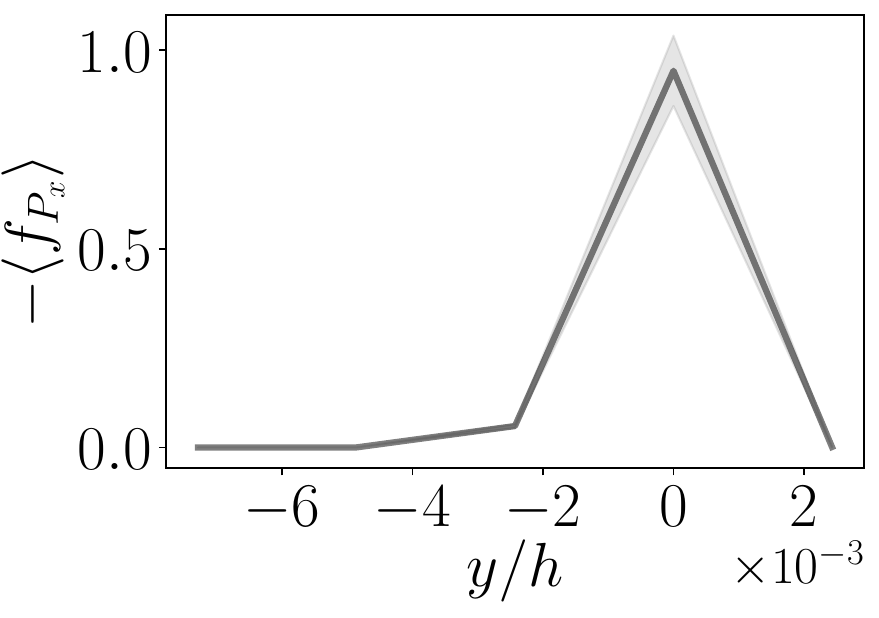}
        \includegraphics[scale=0.31]{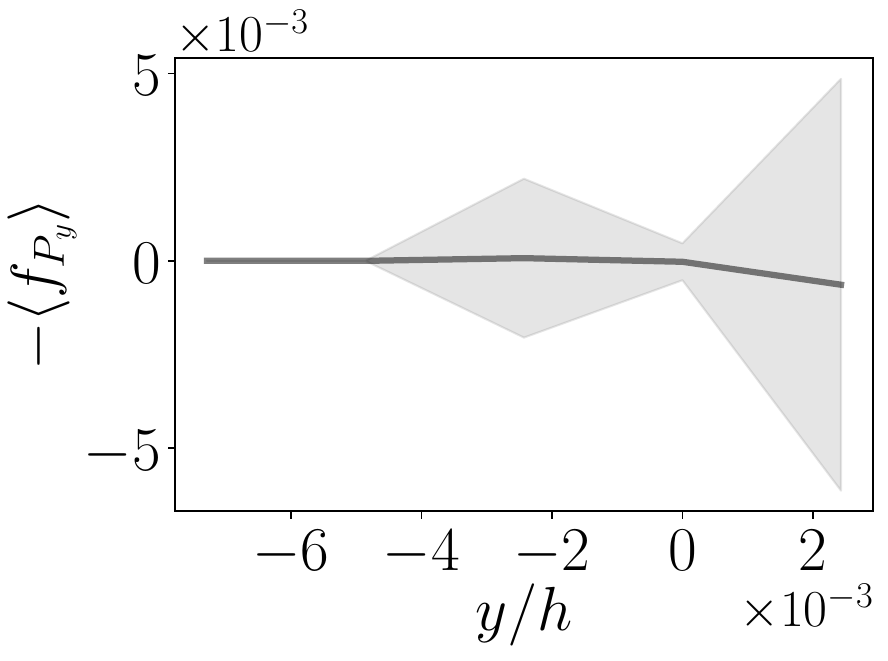}
        \includegraphics[scale=0.31]{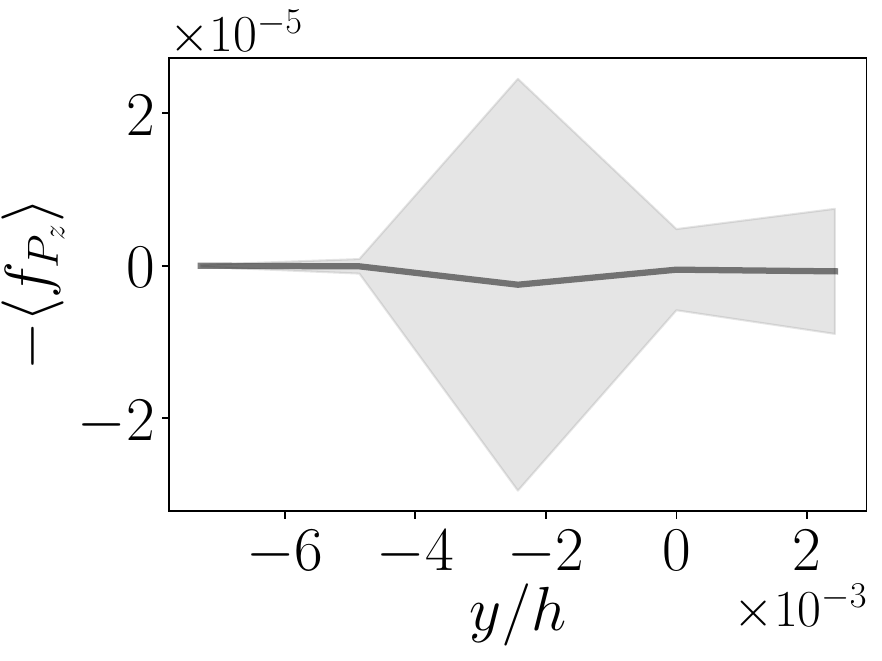}   
    \end{tabular}
    \caption{Averaged penalty source term $\langle \boldsymbol{f}_P \rangle$ of the database employed for training. In light grey, we represent the standard deviation.}
    \label{fig:coefficientsF}
\end{figure}

The model performance is assessed via two metrics utilising $p = 20\%$ of the entire training dataset $m_T$: the coefficient of determination $\mathcal{R}^2$ to evaluate how well the model captures the variance in data and the Normalised Root Mean Square Error (NRMSE) to provide insight into the average of the prediction errors. For normalisation, selecting the interquartile range---$Q_1$ and $Q_3$ representing the $25$\textsuperscript{th} and $75$\textsuperscript{th} percentiles of the actual penalty force $\boldsymbol{f}_P$---is advantageous due to its robustness against outliers. Their expressions are respectively:

\begin{eqnarray}
    \mathcal{R}^2 &=& 1 - \frac{\sum_{i=1}^p (\mathcal{Y}_i - \hat{\mathcal{Y}_i})^2} {\sum_{i=1}^p (\mathcal{Y}_i - \overline{\mathcal{Y}})^2} \\
    \textrm{NRMSE} &=& \frac{1}{Q_3 - Q_1} \sqrt{\frac{\sum_{i=1}^p (\mathcal{Y}_i - \hat{\mathcal{Y}}_i)^2}{p}}
\end{eqnarray}
Preliminary investigations and tests dealing with the ML training show that the convergence of the obtained black-tool IBM depends on the volume forcing component. More precisely, the forcing component working in the streamwise $x$ direction, which is the largest in magnitude, exhibits significantly better diagnostics than the other two components. 
This suggests a significant presence of noise in the samples in the wall-normal $y$ and spanwise $z$ directions. This observation is corroborated by Figure \ref{fig:coefficientsF}, where several lower orders of magnitude are evident for the source term in these directions compared to the streamwise value. 
Therefore, the decision is made to solely predict the streamwise component of the forcing, yielding $\hat{\boldsymbol{f}}_P = (\hat{f}_{P_x}, 0, 0)$, defining this particular simulation as DNS-IBM-ML\textsubscript{p.o.}.


\begin{table}
  \begin{center}
\def~{\hphantom{0}}
\scalebox{1}{
  \begin{tabular}{lccc}
       &
       \multicolumn{3}{c}{IBM\textsubscript{RFR}} \\
       & $\hat{f}_{P_x}$ & $\hat{f}_{P_y}$ & $\hat{f}_{P_z}$ \\
       & & & \\
       $\mathcal{R}^2$ & $99.72\%$ & $67.02\%$ & $88.28\%$ \\
       NRMSE & $25.38\%$ & $324.40\%$ & $55.71\%$
  \end{tabular}}
  \caption{Metrics computed for each component of the forcing term $\hat{\boldsymbol{f}}_P$.}
  \label{tab:metrics}
  \end{center}
\end{table}

This model incorporates the normalisation of the features and the labelled data, ensuring its applicability across diverse conditions that are not only limited to the training dataset. This is performed, first, by scaling the features $(\boldsymbol{\cdot})^\ast$ relative to the global friction velocity determined through DA with parametric optimisation, i.e., $u_{\tau_{\textrm{DA/p.o.}}} = 0.0509$ (refer to Table \ref{tab:summary1}). Consequently, for each local sample indexed by $i$, the feature vector is $\mathcal{X}_i = (u_{x_i}^\ast, u_{y_i}^\ast, u_{z_i}^\ast,  y_i^\ast) = \left(u_{x_i} / {u_{\tau_{\textrm{DA/p.o.}}}}, u_{y_i} / {u_{\tau_{\textrm{DA/p.o.}}}}, u_{z_i} / {u_{\tau_{\textrm{DA/p.o.}}}}, (y_i/h) \,{u_{\tau_{\textrm{DA/p.o.}}}} / \nu  \right)$. Second, the output of the model consists of the forcing term normalised by considering $\boldsymbol{f}_{P_i}^\ast = \boldsymbol{f}_{P_i}\,\delta_{\nu_{\textrm{DA/p.o.}}} / u_{\tau_{\textrm{DA/p.o.}}}^2$.

\begin{figure}
    \centering
    \includegraphics[scale=0.9]{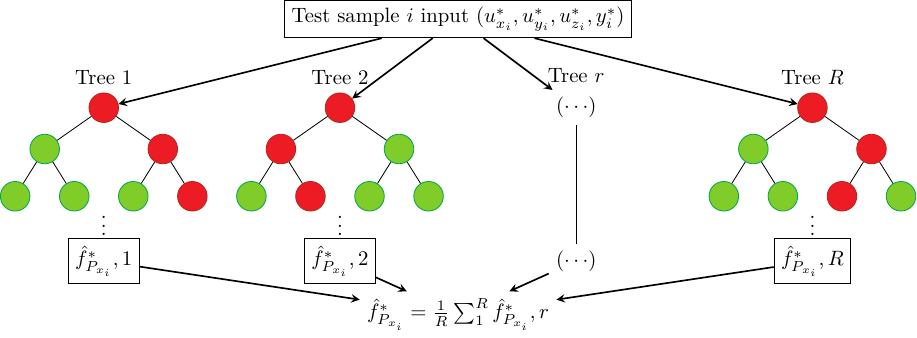}
    \caption{Overview of RFR algorithm for the prediction of the streamwise forcing term $\hat{f}_{P_x}$.}
    \label{fig:RFR_Scheme}
\end{figure}

The generation of the black-box IBM via the RFR model (IBM\textsubscript{RFR}) is shown in Figure \ref{fig:RFR_Scheme} for the simulation DNS-IBM-ML\textsubscript{p.o.}. In this representation, each local sample $i$ of the training dataset includes the instantaneous velocity values and the wall-normal distance at the cell centre of the respective mesh element under consideration. The predicted force $\hat{\boldsymbol{f}}_P^\ast$ is estimated and converted to $\hat{\boldsymbol{f}}_P$ \textit{on the fly} within the CFD formalism (see \ref{sec:RFR_PISO} for more details about the implementation of RFR inside the PISO algorithm).

\subsection{Validation of Random Forest Regression in the training dataset}
\label{sec:conditions_dataset}

\begin{figure}
    \begin{tabular}{ccc}
    \includegraphics[width=0.32\linewidth]{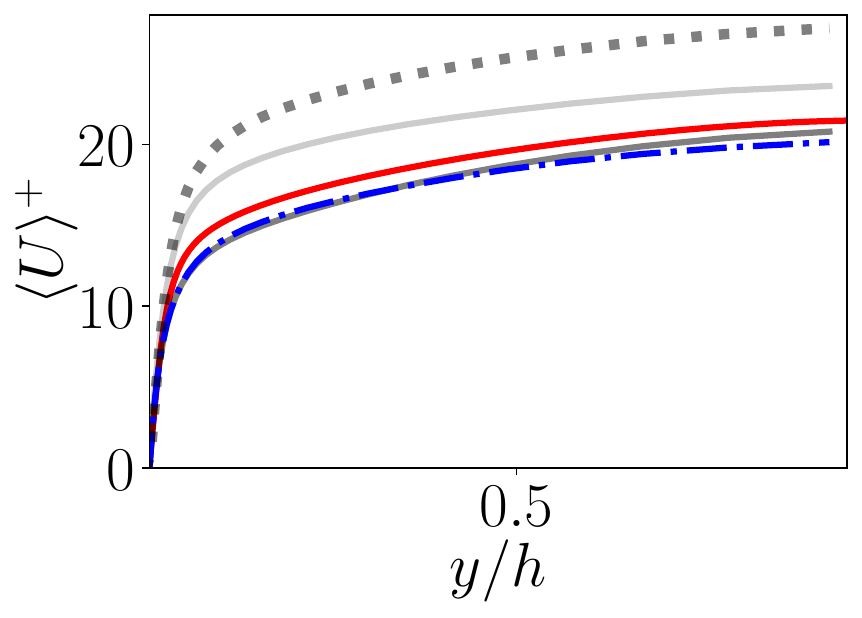} & 
    \includegraphics[width=0.32\linewidth]{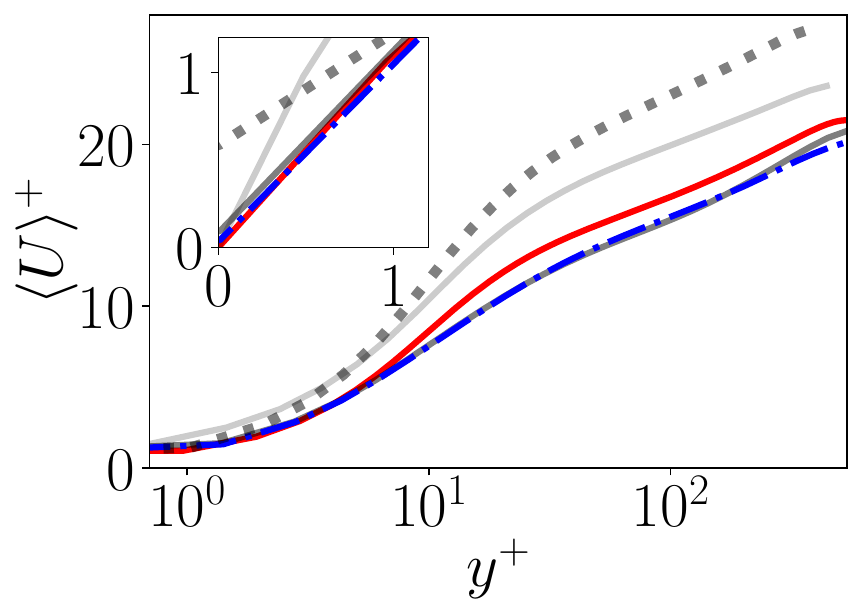} &
    \includegraphics[width=0.32\linewidth]{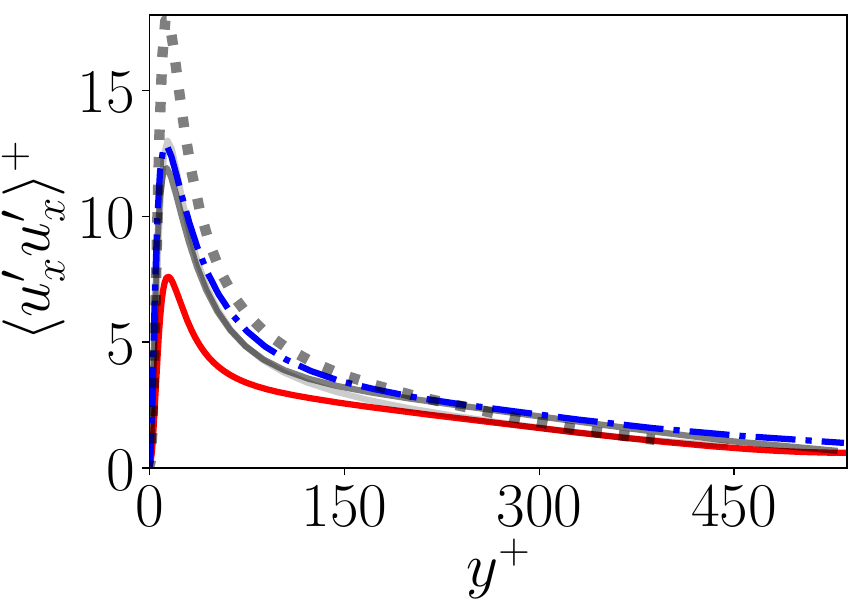} \\
    \textit{(a)} & \textit{(b)} & \textit{(c)} \\
    \includegraphics[width=0.32\linewidth]{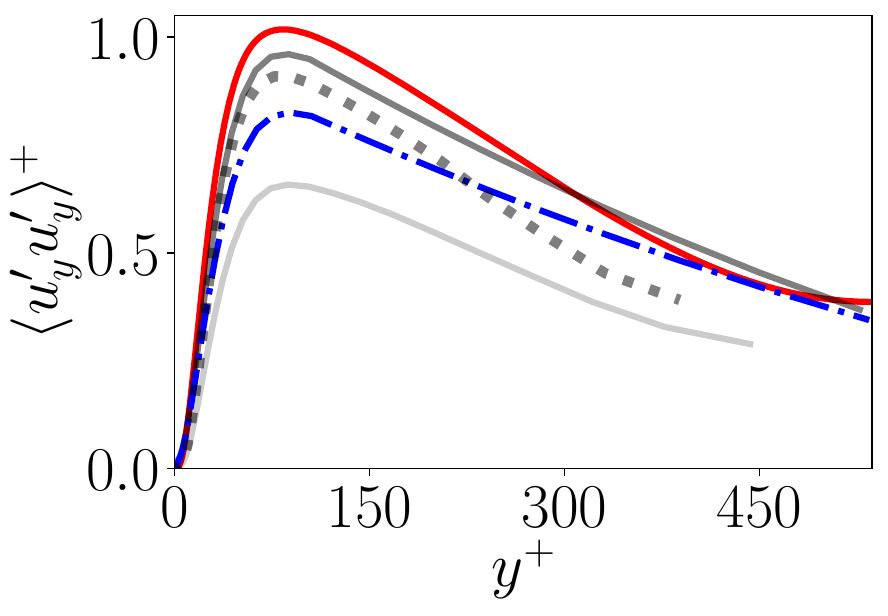} & \includegraphics[width=0.32\linewidth, height=3cm]{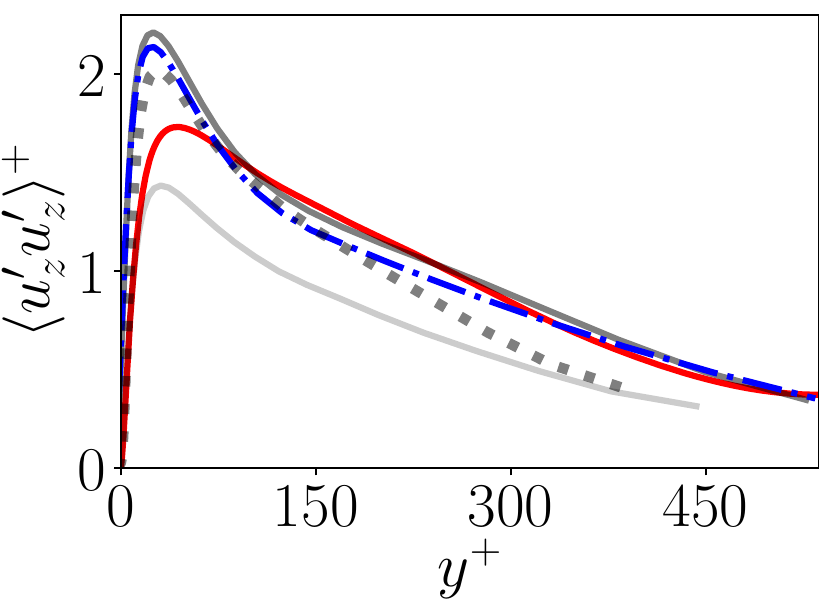} &
    \includegraphics[width=0.32\linewidth]{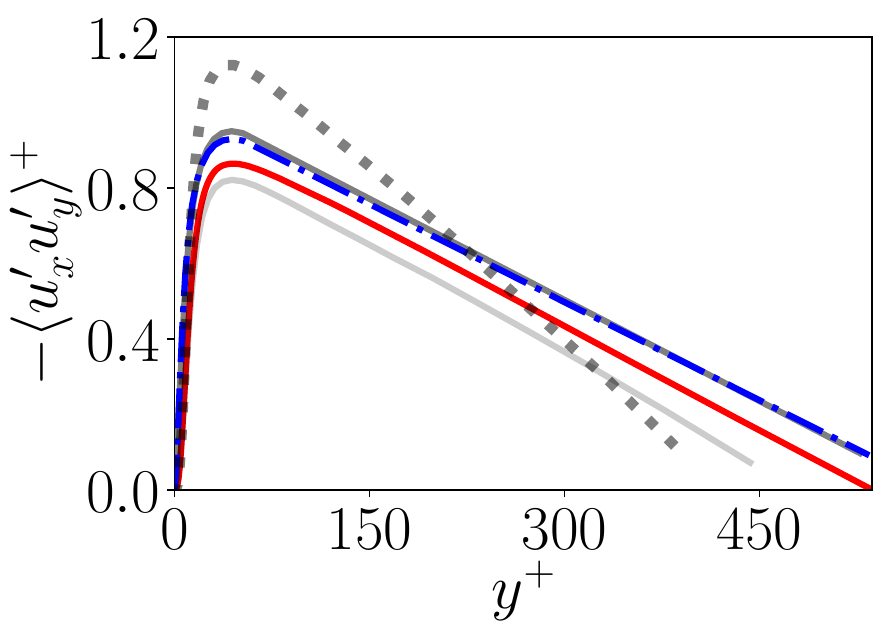} \\
    \textit{(d)} & \textit{(e)} & \textit{(f)}
    \end{tabular}
    \caption{Comparison of the main statistical moments of the velocity field. Results are shown for simulations (\protect\bluelinedasheddotted) DNS-IBM-ML\textsubscript{p.o.}, (\protect\greylinesolidstrong) DNS-IBM-DA\textsubscript{p.o.}, (\protect\greylinesolidsoft) DNS-BF, (\protect\greylinedottedstrong) DNS-IBM, and (\protect\redline) R-DNS-BF.}
    \label{fig:ML_initial}
\end{figure}

The black-box IBM\textsubscript{RFR} is now assessed using conditions consistent with the training data. The analysis of first-order and second-order statistics, which are shown in Figure \ref{fig:ML_initial}, indicates that the performance achieved by the ML-generated model closely matches that of DA with parametric optimisation, albeit with a slight degradation observed for the component $\langle u_y^\prime u_y^\prime \rangle^+$. In any case, the black-box IBM is systematically more accurate than the coarse-grained body-fitted (DNS-BF), and IBM runs for most of the statistical moments. This highlights the excellent convergence of ML training thanks to the richness of data provided by the DA algorithm. In addition, as shown in Table \ref{tab:summary1}, the simulation elapsed time is approximately five times smaller in DNS-IBM-ML\textsubscript{p.o.} compared to DNS-IBM-DA\textsubscript{s.e.}. This discrepancy increases to roughly $5\,(N_e+1)$ when factoring in total CPU time/cost $C.C^\star$. Also, as shown in Figure \ref{fig:ML_initial}\textit{(b)}, the zero-velocity condition at the wall is properly constraint and minimal discrepancies are observed for the prediction of $u_\tau$ (around $1.38\%$) when compared with the run DNS-IBM-DA\textsubscript{s.e.}.

\begin{figure}
\centering
\begin{tabular}{cc}
    \includegraphics[width=0.5\linewidth, trim = {5cm 1cm 17cm 1cm}, clip]{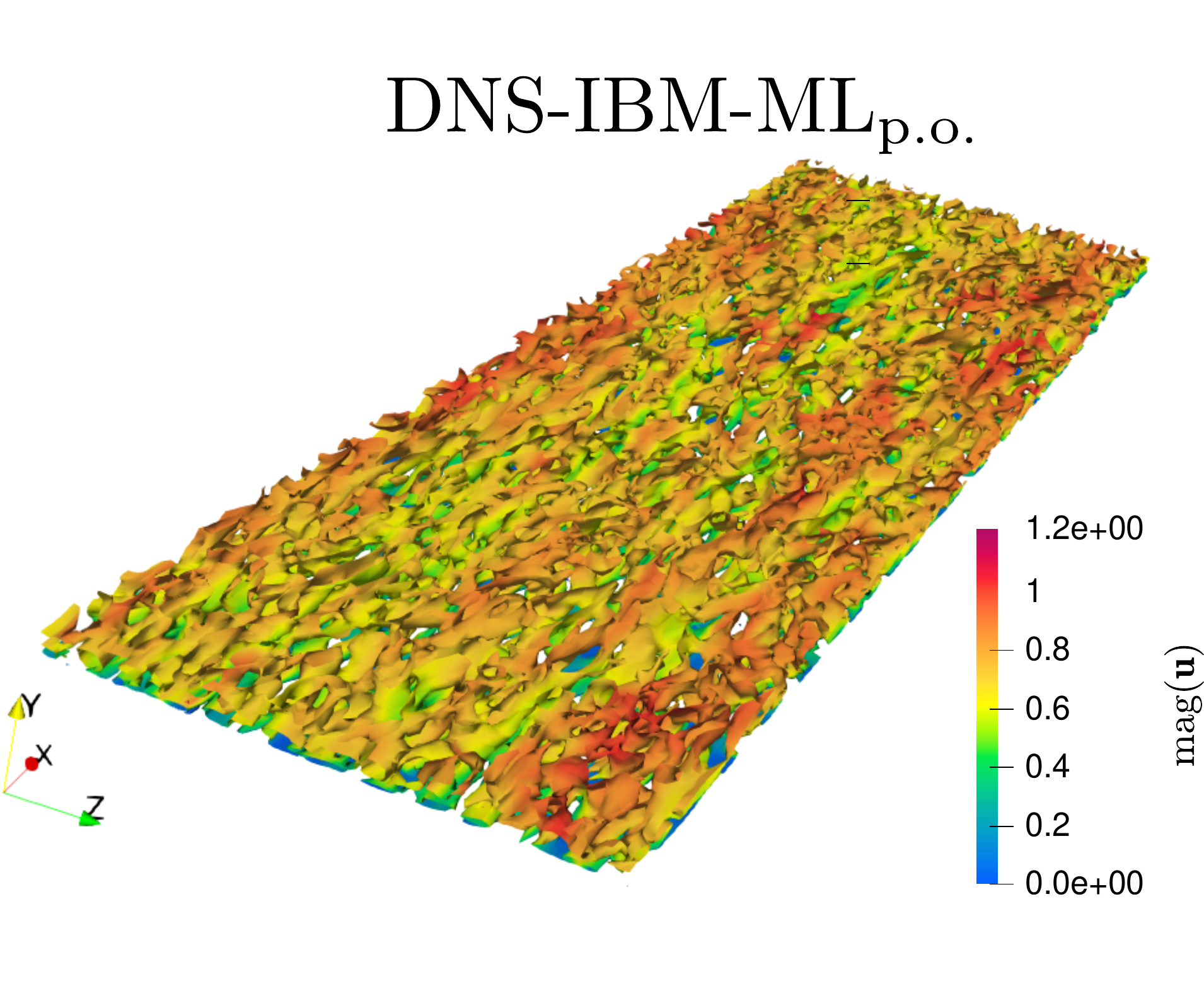} & 
    \includegraphics[width=0.5\linewidth, trim = {5cm 1cm 17cm 1cm}, clip]{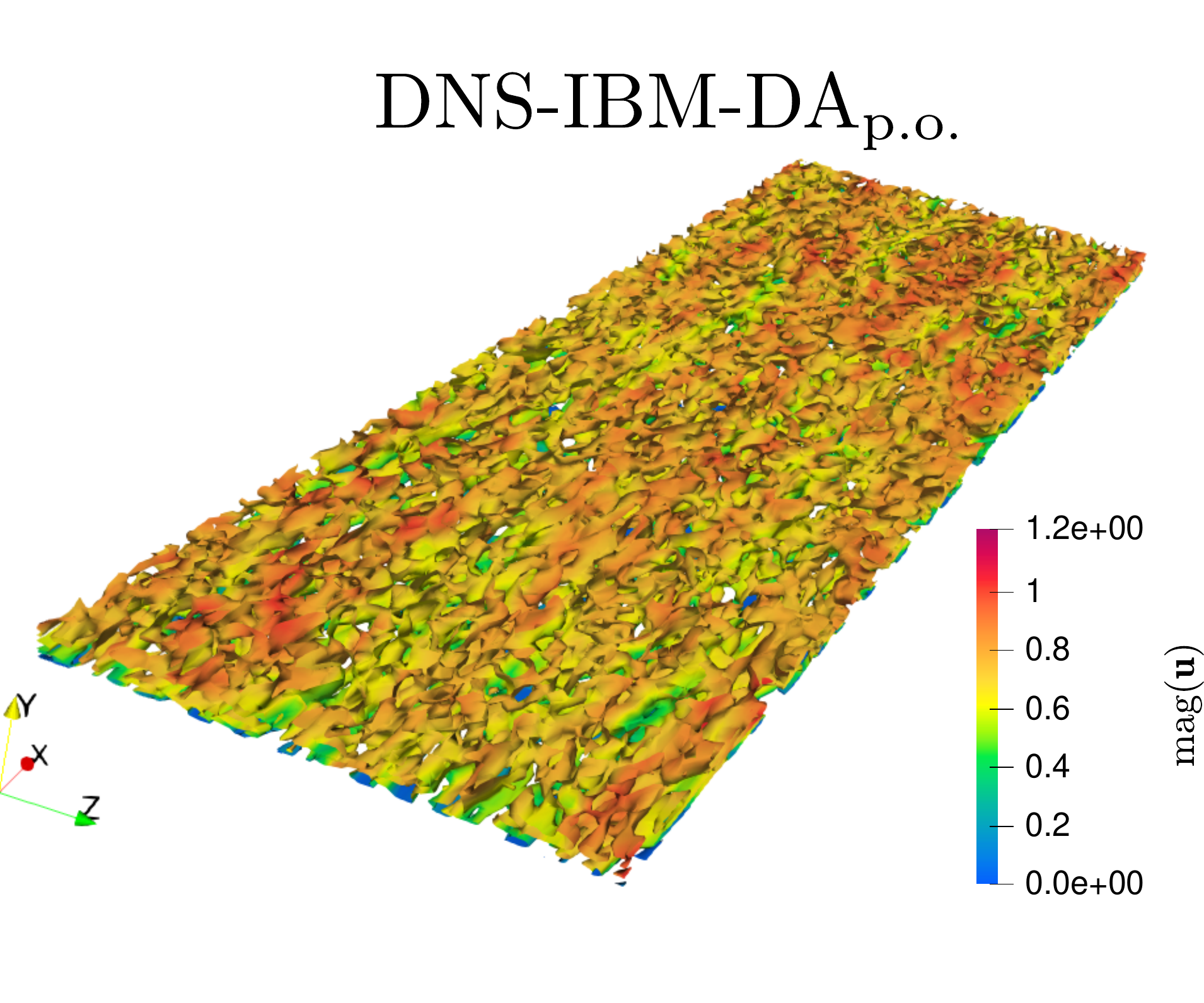} \\
    \textit{(a)} & \textit{(b)} \\
    \includegraphics[width=0.5\linewidth, trim = {5cm 1cm 17cm 1cm}, clip]{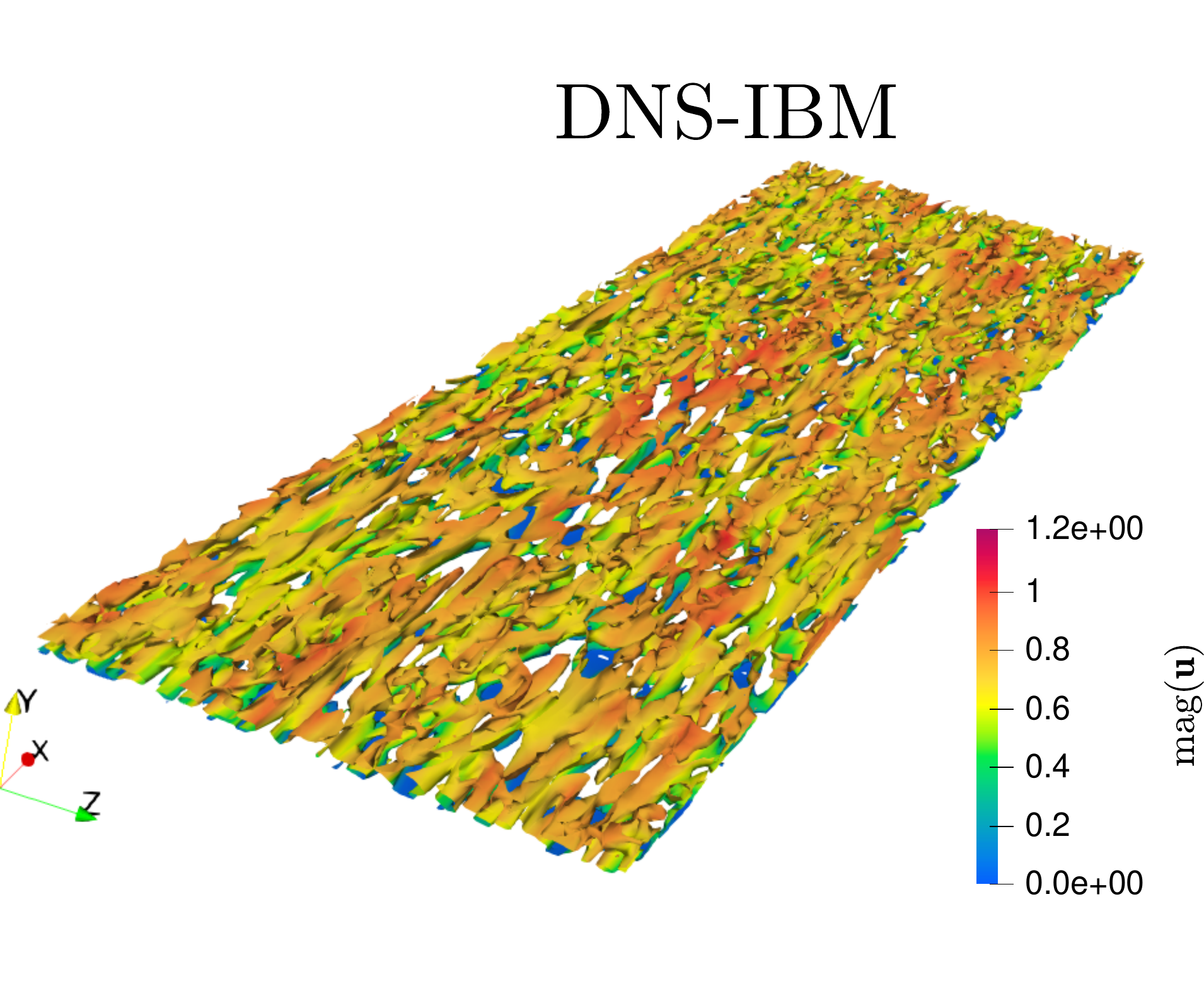} & 
    \includegraphics[width=0.5\linewidth, trim = {5cm 1cm 17cm 1cm}, clip]{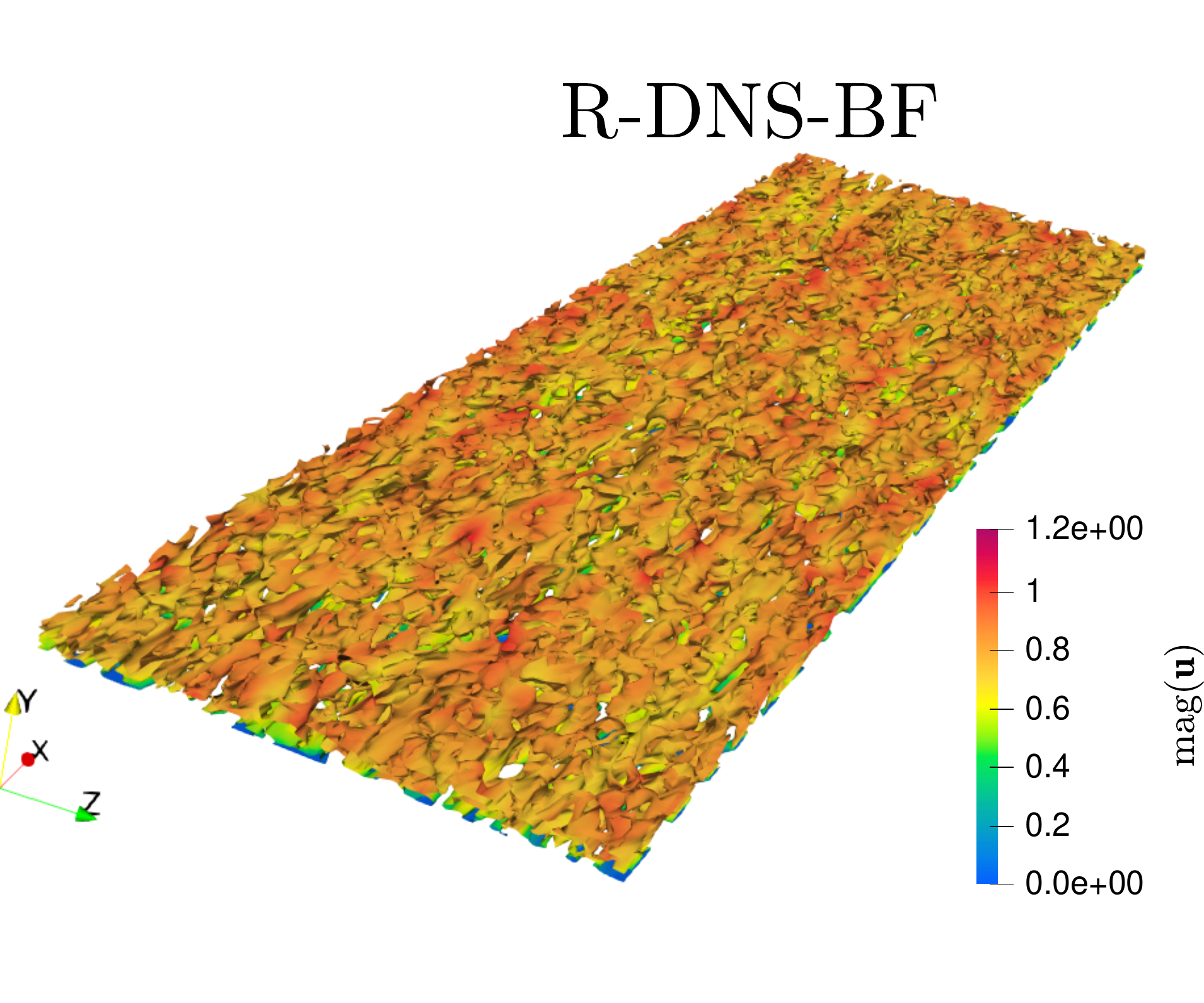} \\
    \textit{(c)} & \textit{(d)} 
    \end{tabular} 
\caption{Isocontours of Q-criterion calculated for $y/h = 0.18$ ($\Delta y^\star \approx 100$). The outcomes are presented for simulations: \textit{(a)} DNS-IBM-ML\textsubscript{p.o.}, \textit{(b)} DNS-IBM-DA\textsubscript{p.o.}, \textit{(c)} DNS-IBM, and \textit{(d)} R-DNS-BF.}
\label{fig:Qcriterion}
\end{figure}

The investigation of the instantaneous flow patterns shown in Figure \ref{fig:Qcriterion}, highlights a close similarity between the Q-criterion near-wall isocontours of DNS-IBM-ML\textsubscript{p.o.} and those of the DA strategy. The training performed using instantaneous flow data allows the IBM\textsubscript{RFR} model to obtain a qualitative correct representation of the main flow structures. 

\begin{figure}
    \centering
    \begin{tabular}{ccc}
    \includegraphics[width=0.32\linewidth]{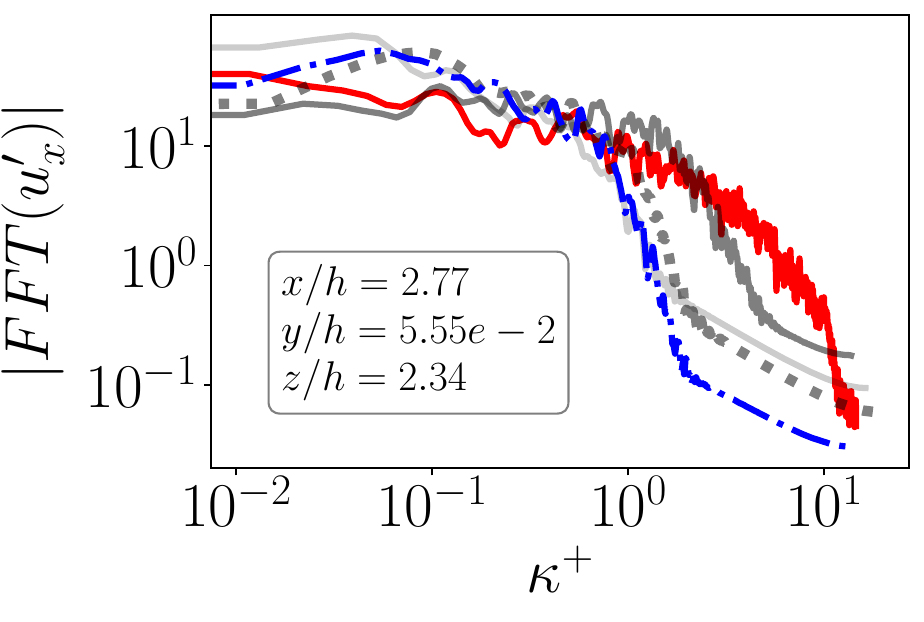} & \includegraphics[width=0.32\linewidth]{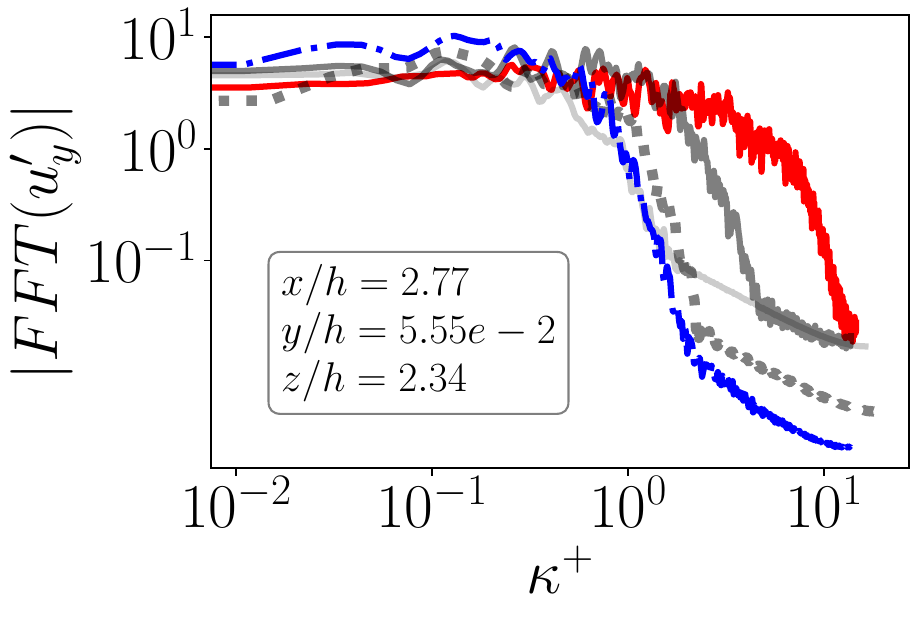} &
    \includegraphics[width=0.32\linewidth]{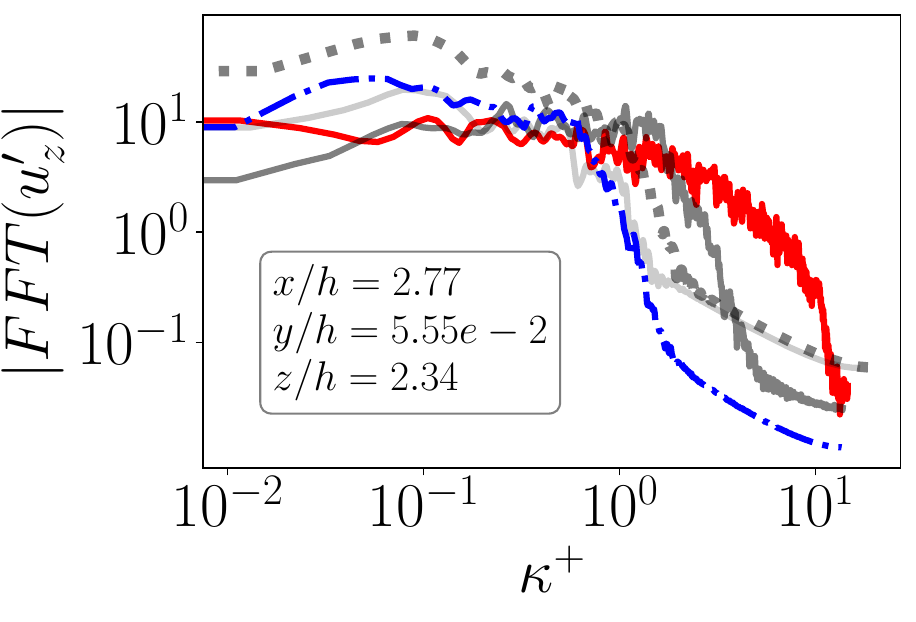} \\
    \textit{(a)} & \textit{(b)} & \textit{(c)} \\
    \includegraphics[width=0.32\linewidth]{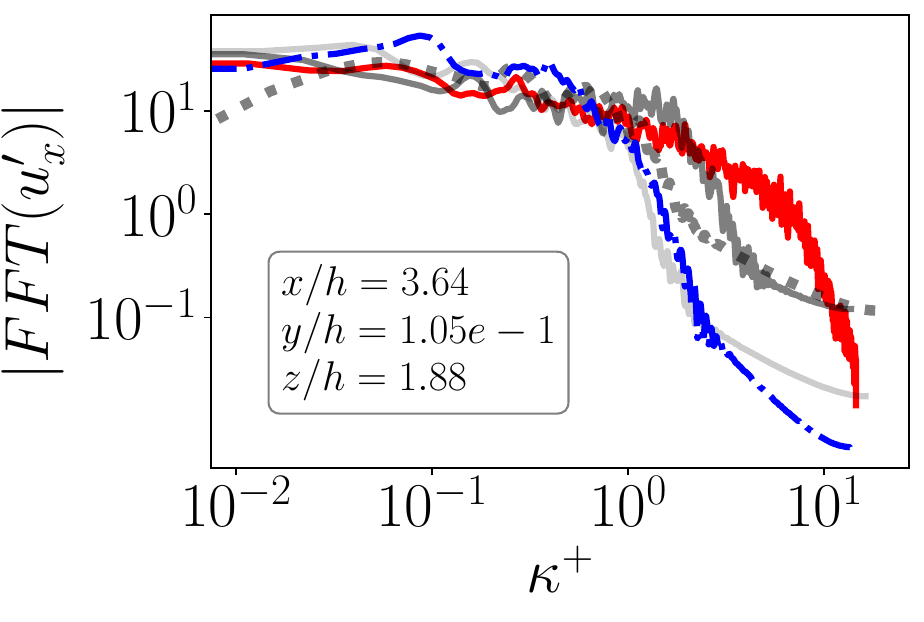} & \includegraphics[width=0.32\linewidth]{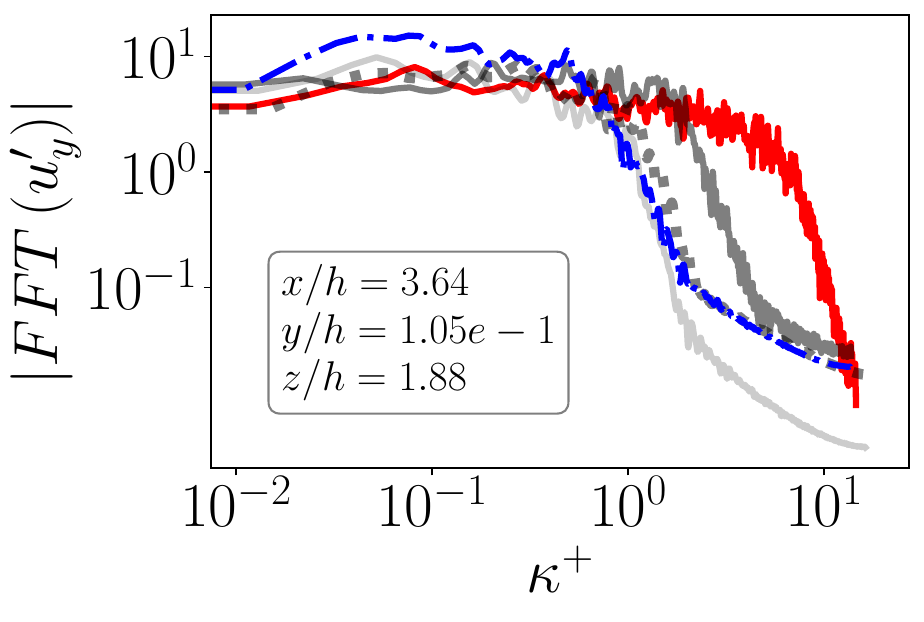} &
    \includegraphics[width=0.32\linewidth]{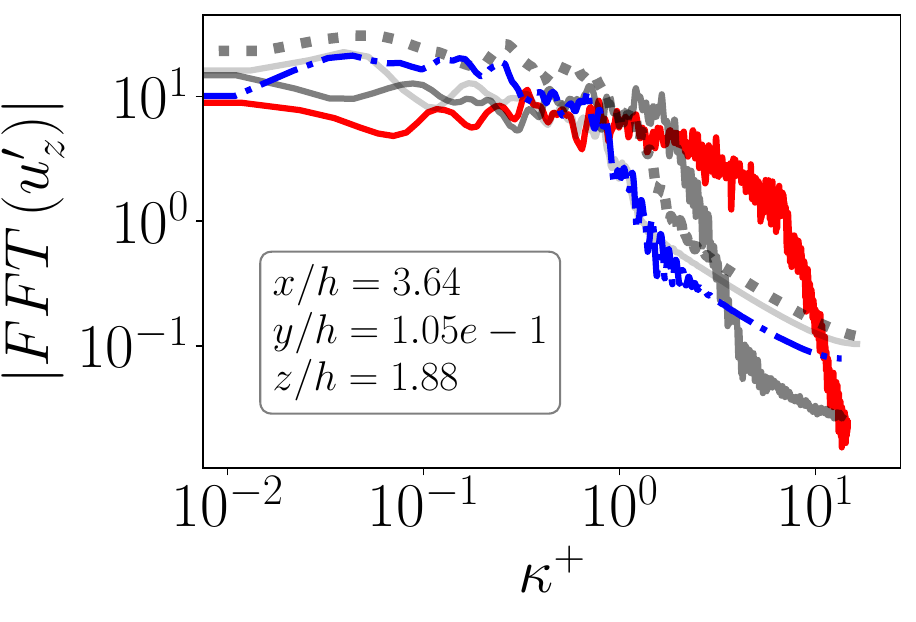} \\
    \textit{(d)} & \textit{(e)} & \textit{(f)}
    \end{tabular}
    \caption{$FFT$ calculated by sampling the fluctuating velocity field $\boldsymbol{u}^\prime$ located at (first row) $\Delta y^\star \approx 30$ and  (second row) $\Delta y^\star \approx 56$. Results are shown for the simulations (\protect\bluelinedasheddotted) DNS-IBM-ML\textsubscript{p.o.}, (\protect\greylinesolidstrong) DNS-IBM-DA\textsubscript{p.o.}, (\protect\greylinedottedstrong) DNS-IBM-CF, (\protect\greylinesolidsoft) DNS-BF, and (\protect\redline) R-DNS-BF.}
    \label{fig:Spectra}
\end{figure}

However, the analysis of the time spectra of the fluctuating velocity $\boldsymbol{u}^\prime$ at different localised sensors, which are obtained via Fast Fourier Transform [$FFT$, \cite{Press2017_cambridge}]  shows some limitations of IBM\textsubscript{RFR}. Here, the dimensionless wavenumber $\kappa^+ = 2\pi f \nu / (U_c \, u_\tau)$ is estimated from the $FFT$ frequency $f$ by assuming Taylor's frozen turbulence hypothesis \cite{Taylor1938_prsl}, consistent with the approach in \cite{Valero2023}. The results, which are presented in Figure \ref{fig:Spectra}, show that the predicted spectrum is very similar to the one obtained by the runs DNS-IBM and DNS-BF. The prediction for this quantity is, therefore, significantly less accurate than the run IBM-DNS-DA\textsubscript{p.o.} run, which systematically gets closer to the reference results of simulation R-DNS-BF. Several factors may contribute to the ML model's degradation relative to the simulation used for training. First, the $u_\tau$ value used for normalisation during the training task is the one obtained in the DA run, but, as shown in Table \ref{tab:summary1}, a small yet non-negligible discrepancy is observed for the prediction obtained with the IBM\textsubscript{RFR}. In addition, the choice to simplify the black-box volume forcing term so that  $\hat{\boldsymbol{f}}_P = (\hat{f}_{P_x}, 0, 0)$, may provide some sources of inaccuracy in developing instantaneous physical strategies. However, one can see from the comprehensive analysis of the results that the IBM\textsubscript{RFR} globally behaves like the DA prediction used for training, and its behaviour is never worse than the simulations DNS-BF and DNS-IBM performed on the same grid. 

This investigation validates the possibility of using instantaneous data to train a black-box IBM model. This task is of moderate difficulty because, even if the coefficients $D_{ij}$ are varying in the $y$ direction for the DA algorithm, the volume forcing is still a linear function of the velocity field. In the next section, a more complex ML tool is developed, in order to capture the non-linear dynamics associated with the field update of state estimation. 



\subsection{ Strategy two: Machine Learning experiment to obtain a DA state estimation surrogate tool}
\label{sec:ML_parameterOpt}

The analysis of the DA performance, when applied to the turbulent plane channel flow, highlighted the importance of the flow update due to state estimation in order to obtain accurate results. However, it was also shown that this task can be performed only when observation is available, therefore precluding systematic application in particular if data is not available for a long time window. Here, RFR techniques are used to obtain a model mimicking the flow correction of state estimation. The training of such a model is performed by sampling flow information \emph{on the fly} while applying the DA algorithm. This case is significantly more complex than the generation of the black-box IBM. First, from a physical perspective, the model itself is more intricate since it involves the representation of the Kalman gain matrix $\boldsymbol{K}_k$ from the DA algorithm, which captures the optimal non-linear correlation between the state and high-fidelity observations. This task is essential for an accurate representation of non-linear dynamics. 
However, an important simplification is that the test case investigated here is statistically steady. This implies that once the EnKF dissipates initial discrepancies due to the selected prior state, the action of the DA tool is statistically the same over the observation window. In particular, the optimised coefficients for $D_{ij}$ are stationary in time, which simplifies the task of setting the boundaries for the observation window used for the ML training.

\begin{figure}[h!]
    \centering
    \includegraphics[scale=0.55, trim={2cm 0 1cm 0}, clip]{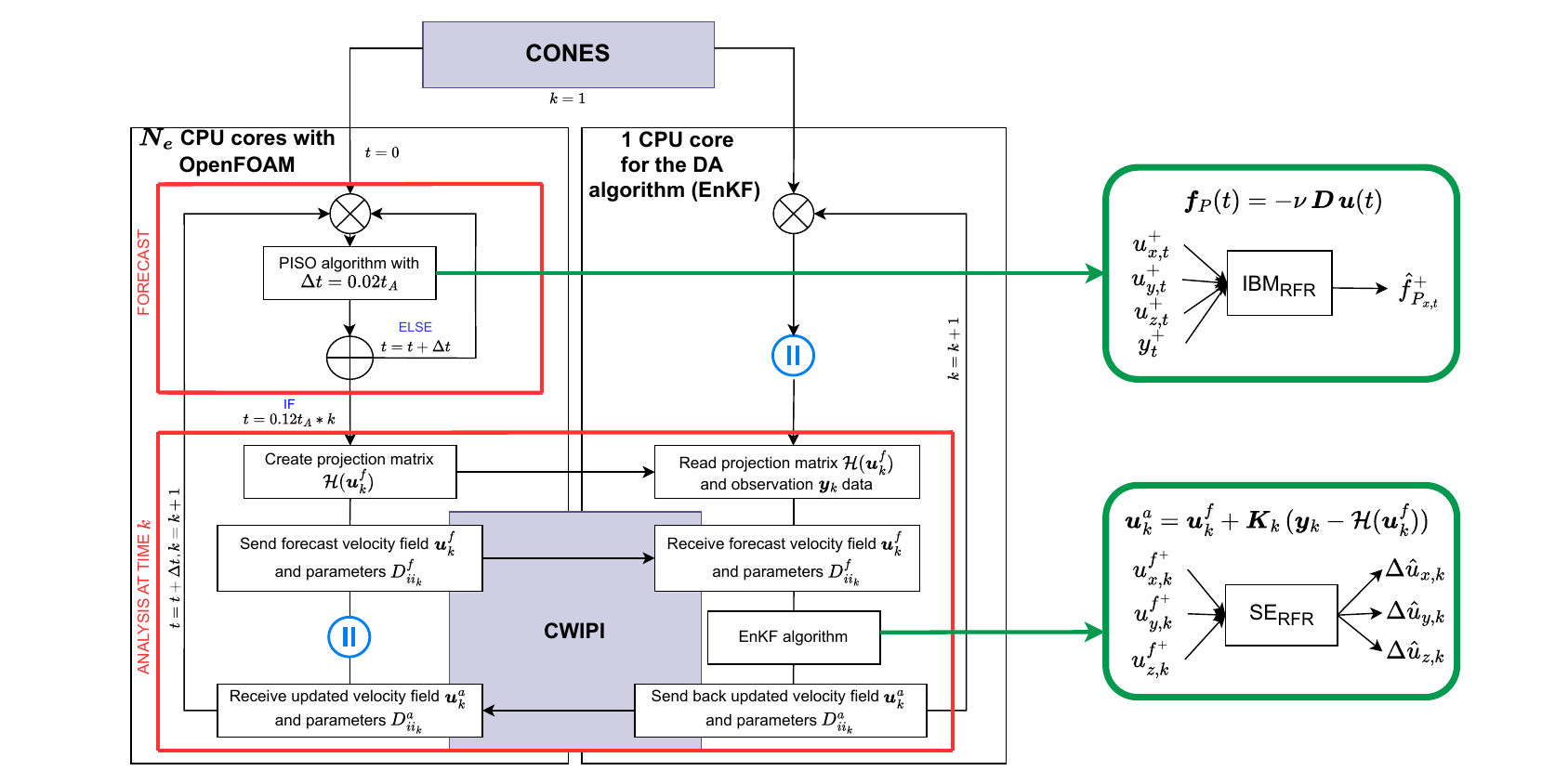}
    \caption{Overview of the DA code CONES employed for state estimation and parameter optimisation, highlighting the components that are modelled by either IBM\textsubscript{RFR} or SE\textsubscript{RFR}.}
    \label{fig:RFR_Scheme_insideEnKF}
\end{figure}

Figure \ref{fig:RFR_Scheme_insideEnKF} shows the scheme of the tasks performed. CONES \cite{Villanueva2024_cof} is employed to establish a connection between CFD, DA, and ML tools for the case of investigation, which is the turbulent plane channel flow previously investigated. The black-box IBM\textsubscript{RFR} $\hat{\boldsymbol{f}}_P$ derived in \S\ref{sec:ML_parameterOpt}, is included within the Navier--Stokes equations through the PISO algorithm (see \ref{sec:RFR_PISO}). The second ML algorithm aims to create a generative model mimicking the correction via state estimation. This tool is referred to as RFR for state estimation (SE\textsubscript{RFR}).

The input and output features used for the training of SE\textsubscript{RFR} are now discussed. The feature matrix and the dependent values are defined, respectively, as $\mathcal{X} = \left(u_{x,k}^{f^+}, u_{y,k}^{f^+}, u_{z,k}^{f^+} \right) = \left(u_{x,k}^f / u_{\tau_k}, u_{y,k}^f/ u_{\tau_k}, u_{z,k}^f/ u_{\tau_k} \right)$, and $\mathcal{Y} = \left(\Delta u_{x,k}^{+}, \Delta u_{y,k}^+, \Delta u_{z,k}^+ \right) = \left((u_{x,k}^a - u_{x,k}^f) / u_{\tau_k}, (u_{y,k}^a - u_{y,k}^f) / u_{\tau_k}, (u_{z,k}^f - u_{z,k}^f) / u_{\tau_k} \right)$, where the instantaneous value of $u_{\tau_k}$ is employed for normalisation in the features. This step clearly shows the contribution of the black box model as an additive correction term, taking the forecast as input and providing as output the difference with the field produced by the analysis phase. $k$ stands for every analysis phase where data is sampled in the observation window. In this setup, data from one of the EnKF members $n_e$  is collected in the form of dimensionless forecast velocities $\boldsymbol{u}_{k, n_e}^{f^+}$ and dimensionless analysis velocities $\boldsymbol{u}_{k, n_e}^{a^+}$ velocities at the $10$ near-wall layers located either in the solid $\Omega_b$ or interface $\Sigma_b$ regions ($5$ at the top wall and $5$ and the bottom wall) over a time period of $1.8\,t_A$, which corresponds to $k_{\textrm{max}} = 15$ consecutive analysis phases of the EnKF. The total amount of data is $m_T = 1\,228\,800$ sampling data points, selected to provide significant representation while optimising computational efficiency and minimising RAM usage, requiring only $7.5$ GB of memory. The same approach is applied for the training of the IBM\textsubscript{RFR} model, where the features are also normalised with respect to the instantaneous $u_{\tau_k}$. Data is collected at the same time instances, resulting in the same number of sampling data points as in the SE\textsubscript{RFR} model while reducing RAM usage to just $2.1$ GB of memory. In both models, the hyperparameters during training are consistent with those discussed in \S\ref{sec:ML_parameterOpt}. One can see in Table \ref{tab:metrics_2} that the metrics $\mathcal{R}^2$ and NRMSE for all prediction show an excellent rate of convergence. 

\begin{table}
  \begin{center}
\def~{\hphantom{0}}
\scalebox{1}{
  \begin{tabular}{lcccc}
       & $\hat{f}_{P_x}$ & $\Delta \hat{u}_x$ & $\Delta \hat{u}_y$ & $\Delta \hat{u}_z$ \\
       & & & & \\
       $\mathcal{R}^2$ & $99.91\%$ & $99.98\%$ & $99.97\%$ & $99.98\%$ \\
       NRMSE & $15.59\%$ & $7.73\%$ & $3.16\%$ & $1.49\%$ \\
  \end{tabular}}
  \caption{Metrics computed for all dependent variables estimated via either IBM\textsubscript{RFR} or SE\textsubscript{RFR}.}
  \label{tab:metrics_2}
  \end{center}
\end{table}

\begin{figure}
    \begin{tabular}{ccc}
    \includegraphics[width=0.32\linewidth]{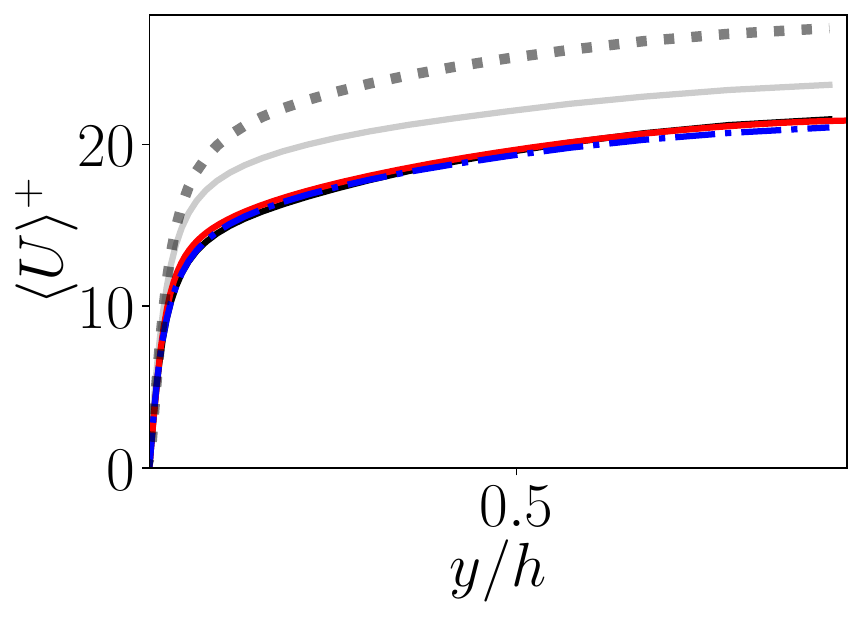} & 
    \includegraphics[width=0.32\linewidth]{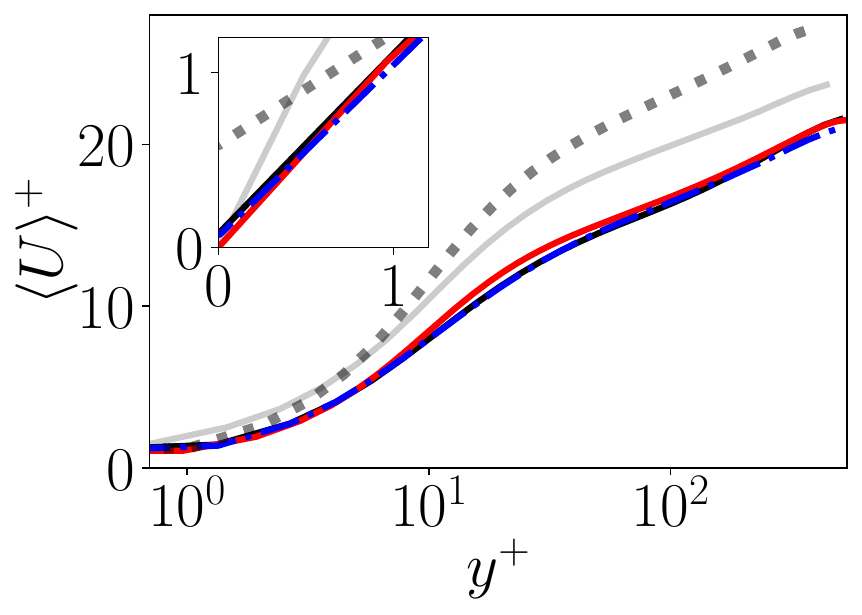} &
    \includegraphics[width=0.32\linewidth]{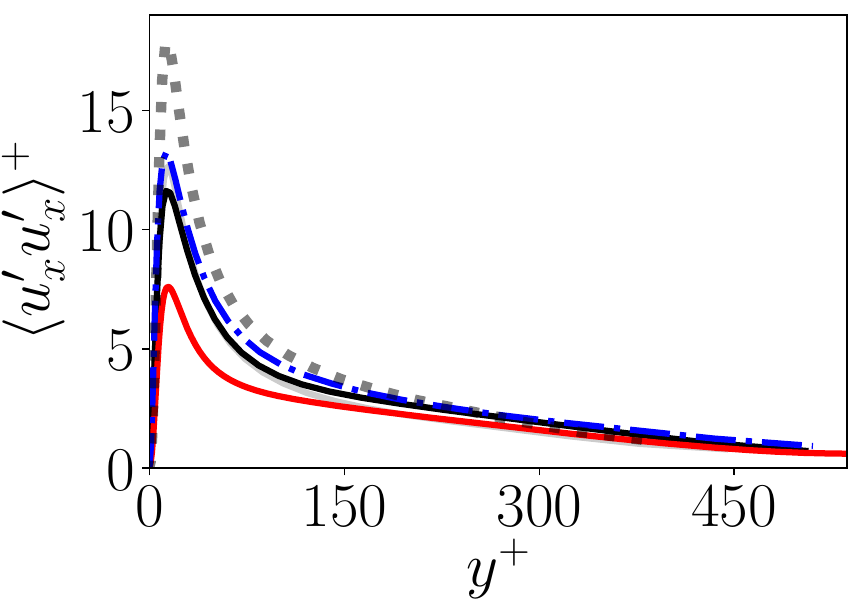} \\
    \textit{(a)} & \textit{(b)} & \textit{(c)} \\
    \includegraphics[width=0.32\linewidth]{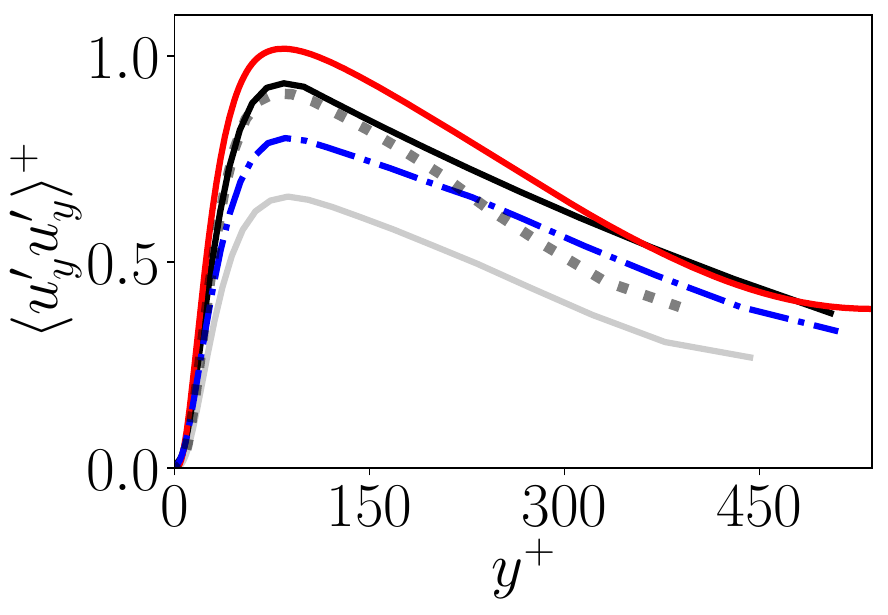} & \includegraphics[width=0.32\linewidth, height=3cm]{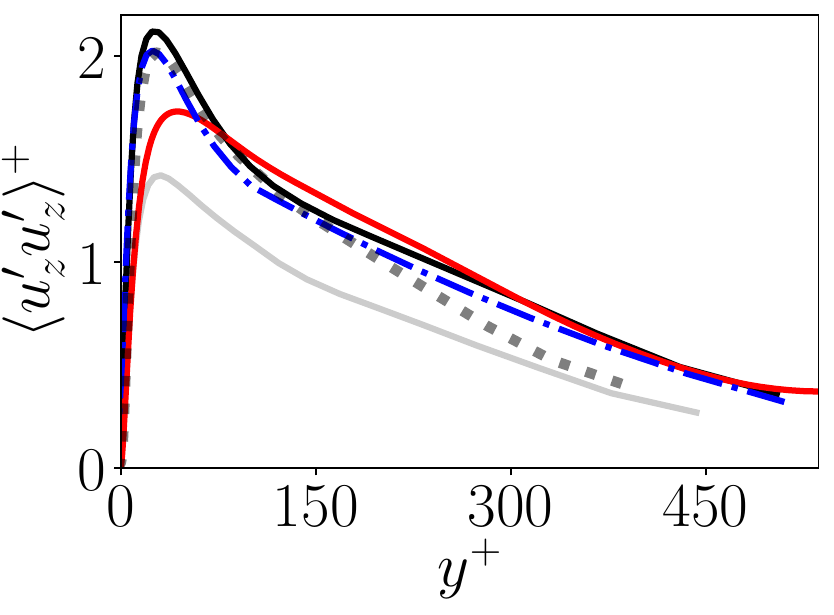} &
    \includegraphics[width=0.32\linewidth]{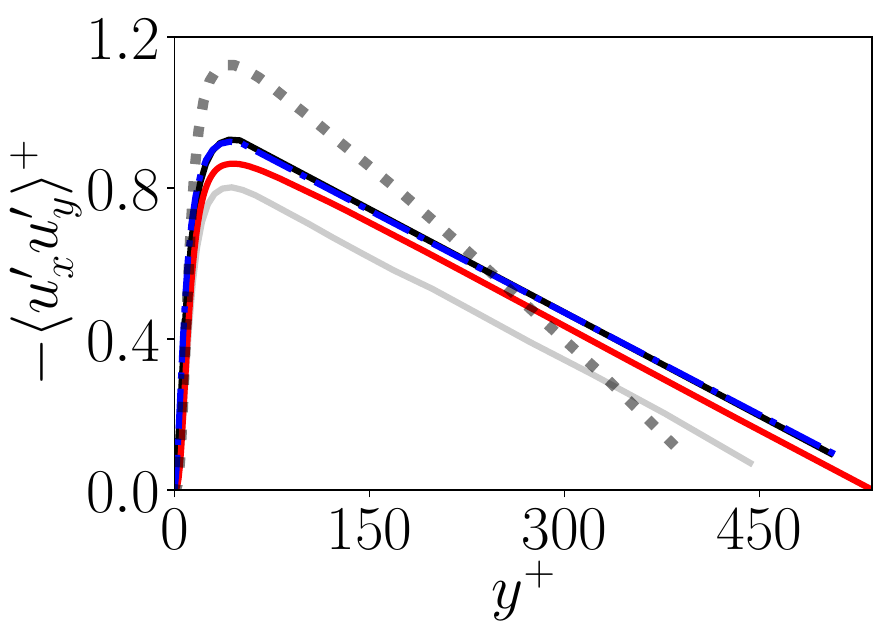} \\
    \textit{(d)} & \textit{(e)} & \textit{(f)}
    \end{tabular}
    \caption{Comparison of the main statistical moments of the velocity field. Results are shown for simulations (\protect\bluelinedasheddotted) DNS-IBM-ML\textsubscript{s.e.}, (\protect\blacklinesolid) DNS-IBM-DA\textsubscript{s.e.}, (\protect\greylinesolidsoft) DNS-BF, (\protect\greylinedottedstrong) DNS-IBM, and (\protect\redline) R-DNS-BF.}
    \label{fig:streaming}
\end{figure}

Figure \ref{fig:streaming} shows the results of the simulation DNS-IBM-ML\textsubscript{s.e.}, which integrates both RFR models (IBM\textsubscript{RFR} and SE\textsubscript{RFR}) under the same conditions as the training dataset. The first-order statistics now perfectly align with the homologous DNS-IBM-DA\textsubscript{s.e.} and the R-DNS-BF runs, with a discrepancy of $2.71\%$ for the value of $u_\tau$ compared to the reference fine-grained simulation and computational costs $C.C^\star = 8.78$, matching the order of magnitude of the coarse-grained body-fitted simulation (DNS-BF) as detailed in Table \ref{tab:summary1}. This demonstrates the critical role of state estimation in unsteady simulations in obtaining accurate field predictions. Additionally, all Reynolds stress tensor components, except for $\langle u_y^\prime u_y^\prime \rangle^+$, show strong consistency with the curves obtained using DA. 

\begin{figure}
    \begin{tabular}{ccc}
    \includegraphics[width=0.5\linewidth, trim = {5cm 1.5cm 0 1.5cm}, clip]{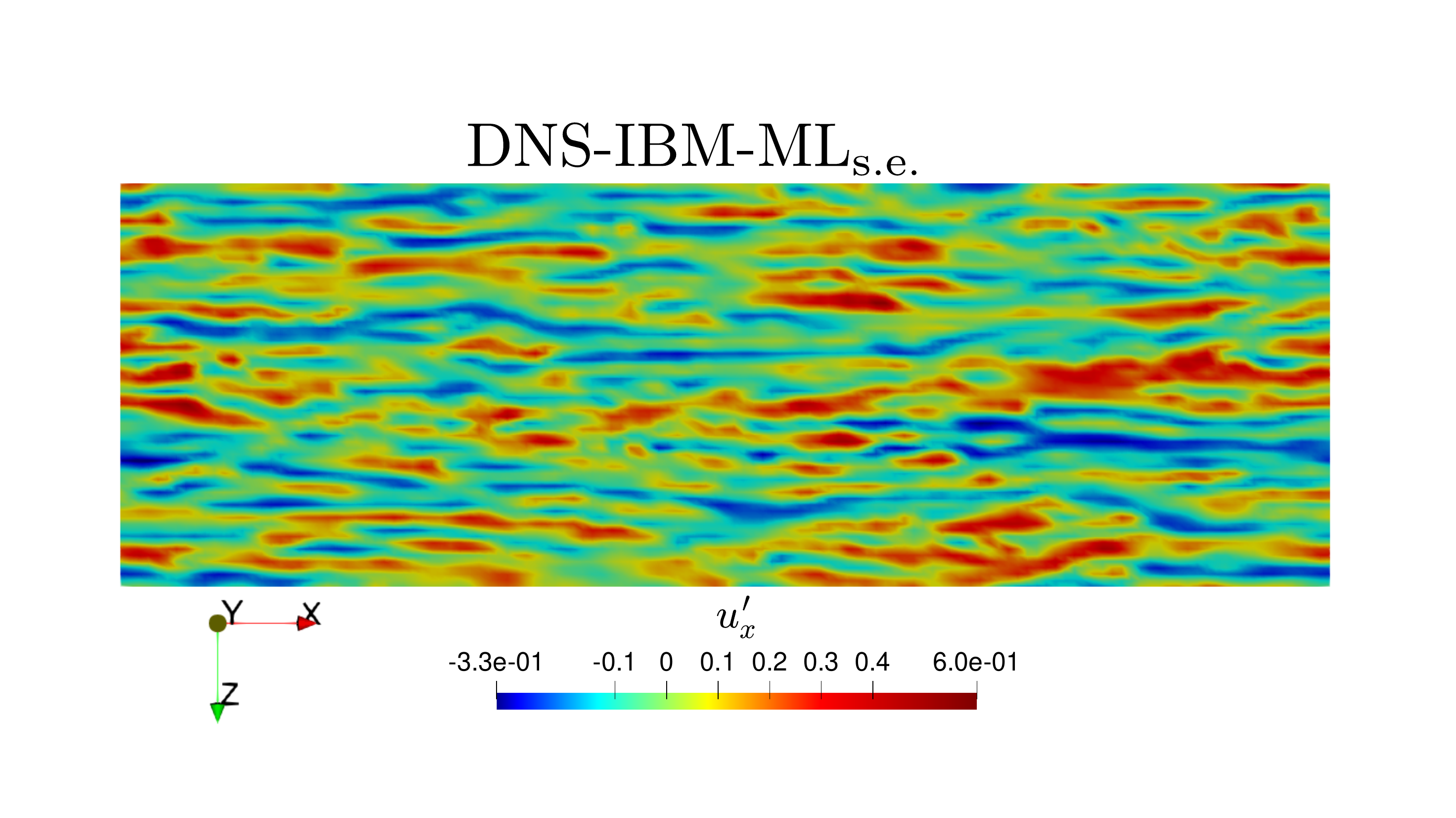} & 
    \includegraphics[width=0.5\linewidth, trim = {5cm 1.5cm 0 1.5cm}, clip]{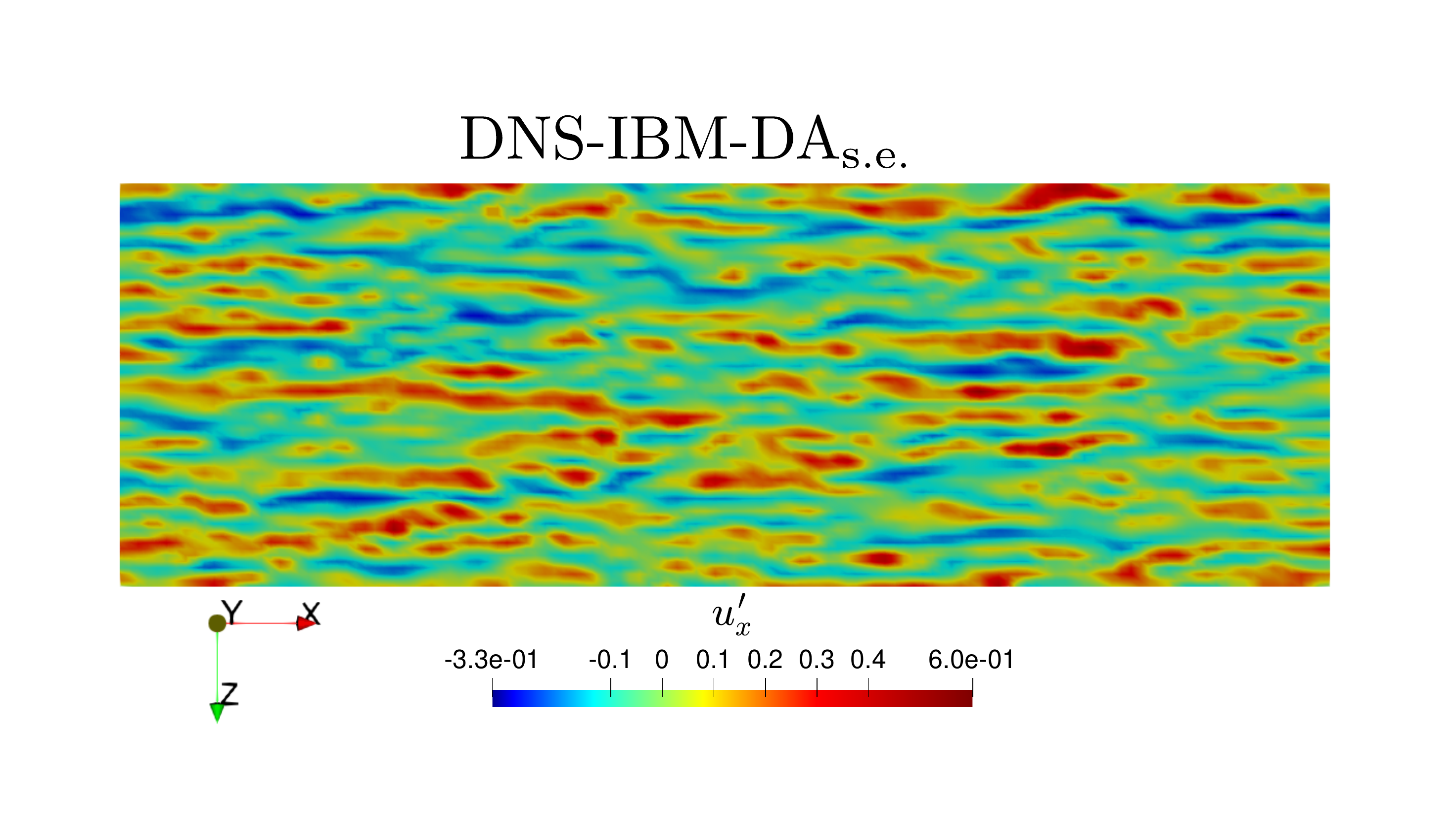} \\
    \textit{(a)} & \textit{(b)} \\
    \includegraphics[width=0.5\linewidth, trim = {5cm 1.5cm 0 1.5cm}, clip]{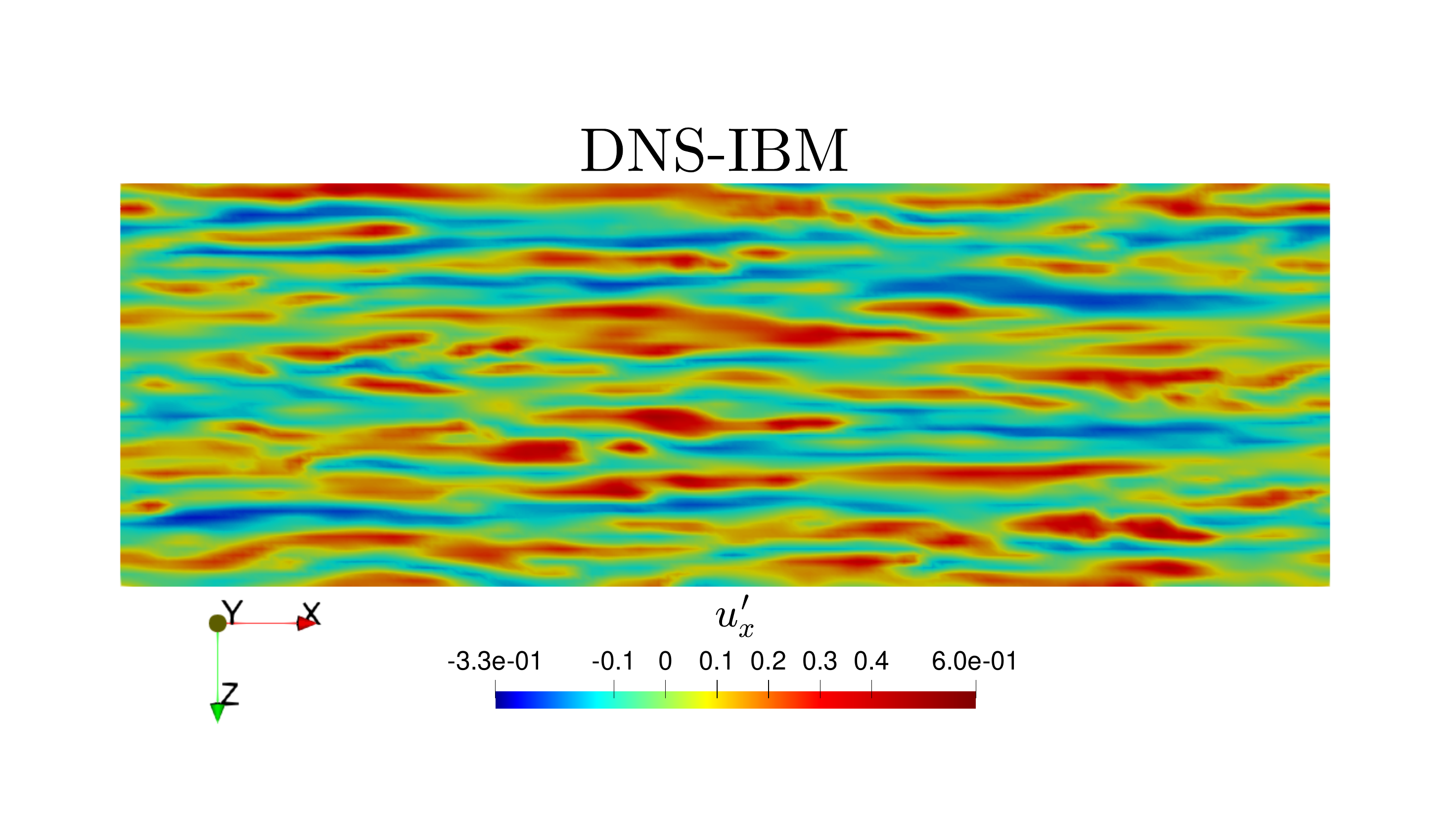} & 
    \includegraphics[width=0.5\linewidth, trim = {5cm 1.5cm 0 1.5cm}, clip]{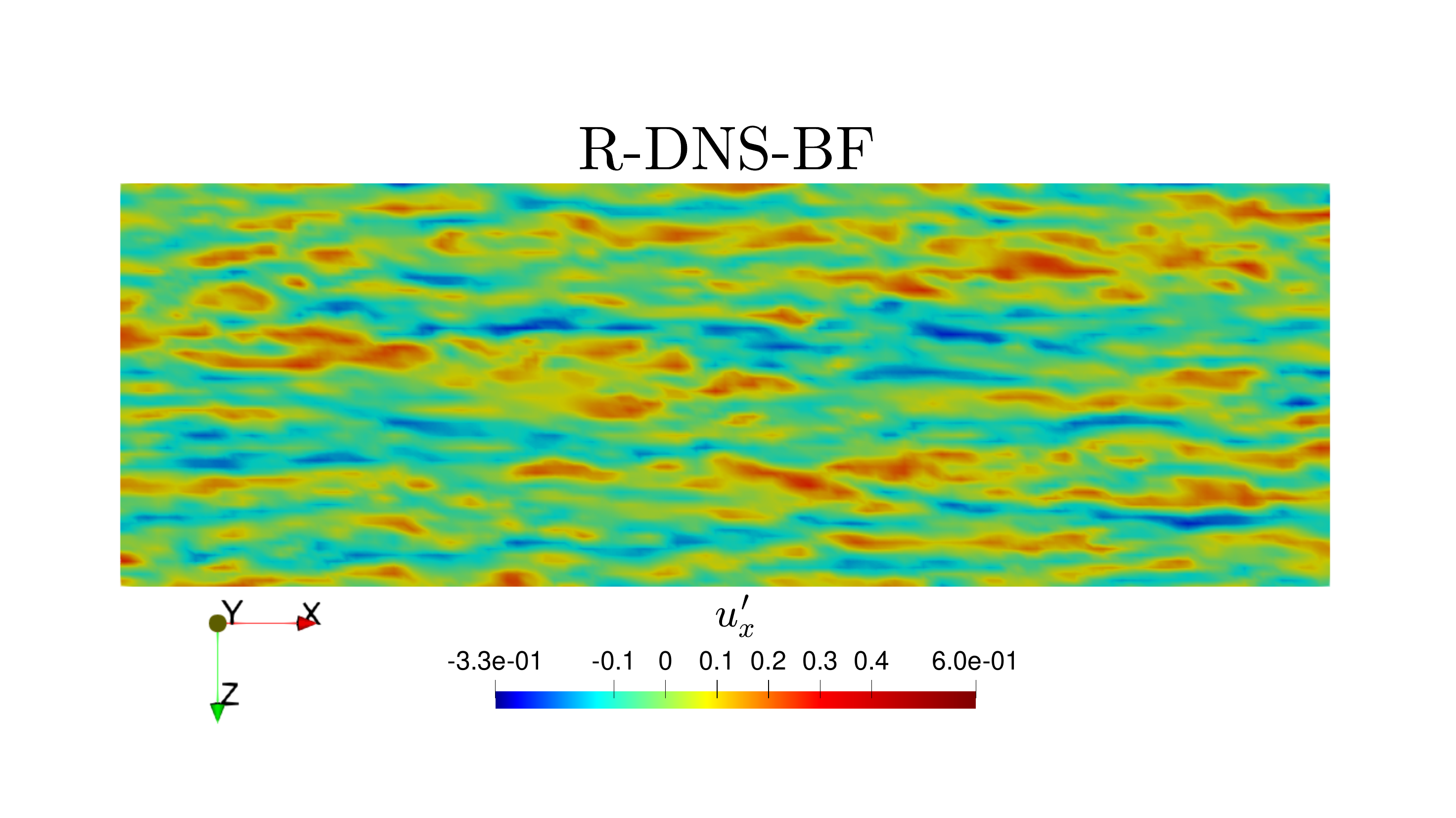} \\
    \textit{(c)} & \textit{(d)} \\
    \end{tabular}
    \caption{Visualisation of the near-wall streaks for $y/h = 0.03$ ($\Delta y^\star \approx 15$). Results are shown for simulations \textit{(a)} DNS-IBM-ML\textsubscript{s.e.}, \textit{(b)} DNS-IBM-DA\textsubscript{s.e.}, \textit{(c)} DNS-IBM, and \textit{(d)} R-DNS-BF.}
    \label{fig:streaks}
\end{figure}

Instantaneous features are shown in Figure \ref{fig:streaks} for the spatial structure of the streaks in a plane parallel to the wall. The R-DNS-BF simulation exhibits the shortest streaks and the smallest velocity fluctuations, whereas the DNS-IBM simulation produces the longest streaks with the highest velocity fluctuations. The DNS-IBM-DA\textsubscript{s.e.} and DNS-IBM-ML\textsubscript{s.e.} simulations fall between these extremes, even if the mesh resolution is identical to the one for the run DNS-IBM. For brevity, the spectral analysis is omitted, as it reveals no significant deviations from the curves obtained for the DNS-IBM-ML\textsubscript{p.o.} simulation.

These results highlight the potential for simultaneously applying DA and ML algorithms \emph{on the fly}, facilitating advanced applications in complex flow scenarios. For instance, in DTs, real-world physical systems could provide sparse spatial and temporal measurements obtained experimentally and integrate them into a coarse-grained/low-fidelity simulation by means of a sequential DA algorithm to produce an optimised continuous state. Subsequently, an ML tool could leverage this extensive high-fidelity dataset to accurately refine operational parameters in real time, enhancing system performance and preventing premature failure, especially under extreme conditions.

However, one could argue about how the model behaves under circumstances that deviate from those used for sampling for the training dataset. This important point is discussed and investigated in \S\ref{sec:different_training}. 

\section{Beyond training conditions: different grid refinement and $Re_\tau$ numbers}
\label{sec:different_training}

In this section, two open questions about the application of DA and ML tools outside the training conditions are studied. Firstly, we investigate the impact of altering mesh refinement to achieve a wall resolution of $y^+ \approx 1$, corresponding to spatial interpolation scenarios. Secondly, we explore the effect of varying $Re_\tau$ and, consequently, modifying the near-wall physics.


For the test cases with different grid refinement, we investigate two configurations with higher resolution near the wall. In the first arrangement, labelled as IBM-ML-grid 1, the grid presents fewer mesh elements than the one with the training conditions (IBM-DNS-ML\textsubscript{s.e.}) by coarsening the cells in the streamwise and spanwise directions by a factor of two, while simultaneously refining the wall-normal direction by the same factor. In contrast, the second configuration, denoted as IBM-ML-grid 2, contains more mesh elements by refining only the wall-normal direction by a factor of two. Table \ref{tab:summary2} provides a summary of these two complementary simulations, where it is important to note that, in these two runs, all quantities $\Delta (\cdot)^\star$ are normalised with respect to $u_\tau$ from R-DNS-BF.

\begin{table}
  \begin{center}
\def~{\hphantom{0}}
\scalebox{0.8}{
  \begin{tabular}{lccccccccc}
       & $N_x \times N_y \times N_z$ & $\Delta x^\star$ &  $\Delta z^\star$ & $\Delta y^\star_{\text{min}}$ & $\Delta y^\star_{\text{max}} $ & $L_x/h$ & $L_z/h$ & $u_\tau$ \\
       & & & & & & & & \\
       DNS-IBM-ML\textsubscript{s.e.} & $128 \times 64 \times 64$ & $39.5$ & $26.3$ & $1.3$ & $67.6$ & $3\pi$ & $\pi$ & $0.0493$ \\
       & & & & & & & & \\
       IBM-ML-grid 1 & \multirow{3}{*}{$64 \times 128 \times 32$} & \multirow{3}{*}{$79.0$} & \multirow{3}{*}{$52.7$} & \multirow{3}{*}{$1$} & \multirow{3}{*}{$30$} & \multirow{3}{*}{$3\pi$} & \multirow{3}{*}{$\pi$} & $0.0464$ \\
       IBM-DA-grid 1 & & & & & & & & $0.0454$ \\
       IBM-grid 1 & & & & & & & & $0.0312$ \\
       BF-grid 1 & $64 \times 122 \times 32$ & $79.0$ & $52.7$ & $0.82$ & $24.7$ & $3\pi$ & $\pi$ & $0.0356$ \\
       & & & & & & & & \\
       IBM-ML-grid 2 & \multirow{3}{*}{$128 \times 128 \times 64$} & \multirow{3}{*}{$39.5$} & \multirow{3}{*}{$26.3$} & \multirow{3}{*}{$1$} & \multirow{3}{*}{$30$} & \multirow{3}{*}{$3\pi$} & \multirow{3}{*}{$\pi$} & $0.0483$ \\
       IBM-DA-grid 2 & & & & & & & & $0.0458$ \\
       IBM-grid 2 & & & & & & & & $0.0329$ \\
       BF-grid 2 & $128 \times 122 \times 64$ & $39.5$ & $26.3$ & $0.82$ & $24.7$ & $3\pi$ & $\pi$ & $0.0361$ \\
       & & & & & & & & \\
       ML-ReTau395 & \multirow{2}{*}{$256 \times 64 \times 128$} & \multirow{2}{*}{$30.6$} & \multirow{2}{*}{$20.4$} & \multirow{2}{*}{$1.01$} & \multirow{2}{*}{$52.6$} & \multirow{2}{*}{$6\pi$} & \multirow{2}{*}{$2\pi$} & $0.0496$ \\
       IBM-ReTau395 & & & & & & & & $0.0392$ \\
       BF-ReTau395 & $256 \times 58 \times 128$ & $30.6$ & $20.4$ & $0.84$ & $43.9$ & $6\pi$ & $2\pi$ & $0.0440$ \\
       Moser \textit{et al.} \cite{Moser1999_pof} & $384 \times 257 \times 384$ & $9.7$ & $4.8$ & $0.03$ & $7.2$ & $2\pi$ & $\pi$ & $0.0508$ \\
       & & & & & & & & \\
       ML-ReTau950 & \multirow{2}{*}{$64 \times 64 \times 32$} & \multirow{2}{*}{$83.7$} & \multirow{2}{*}{$55.8$} & \multirow{2}{*}{$2.8$} & \multirow{2}{*}{$143.8$} & \multirow{2}{*}{$\frac{3\pi}{2}$} & \multirow{2}{*}{$\frac{\pi}{2}$} & $0.0454$ \\
       IBM-ReTau950 & & & & & & & & $0.0312$ \\
       BF-ReTau950 & $64 \times 58 \times 32$ & $83.7$ & $55.8$ & $2.3$ & $120.1$ & $\frac{3\pi}{2}$ & $\frac{\pi}{2}$ & $0.0361$ \\
       Hoyas \textit{et al.} \cite{Hoyas2008_pof} & $3\,072 \times 385 \times 2\,304$ & $7.6$ & $3.8$ & $0.03$ & $7.6$ & $8\pi$ & $3\pi$ & $0.0454$ \\
       & & & & & & & &
  \end{tabular}}
  \caption{Overview of supplementary simulations conducted to validate the ML model.}
  \label{tab:summary2}
  \end{center}
\end{table}

These simulations are compared with some runs featuring the same grid refinement for a classical penalty IBM (IBM-grid 1 and IBM-grid 2, respectively), body-fitted simulations (BF-grid 1 and BF-grid 2, respectively), and the combination of parametric optimisation and state estimation by DA using the methodology already explained in \S\ref{sec:IBM_DA} (IBM-DA-grid 1 and IBM-DA-grid 2, respectively). The \emph{prior} for the latter simulations is based on the field at $300\,t_A$ and the optimised parameters $D_{ij}$ from the DNS-IBM-DA\textsubscript{s.e.} simulation. A normal distribution with a standard deviation of $1\%$ is applied to each of the $40$ ensemble members to introduce variability into the system. Furthermore, multiplicative inflation $\lambda = 1.01$ is set to the model parameters after each analysis phase to further increase the variability of the ensemble, and clipping is performed to the mesh elements near the wall located either in the body $\Omega_b$, fluid-solid interface $\Sigma_b$ regions or the viscous sublayer of the fluid, i.e., $y^+ < 5$ (see \ref{sec:EnKF} for the definitions and \cite{Valero2023} for the practical implementations). It is insightful to compare the RFR predictions with those from DA, considering \textit{a priori} the latter as the best-case scenario for the ML tool. This is because the data used to train the RFR model is derived from the enhanced predictions generated by DA.

\begin{figure}
    \begin{tabular}{ccc}
    \includegraphics[width=0.32\linewidth]{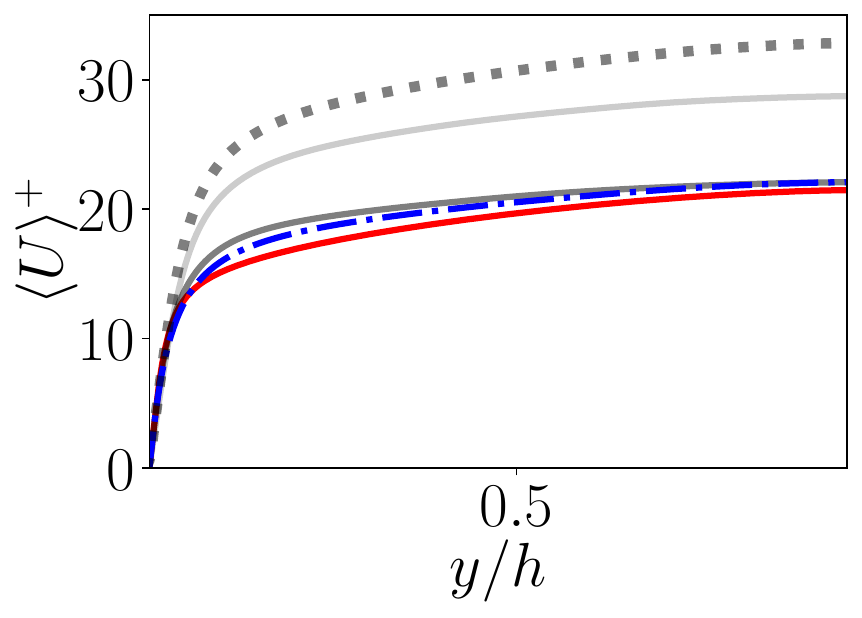} & 
    \includegraphics[width=0.32\linewidth]{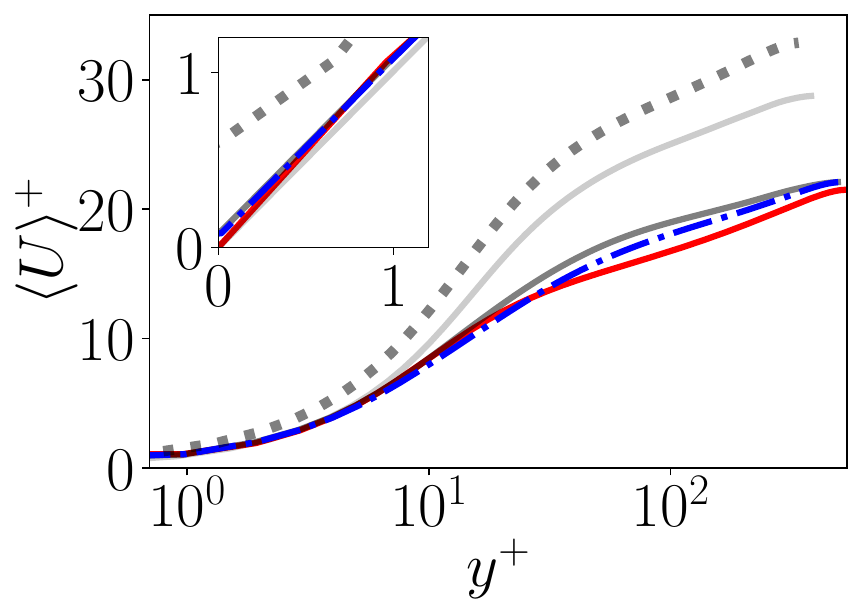} &
    \includegraphics[width=0.32\linewidth]{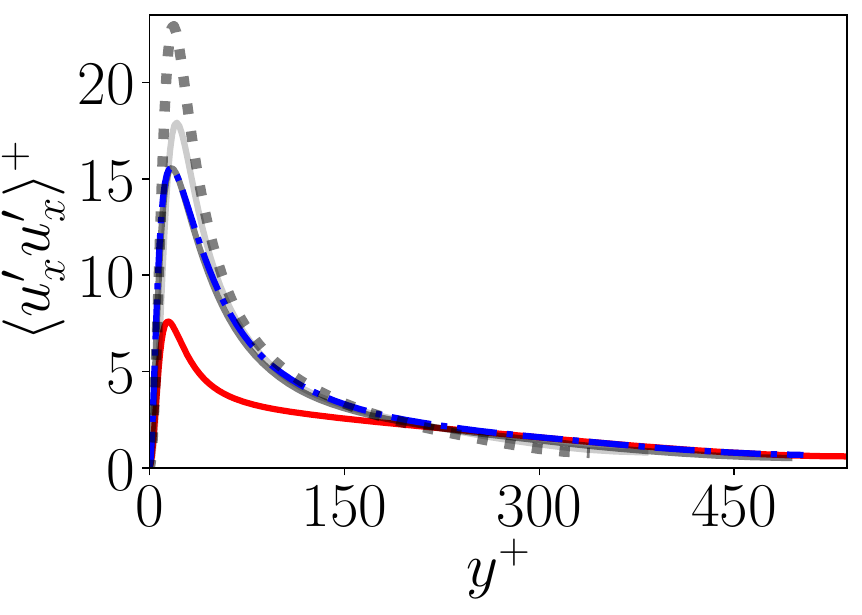} \\
    \textit{(a)} & \textit{(b)} & \textit{(c)} \\
    \includegraphics[width=0.32\linewidth]{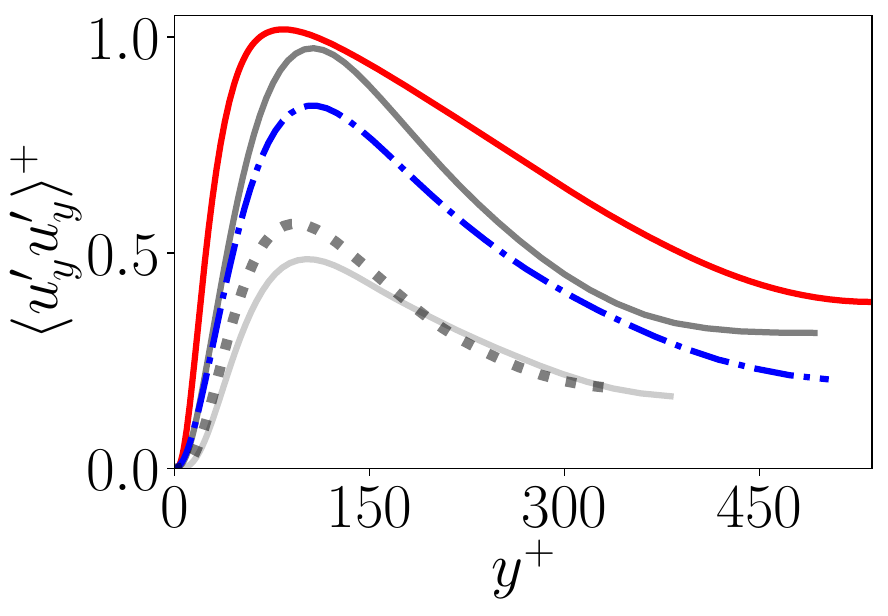} & \includegraphics[width=0.32\linewidth, height=3cm]{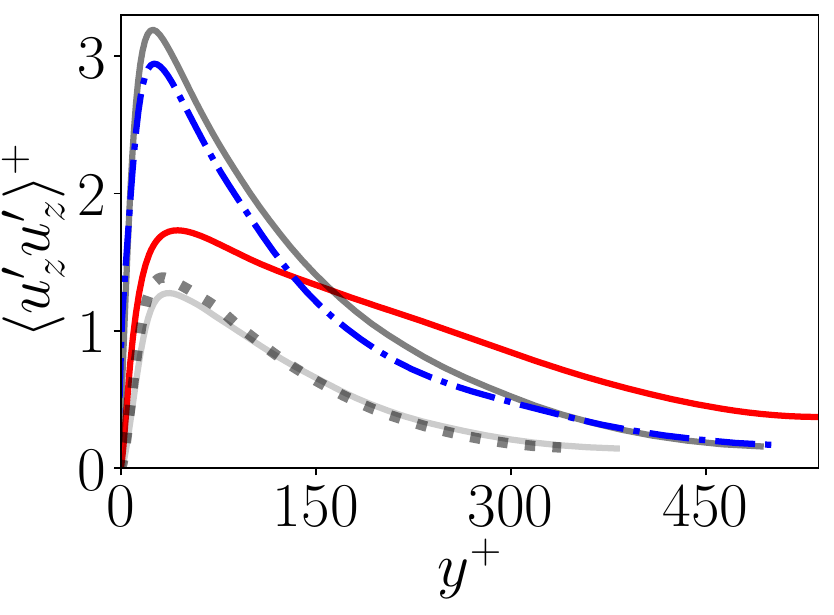} &
    \includegraphics[width=0.32\linewidth]{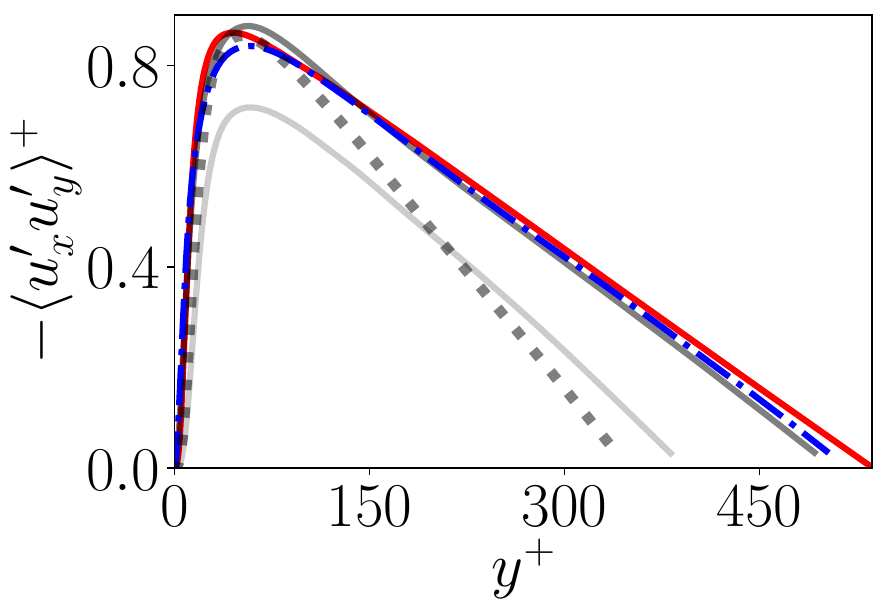} \\
    \textit{(d)} & \textit{(e)} & \textit{(f)}
    \end{tabular}
    \caption{Comparison of the main statistical moments of the velocity field. Results are shown for simulations (\protect\bluelinedasheddotted) IBM-ML-grid 1, (\protect\greylinesolidstrong) IBM-DA-grid 1, (\protect\greylinesolidsoft) BF-grid 1, (\protect\greylinedottedstrong) IBM-grid 1, and (\protect\redline) R-DNS-BF.}
    \label{fig:ML_grid1}
\end{figure}

The R-DNS-BF run is again used as a reference for comparison. In Figure \ref{fig:ML_grid1}, the main statistical moments of the first investigation are shown, revealing that the simulations employing DA and ML, i.e., IBM-DA-grid 1 and IBM-ML-grid 1, respectively, achieve an extremely close overall agreement with the reference fine simulation, especially for the first-order statistics. The RFR model built with data using the original grid refinement demonstrates a good agreement with the DA results for all Reynolds stress tensor components, in particular for $\langle u_x^\prime u_x^\prime \rangle^+$ and $\langle u_x^\prime u_y^\prime \rangle^+$. While slight degradation is observed in $\langle u_y^\prime u_y^\prime \rangle^+$ compared to DA, it is still significantly more accurate than the classical body-fitted simulation (BF-grid 1) and even more so than the classical penalisation IBM (IBM-grid 1). Furthermore, $\langle u_z^\prime u_z^\prime \rangle^+$ is closer to the reference simulation curve than the one obtained via DA. The accuracy of the prediction is also observed for the friction velocity $u_\tau$. The ML model can predict its value with a discrepancy of $3.33\%$ against the value estimated by the reference body-fitted simulation.

\begin{figure}
    \begin{tabular}{ccc}
    \includegraphics[width=0.32\linewidth]{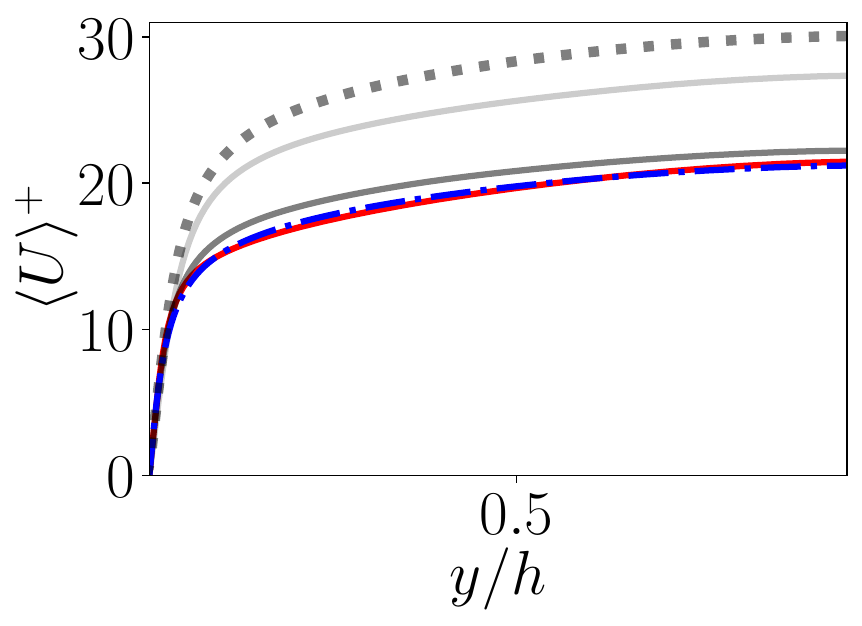} & 
    \includegraphics[width=0.32\linewidth]{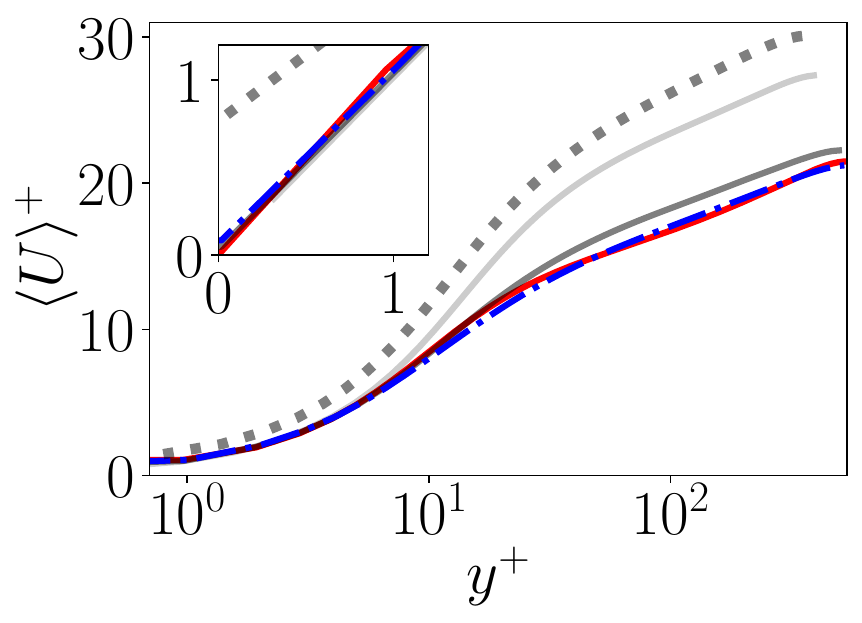} &
    \includegraphics[width=0.32\linewidth]{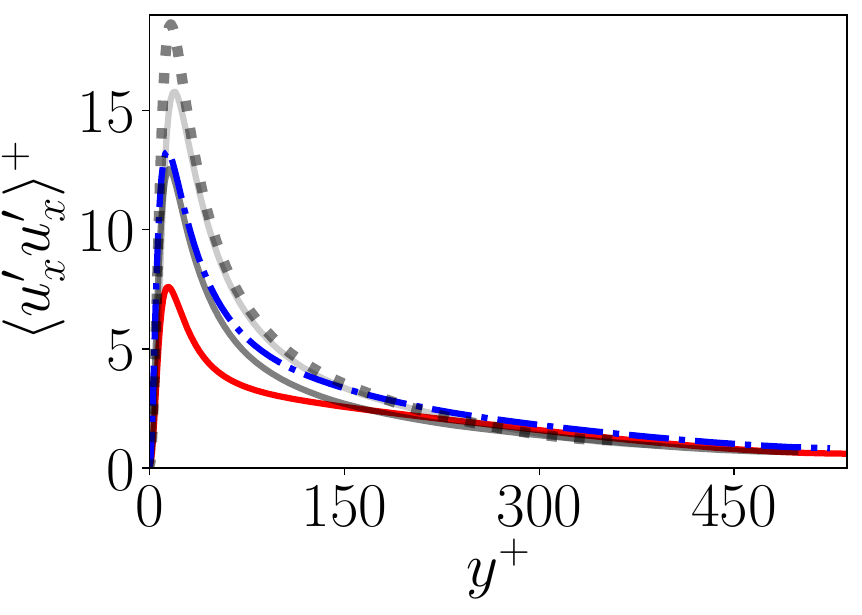} \\
    \textit{(a)} & \textit{(b)} & \textit{(c)} \\
    \includegraphics[width=0.32\linewidth]{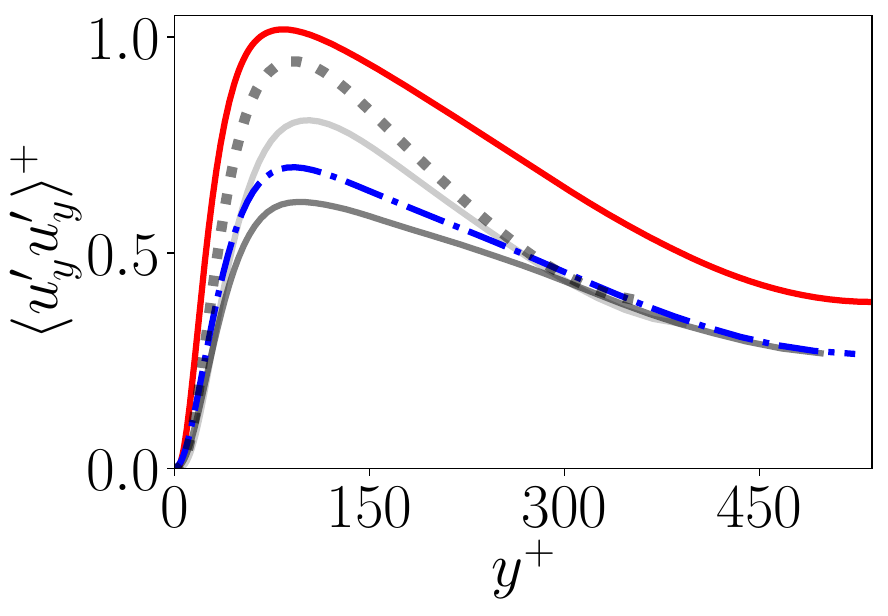} & \includegraphics[width=0.32\linewidth, height=3cm]{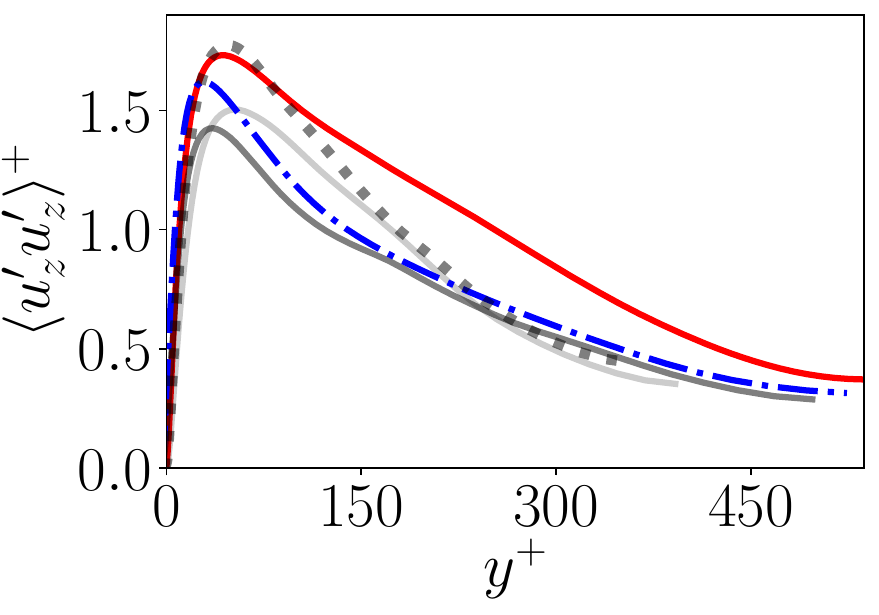} &
    \includegraphics[width=0.32\linewidth]{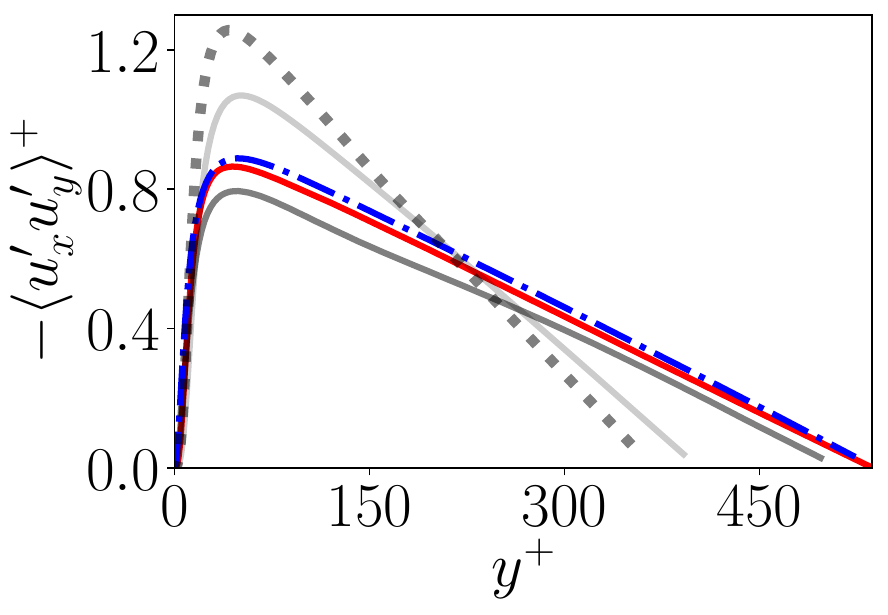} \\
    \textit{(d)} & \textit{(e)} & \textit{(f)}
    \end{tabular}
    \caption{Comparison of the main statistical moments of the velocity field. Results are shown for simulations (\protect\bluelinedasheddotted) ML-grid 2, (\protect\greylinesolidstrong) IBM-DA-grid 2, (\protect\greylinesolidsoft) BF-grid 2, (\protect\greylinedottedstrong) IBM-grid 2, and (\protect\redline) R-DNS-BF.}
    \label{fig:ML_grid2}
\end{figure}

For the second configuration, the main statistical moments represented in Figure \ref{fig:ML_grid2} demonstrate that the grid resolution in the streamwise and spanwise directions is crucial for accurately predicting flow statistics. The DA and ML runs (IBM-DA-grid 2 and IBM-ML-grid 2, respectively) yield results closely aligned with the R-DNS-BF benchmark, with the ML simulation nearly overlapping the reference curves when predicting the first-order statistics and $\langle u_x^\prime u_y^\prime \rangle^+$. Also, one can see that, compared with the DA methodology, the ML simulation performs similarly or slightly better across all the Reynolds stress tensor components. As observed, the ML run outperforms the homologous BF-grid 2 simulation except for $\langle u_y^\prime u_y^\prime \rangle^+$. Significant improvements are also evident when compared to the IBM-grid 2 run, although the latter predicts the Reynolds stress tensor components $\langle u_y^\prime u_y^\prime \rangle^+$ and $\langle u_z^\prime u_z^\prime \rangle^+$ more accurately. The prediction of the wall shear stress is also excellent, with only a discrepancy in the $u_\tau$ value of $0.63\%$ against the R-DNS-BF run.

The second analysis using the ML model comes at its extension and validation across different $Re_\tau$ values. In particular, two conditions for the Reynolds number, namely $Re_\tau = 395$ and $Re_\tau = 950$, are investigated with the same grid resolution as in DNS-IBM-ML\textsubscript{s.e.}. These configurations are selected, first of all, because high-fidelity databases are available for these Reynolds numbers, allowing for accurate normalisation of the quantities $\Delta (\cdot)^\star$ in Table \ref{tab:summary2} against $u_\tau$ from the reference data. Additionally, the different near-wall physics at these $Re_\tau$ provide valuable insights. In comparison with the R-DNS-BF run with $Re_\tau \approx 550$, the configuration with $Re_\tau \approx 395$ exhibits milder velocity gradients in the proximity of the wall. On the contrary, a configuration with $Re_\tau \approx 950$ presents a significantly thinner region with high-velocity gradients, and the influence of the outer flow on the near-wall region becomes stronger, increasing the complexity of the flow. Also, to ensure a comparable number of flow structures in these additional simulations, the size of the computational domain is halved in the streamwise and spanwise directions for the configurations with $Re_\tau \approx 395$, and doubled for the simulations with $Re_\tau \approx 950$. The bulk velocity and kinematic viscosity $\nu$ are recalculated to ensure that the centreline velocity $U_c$ remains unchanged.

\begin{figure}
    \begin{tabular}{ccc}
    \includegraphics[width=0.32\linewidth]{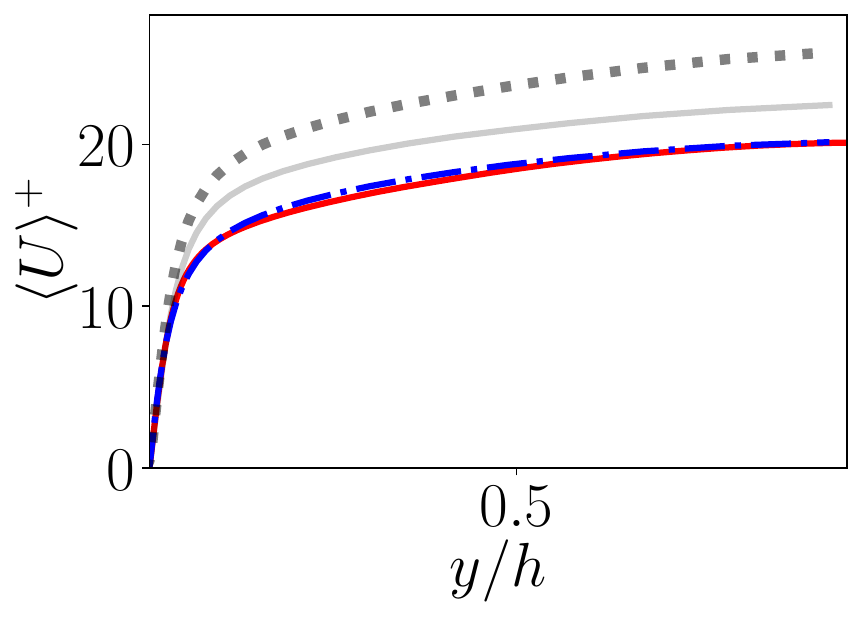} & 
    \includegraphics[width=0.32\linewidth]{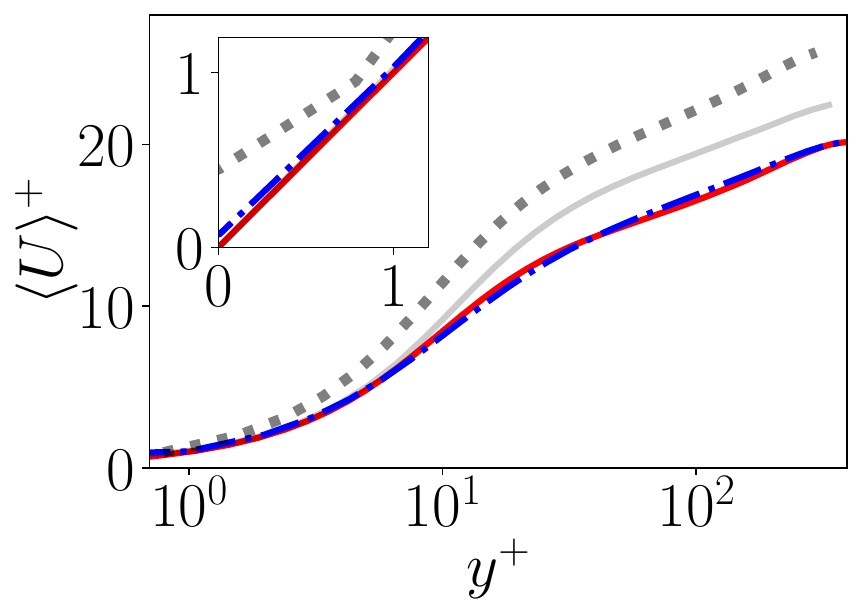} &
    \includegraphics[width=0.32\linewidth]{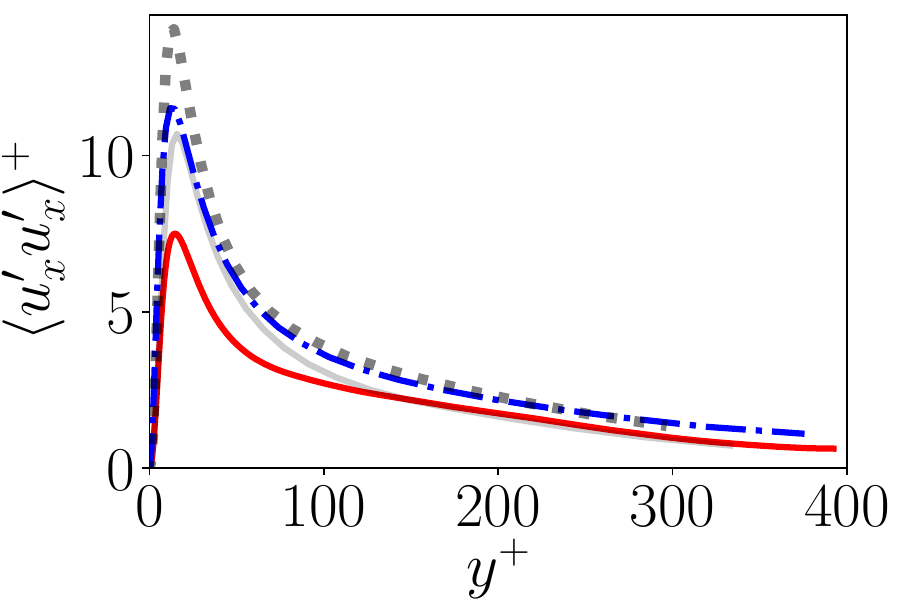} \\
    \textit{(a)} & \textit{(b)} & \textit{(c)} \\
    \includegraphics[width=0.32\linewidth]{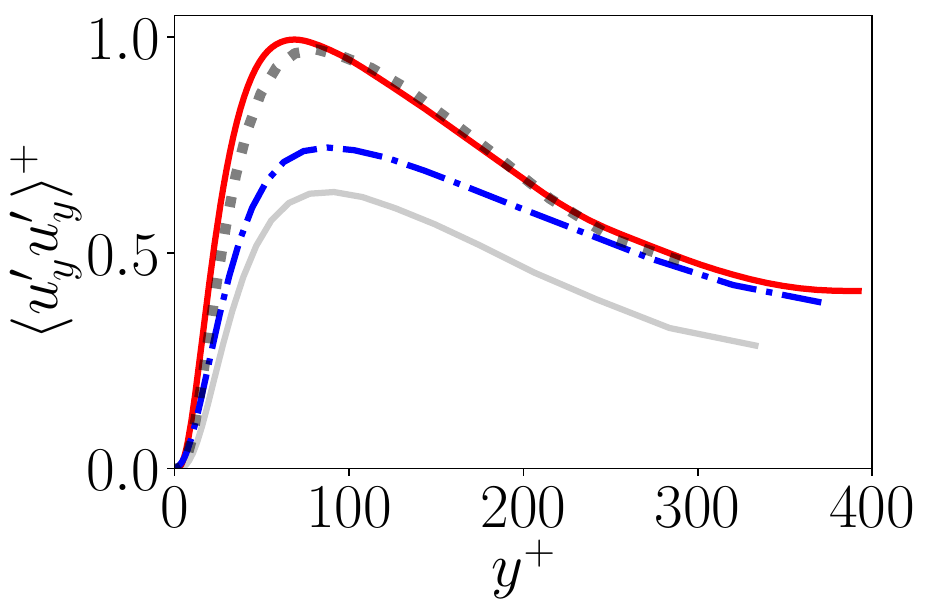} & \includegraphics[width=0.32\linewidth, height=3cm]{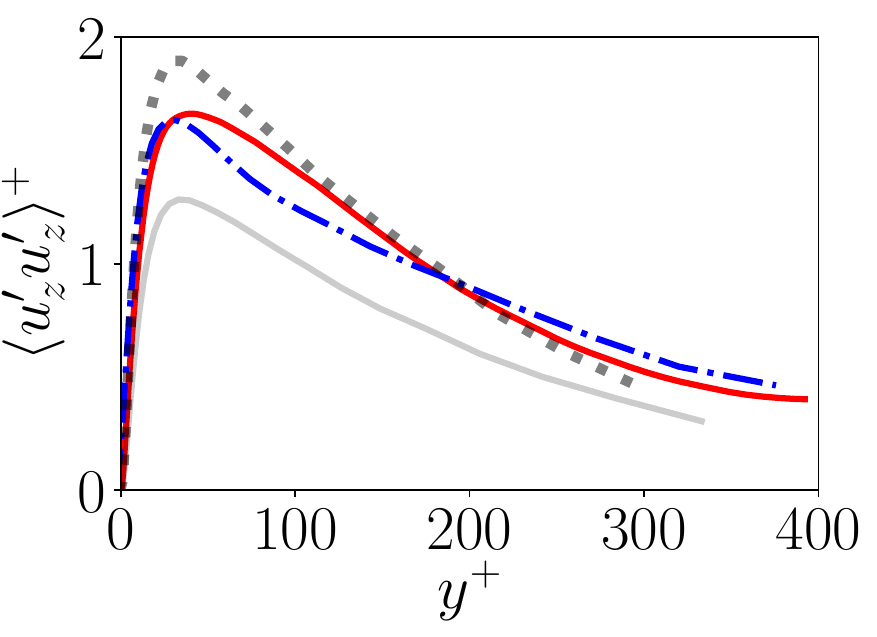} &
    \includegraphics[width=0.32\linewidth]{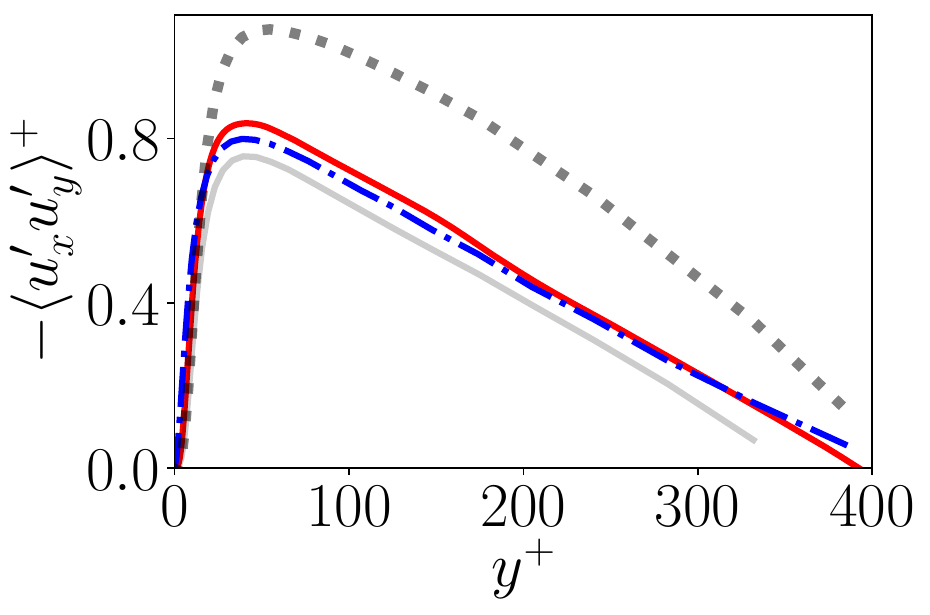} \\
    \textit{(d)} & \textit{(e)} & \textit{(f)}
    \end{tabular}
    \caption{Comparison of the main statistical moments of the velocity field for $Re_\tau = 395$. Results are shown for simulations (\protect\bluelinedasheddotted) ML-ReTau395, (\protect\greylinesolidsoft) BF-ReTau395, (\protect\greylinedottedstrong) IBM-ReTau395, and (\protect\redline) Moser \textit{et al.} \cite{Moser1999_pof}.}
    \label{fig:ReTau_395}
\end{figure}

Given the differences in the physics for these cases compared to the original ML model and the lack of instantaneous high-fidelity data to be observed, DA runs were not performed for this analysis. Even if data was available, it is highly possible that the new DA-optimised parametric configuration would exhibit differences from the results obtained for the run DNS-IBM-DA\textsubscript{s.e.}, establishing difficulties for the comparisons between the tools. The DNS database by Moser \textit{et al.} \cite{Moser1999_pof} serves as the reference for the coarse-grained simulations with $Re_\tau \approx 395$. One can see in Figure \ref{fig:ReTau_395} that the ML model, alluded to as ML-ReTau395, matches the reference data when predicting the first-order statistics and shows a relatively close match in predicting $\langle u_z^\prime u_z^\prime \rangle^+$ and $\langle u_x^\prime u_y^\prime \rangle^+$. Overall, the ML-ReTau395 run predicts the flow features more accurately than their homologous body-fitted (BF-ReTau395) and classical penalisation IBM (IBM-ReTau395) runs with the same grid resolution. Besides, the prediction of the $u_\tau$ is notably accurate, with only a slight discrepancy of $2.36\%$ with respect to the DNS data. This suggests that the ML model can effectively predict wall-bounded turbulent boundary layers when the wall resolution is better than the one used for the training. This finding is expected considering that the grids used for the case at $Re_{\tau}=550$ were already able to resolve the viscous sublayer of the flow. Therefore, no modification in the physical behaviour is observed.

\begin{figure}
    \begin{tabular}{ccc}
    \includegraphics[width=0.32\linewidth]{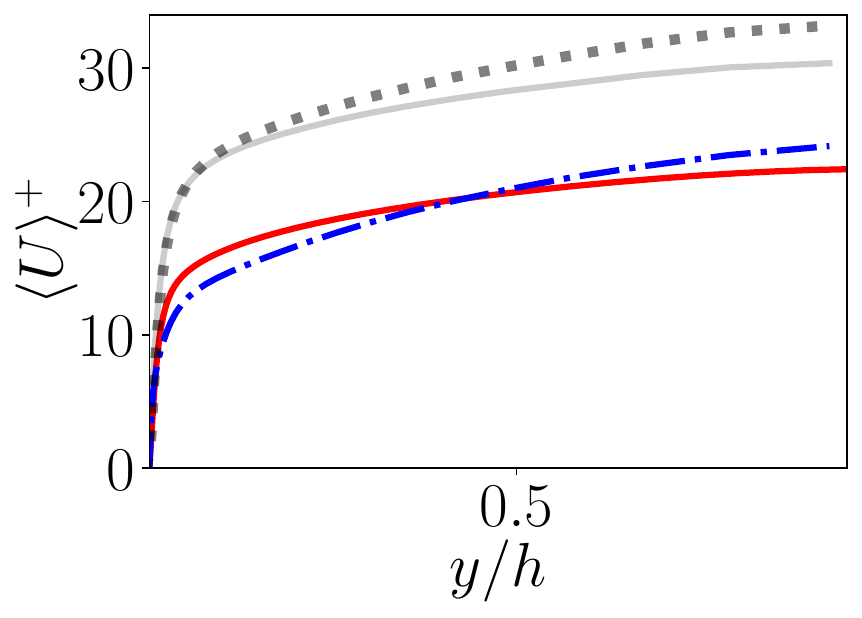} & 
    \includegraphics[width=0.32\linewidth]{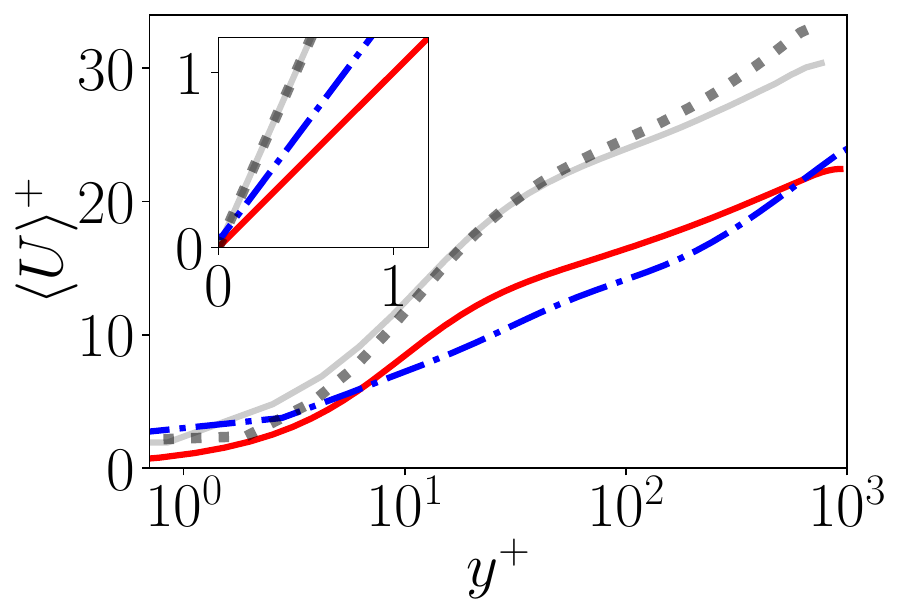} &
    \includegraphics[width=0.32\linewidth]{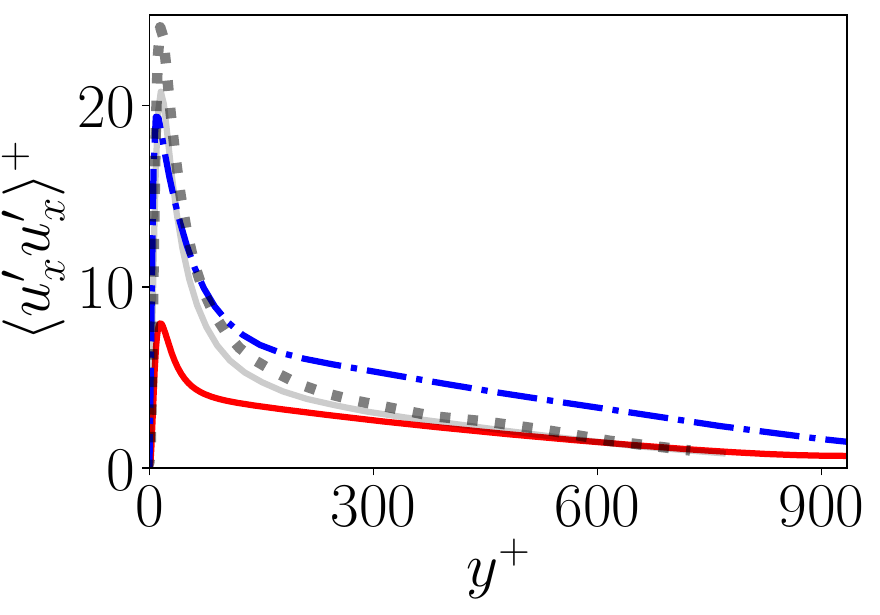} \\
    \textit{(a)} & \textit{(b)} & \textit{(c)} \\
    \includegraphics[width=0.32\linewidth]{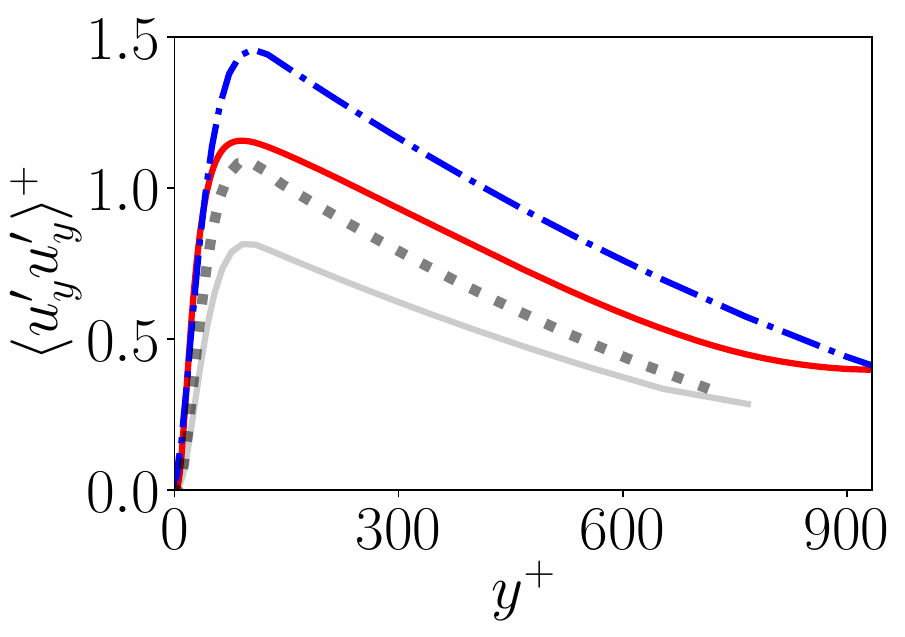} & \includegraphics[width=0.32\linewidth, height=3cm]{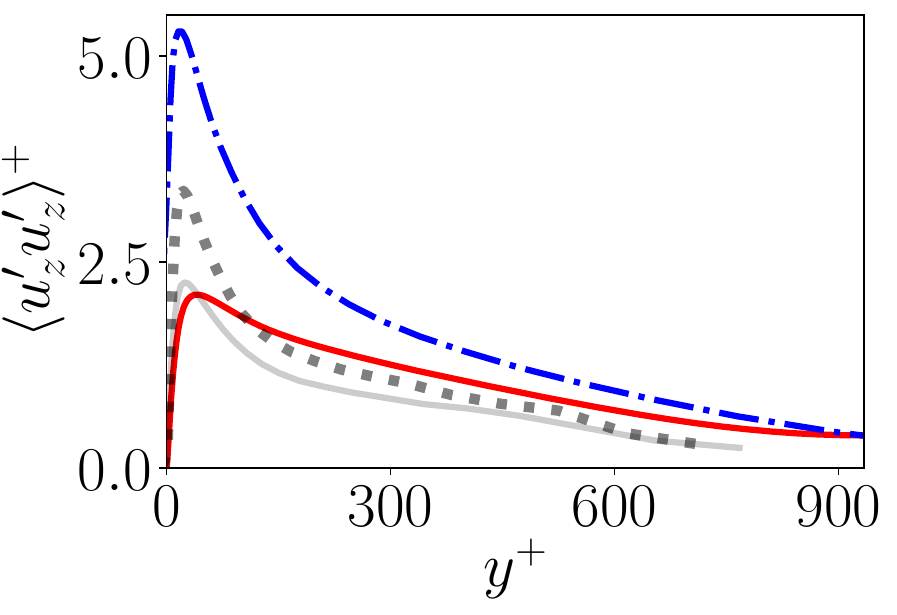} &
    \includegraphics[width=0.32\linewidth]{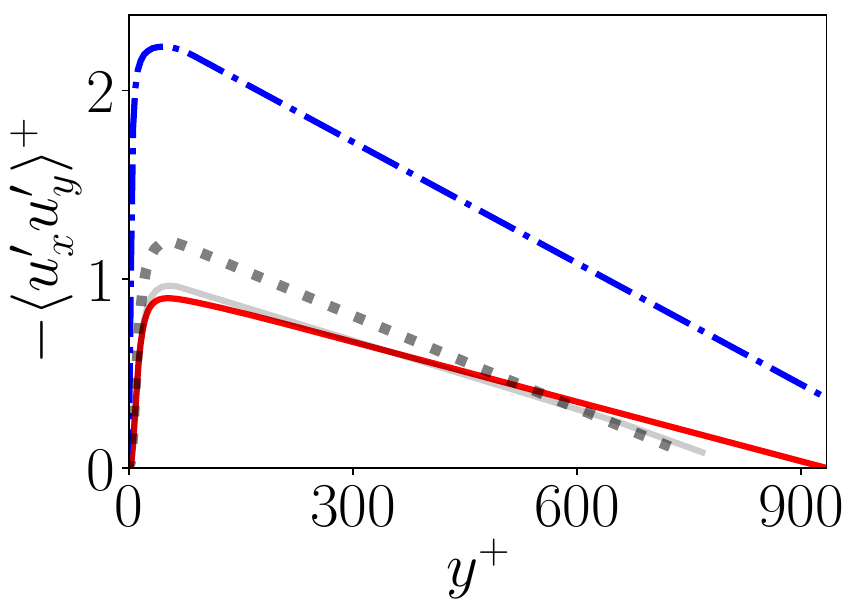} \\
    \textit{(d)} & \textit{(e)} & \textit{(f)}
    \end{tabular}
    \caption{Comparison of the main statistical moments of the velocity field. Results are shown for simulations (\protect\bluelinedasheddotted) ML-ReTau950, (\protect\greylinesolidsoft) BF-ReTau950, (\protect\greylinedottedstrong) IBM-ReTau950, and (\protect\redline) Hoyas \textit{et al.} \cite{Hoyas2008_pof}.}
    \label{fig:ReTau_950}
\end{figure}

At last, the case for $Re_\tau \approx 950$ is considered. The DNS database by Hoyas \textit{et al.} \cite{Hoyas2008_pof} is used for comparison, with the corresponding curves represented in Figure \ref{fig:ReTau_950}. This case highlights the limitations of the ML model (ML-ReTau950) when applied to flow conditions that differ significantly from the training environment. Although the predicted first-order statistics and $u_\tau$ show a reasonable agreement with the reference data, second-order statistics exhibit poorer resolution compared to the original body-fitted and IBM approaches, particularly for $\langle u_y^\prime u_y^\prime \rangle^+$ and $\langle u_z^\prime u_z^\prime \rangle^+$. The main problem here is associated with previous considerations for the case $Re_\tau \approx 395$. The ML model is trained using data from wall-resolved simulation while, in this case, the first grid point is already in the buffer layer. Therefore, the physics learnt during the training does not match what the algorithm needs to replicate in this case. To accurately capture the flow features under these conditions, the ML model would either need to incorporate wall modelling physics as already performed with physics-informed neural networks (PINNs) for LES models \cite{Maejima2024_prf, Yang2019_prf}, quantify the uncertainty in the coarse-grained DNS simulations to bound all physically realisable states, like in physics-informed RFRs for RANS simulations \cite{Kaandorp2020_cf, Wang2017_jcp, Heyse2021}), or be trained on a broader database that includes higher $y^+$ values.


\section{Conclusions}
\label{sec:conclusions}

The research activities in this work propose using Machine Learning tools to learn and mimic the state estimation corrections performed in the analysis phase of DA strategies. This research advancement opens perspectives of more systematic application of such state corrections, which are limited to the data availability in classical DA procedures. The application here is performed by combining the Ensemble Kalman Filter, a sequential Data Assimilation algorithm, and Random Forest Regression, an ensemble-learning Machine Learning method. The generated black-box models consist of a correction term accounting for DA's state estimation as well as a surrogate IBM. For the latter, the training data is taken from a classical penalisation IBM, whose free coefficients are optimised by the EnKF. The strategy is assessed by investigating the turbulent plane channel flow test case. The analysis of the flow statistics, including the friction velocity $u_\tau$, the mean streamwise velocity $\langle U \rangle^+$ and the Reynolds stress tensor components $\langle u_i^\prime u_j^\prime \rangle^+$, indicates that the black-box models are able to obtain a predictive accuracy of the same level of the DA approaches. However, the analysis of the spectra of the fluctuating component of the velocity field shows some degradation of the results, whose accuracy is in line with the prior models used in the investigation. When applied to configurations outside of the training range, such as those due to variations in wall grid resolution and $Re_\tau$, the ML black-box tools show good features of robustness, computational efficiency, and adaptability to similar physical configurations. However, limitations are clearly observed when the models are used to extrapolate physical information which is not available in the training set.

The hybrid approach combining DA with ML presented in this work shows significant potential for applications involving parametric optimisation and state estimation via on-the-fly training. The team's current research explores this DA-ML combination in statistically unsteady cases, such as an oscillatory cylinder with horizontal sinusoidal motion. This strategy holds promise for delivering fast, accurate, and interpretable results in various applications, suggesting a valuable pathway for future work in the field where both methodologies could be employed with streaming data in real industrial configurations.

\section*{Acknowledgments}

The authors would like to thank Prof. Nicolas Hascoët for his initial insights into the Machine Learning models used in this study. This project was provided with computer and storage resources by GENCI at TGCC thanks to the grant 2023-A0142A01741 on the supercomputer Joliot Curie's SKL partition. This research was conducted in the framework of the ANR-JCJC-2021 IWP-IBM-DA project.

\appendix

\section{Ensemble Kalman Filter algorithm}
\label{sec:EnKF}

The Ensemble Kalman Filter (EnKF) is a Monte-Carlo approach of the Kalman Filter (KF) \cite{Kalman1960_jbs}, where both the physical system state $\boldsymbol{u}$ (typically the velocity field discretised over the computational domain in CFD applications) and the covariance error matrix $\boldsymbol{P}$---which quantifies the discrepancy $\boldsymbol{e}$ between the system state predicted by a dynamic model $\mathcal{M}(\theta)$ defined by some parameters $\theta$, and its true state (i.e., $\boldsymbol{P} = \mathbb{E}[\boldsymbol{e} (\boldsymbol{e})^T ]$)---are propagated in time. This propagation is achieved by defining an ensemble of realisations called ensemble members. Each ensemble member has distinct state vectors $\boldsymbol{u}_i$ ($i \in \left[1, N_e \right]$, where $N_e$ is the number of ensemble members) to represent the system state. The average of these vectors $\boldsymbol{\overline{u}}$ is assumed to correspond to the true state, while their variability reflects the error covariance matrix $\boldsymbol{P}$. To enhance performance, the \emph{extended-state} EnKF approach \cite{Asch2016_siam} can be employed. In this method, certain model parameters $\theta$ are also propagated alongside the system state to reduce model bias. Hence, the system state $\boldsymbol{u}^\prime = [\boldsymbol{u} \,\theta]$.

The EnKF operates in a two-step procedure. The first step corresponds to a time advancement in the numerical model $\mathcal{M}(\theta)$, known as forecast, represented as $\boldsymbol{\cdot}^f$. By assuming that this progression is effectuated from time $k-1$ to time $k$, the system state predicted by the model can be estimated as:

\begin{equation}
    \boldsymbol{u}^f_{i,k} = \mathcal{M}(\theta_{i,k:k-1}) \, \boldsymbol{u}_{i,k-1}
    \label{eqn:forecast}
\end{equation}
If observation is available in time $k$, an analysis phase, denoted by $\boldsymbol{\cdot}^a$, is performed, where the EnKF algorithm updates the system state and the model to optimally combine the dynamics of both the model and observations, effectively minimising the covariance error matrix of the updated state $\boldsymbol{P}_k^a$. However, rather than explicitly computing this matrix---which would be computationally prohibitive due to its size of $[n_{DF}, n_{DF}]$, where $n_{DF}$ is the system’s degrees of freedom and equal to $3$ times the number of mesh elements for 3D simulations in CFD applications---the EnKF uses an alternative approach. Instead, the algorithm defines an \emph{anomaly matrix} $\boldsymbol{X}$, which measures the deviation of the ensemble members from the mean $\boldsymbol{\overline{u}}$. If the ensemble is representative enough, and there is statistical independence among the ensemble members, the forecast covariance matrix $\boldsymbol{P}_k^f$ can be approximated by:

\begin{equation}
    \boldsymbol{P}_k^f \approx \boldsymbol{X}_k^f \left(\boldsymbol{X}_k^f \right)^T
\end{equation}
where the anomaly matrix $\boldsymbol{X}_k^f$ has size $[n_{DF}, N_e]$ and is estimated from the state vectors $\boldsymbol{u}_i$ and the mean $\boldsymbol{\overline{u}}$ as:

\begin{eqnarray}
    \boldsymbol{\overline{u}}^f_k &=& \frac{\sum_i^{N_e} \boldsymbol{u}_{i,k}^f}{N_e} \\
    \left[\boldsymbol{X}_k^f \right]_i &=& \frac{\boldsymbol{u}^f_{i,k}-\boldsymbol{\overline{u}}^f_k}{\sqrt{N_e-1}} \label{eqn:anomalies_state} 
\end{eqnarray}
We require an ensemble of $n_o$ observations to ensure dimensional consistency during algebraic operations. The resulting observation matrix $\boldsymbol{Y}_k$, which has dimensions $[n_o,N_e]$, is derived from perturbing the observation vector $\boldsymbol{y}_k$ of size $[n_o,1]$. This perturbation is modelled using Gaussian noise, where the mean corresponds to the actual observation values in $\boldsymbol{y}_k$. The standard deviation reflects the observation uncertainty, encapsulated in a diagonal matrix denoted as $\boldsymbol{R}_k$ with dimensions $[n_o,n_o]$ and called the observation covariance matrix. The diagonal structure of this matrix assumes that uncertainties are Gaussian, unbiased and independent. For an infinite ensemble size, $\lim_{N_e \to +\infty} \mathbb{E}[(\boldsymbol{e} - \mathbb{E}[\boldsymbol{e}]) (\boldsymbol{e} - \mathbb{E}[\boldsymbol{e}])^T] = \boldsymbol{R}$, which becomes time-independent. Thus, each entry of the observation matrix can be expressed as $\left[\boldsymbol{Y}_k \right]_i=\boldsymbol{y}_k+\boldsymbol{\epsilon}_i$, where $\boldsymbol{\epsilon}_i \sim \mathcal{N}(0, \boldsymbol{R})$. 

The EnKF also calls for the use of a sampling matrix $\mathcal{H} (\boldsymbol{u}^f_k)$, where $\mathcal{H}$ acts as an interpolator that projects the forecast system state into the observation framework. Similar to the system's state $\boldsymbol{u}_k^f$ in Equation \ref{eqn:anomalies_state}, this projection is composed by a matrix $\boldsymbol{S}_k^f$ where each column represents a normalised anomaly estimated from the mean $\overline{\mathcal{H}(\boldsymbol{u}_{k}^f)}$ as:

\begin{eqnarray} 
    \overline{\mathcal{H}(\boldsymbol{u}_{k}^f)} &=& \frac{\sum_i^{N_e} \left[\mathcal{H}(\boldsymbol{u}_{k}^f) \right]_i}{N_e} \\
    \left[\boldsymbol{S}_k^f \right]_i &=& \frac{\left[\mathcal{H}(\boldsymbol{u}_{k}^f) \right]_i - \overline{\mathcal{H}(\boldsymbol{u}^f_k)}}{\sqrt{N_e-1}}
    \label{eqn:anomalies_stateSampling}
\end{eqnarray}

The Kalman gain matrix $\boldsymbol{K}_k$ describing the most optimal correlation between the state and the observations can be defined as expressed in Equation \ref{eqn:KalmanGain} \citep{Carrassi2018_WIREs, Hoteit2015_mwr}. In our context, we apply an element-wise multiplication with a covariance localisation matrix $\boldsymbol{L}$. This matrix is uniform for static grids or observation sensors, with coefficients $L_{ij} \in [0,1]$. This approach enhances the significance of coefficients associated with closer distances $\Delta \boldsymbol{r}_{ij} = (\Delta x_{ij}, \Delta y_{ij}, \Delta z_{ij})$ between the centre of each mesh element and the observation location, compared to those further away. The parameters ($\eta_x, \eta_y, \eta_z$) act as tuning factors, balancing the spatial scale of the system's physical representation with the smoothing needed to mitigate discontinuities at the boundaries of the updated region.

\begin{eqnarray}
    \boldsymbol{K}_k &=& \boldsymbol{L} \odot \boldsymbol{X}_k^f \left(\boldsymbol{S}_k^f \right)^T \left[\boldsymbol{S}_k^f \left(\boldsymbol{S}_k^f \right)^T + \boldsymbol{R} \right]^{-1}
    \label{eqn:KalmanGain} \\
    L_{ij} &=& e^{-\left[\left(\frac{\Delta x_{ij}}{\eta_x}\right)^2+\left(\frac{\Delta y_{ij}}{\eta_y}\right)^2+\left(\frac{\Delta z_{ij}}{\eta_z}\right)^2 \right]}
\end{eqnarray}
All the elements described are crucial for calculating the updated system's state $\boldsymbol{u}_k^a$, which is subsequently used to advance the numerical model $\mathcal{M}(\theta_{i, k+1:k})$ in the following iteration $k+1$. Typically, multiplicative inflation $\lambda_k$ is applied to perturb the updated state, as the filter tends to converge on local optima that are closer to the forecast state rather than the desired global optimum.

\begin{eqnarray}
    \boldsymbol{u}_k^a &=& \boldsymbol{u}_k^f + \boldsymbol{K}_k \left( \boldsymbol{Y}_k - \mathcal{H}(\boldsymbol{u}_k^f)\right)
    \label{eqn:updatedState} \\
    \boldsymbol{u}_k^a &=& \overline{\boldsymbol{u}}_k^a + \lambda_k \, (\boldsymbol{u}_k^a - \overline{\boldsymbol{u}}_k^a)
\end{eqnarray}
The step-by-step implementation of the EnKF is detailed in the Algorithm \ref{alg:EnKF}.

\begin{algorithm}
    \caption{Scheme of the EnKF used in the present study.}
    \label{alg:EnKF}
    \textbf{Input:} $\mathcal{M}$, $\boldsymbol{R}$, $\boldsymbol{y}_k$, and a prior/initial state system $\boldsymbol{u}_{i,0}$, where usually $\boldsymbol{u}_{i,0} \sim \mathcal{N}(\boldsymbol{\overline{u}}_0, \sigma_0^2)$ \\
    \For{$k = 1, 2,..., K$}{
        \For{$i = 1, 2, ..., N_e$}{
    \nl Advancement in time of the state vectors:\\
    \qquad $\boldsymbol{u}_{i,k}^f = \mathcal{M}_{k:k-1}\,\boldsymbol{u}_{i,k-1}$ \\
    \nl Generation of an observation matrix from the confidence level given to the observation data:\\
    \qquad$\boldsymbol{Y}_{i,k} = \boldsymbol{y}_k + \boldsymbol{e}_i$, with $\boldsymbol{e}_i \sim \mathcal{N}(0,\boldsymbol{R})$\\
    \nl Estimation of the ensemble means (system state and projection matrix):\\
    \qquad$\boldsymbol{\overline{u}}_k^f = \frac{1}{N_e}\sum_{i = 1}^{N_e}\boldsymbol{u}_{i,k}^f$,\,
    $\overline{\mathcal{H}(\boldsymbol{u}^f_k}) = \frac{1}{N_e}\sum_{i = 1}^{N_e} \mathcal{H}(\boldsymbol{u}_{i,k}^f)$ \\
    \nl Computation of the anomaly matrices (system state and projection matrix):\\
    \qquad$\boldsymbol{X}_k = \frac{\boldsymbol{u}_{i,k}-\boldsymbol{\overline{u}}_k}{\sqrt{N_e-1}}$,\,
    $\boldsymbol{S}_k = \frac{\mathcal{H}(\boldsymbol{u}_{i,k}^f) - \overline{\mathcal{H}(\boldsymbol{u}^f_k)}}{\sqrt{N_e-1}}$ \\
    \nl Calculation of the Kalman gain (with covariance localisation):\\
    \qquad$\boldsymbol{K}_k = \boldsymbol{L} \odot \boldsymbol{X}_k^f(\boldsymbol{S}_k)^T \left[\boldsymbol{S}_k(\boldsymbol{S}_k)^T + \boldsymbol{R}\right]^{-1}$\\
    \nl Update of the state matrix:\\
    \qquad$\boldsymbol{u}_{i, k}^a = \boldsymbol{u}_{i,k}^f + \boldsymbol{K}_k \left(\boldsymbol{Y}_{i,k}- \mathcal{H}(\boldsymbol{u}_{i,k}^f) \right)$ \\
    \nl Application of (deterministic) covariance inflation: \\
    \qquad$ \boldsymbol{u}_{i,k}^a = \boldsymbol{\overline{u}}_k^a + \lambda_k \left(\boldsymbol{u}_{i,k}^a - \boldsymbol{\overline{u}}_k^a   \right)$
    }
    }
\end{algorithm}

\section{Random Forest Regression in PISO algorithm}
\label{sec:RFR_PISO}

The integration of the RFR model into the CFD solver involves modifying the PISO algorithm \citep{ISSA198640} for parametric optimisation and introducing an additional step for state estimation. During this process, the velocity and pressure fields are iteratively updated to satisfy the momentum and Poisson equations. The RFR requires an explicit estimation of the penalisation volume force $\hat{\boldsymbol{f}}_P$ inside the Equation \ref{eqn:NavierStokes_Mom}. At time $t$, with a time step advancement of $\Delta t$, and considering $j \in [1, J]$ as a single iteration within the PISO loop, the following steps are executed:

\begin{enumerate}
   \item Resolution of the momentum equation, from where $\boldsymbol{u}_{t,j}$ is obtained.

   \begin{equation}
       \boldsymbol{A} \,\boldsymbol{u}_{t,j} - \boldsymbol{b} - \hat{\boldsymbol{f}}_{P_{t-\Delta t}} = -\nabla p_{t-\Delta t}
       \label{eqn:momentumMatrix}
   \end{equation}

   \item Estimation of the dimensional forcing term $\hat{\boldsymbol{f}}_{P_{t,j}}$ from $\hat{\boldsymbol{f}}_{P_{t,j}}^{+/\ast}$ with the predicted velocity $\boldsymbol{u}_{t,j}$ by means of RFR. By calling the operator $\mathcal{F}_{\textrm{p.o.}}^r$ to define the prediction of a single decision tree $r$ and normalising with respect to the instantaneous $u_{\tau_{t,j}}$ or the statistical $u_{\tau_{\textrm{DA/p.o.}}}$ friction velocity:

   \begin{equation}
       \hat{\boldsymbol{f}}_{P_{t,j}}^{+/\ast} = \frac{1}{R}\sum_{r = 1}^R \mathcal{F}_{\textrm{p.o.}}^r \left(\boldsymbol{u}_{t,j}^{+/\ast}, y_{t,j}^{+/\ast} \right)
       \label{eqn:prediction_Force_RFR}
   \end{equation}

   \item Estimation of the pressure $p_{t, j+1}$ through the Poisson equation. $A$ is a scalar field calculated from $\boldsymbol{A}$, whereas $\boldsymbol{T}^\prime(\boldsymbol{u}_{t,j})$ is a tensor field containing the discretised form of all the terms on the left side of Equation \ref{eqn:momentumMatrix}. To account for the update of the forcing term $\hat{\boldsymbol{f}}_{P_{t,j}}$, an additional term is included.

   \begin{equation}
       \boldsymbol{\nabla} \cdot \frac{1}{A} \nabla{p_{t, j+1}} = \boldsymbol{\nabla} \cdot \left(\frac{\boldsymbol{T}^\prime(\boldsymbol{u}_{t,j})}{A}\right) - \boldsymbol{\nabla} \cdot  \left(\frac{\hat{\boldsymbol{f}}_{P_{t,j}}}{A} \right)
       \label{eqn:PoissonEq}
   \end{equation}

   \item Update of the velocity field $\boldsymbol{u}_{t, j+1}$ to satisfy the zero-divergence condition.

   \begin{equation}
       \boldsymbol{u}_{t,j+1} = \frac{\boldsymbol{T}^\prime(\boldsymbol{u}_{t,j})}{A} - \frac{1}{A} \nabla p_{t,j+1} - \frac{1}{A} \hat{\boldsymbol{f}}_{P_{t,j}}
       \label{eqn:velocityCorr}
   \end{equation}
\end{enumerate}

Equations \ref{eqn:PoissonEq} and \ref{eqn:velocityCorr} are solved iteratively until reaching convergence. For state estimation at time $t$, when $j = J$, the velocity $\boldsymbol{u}_{t, J}$ corresponds to $\boldsymbol{u}_k^f$, which is equivalent to the forecast velocity field at the $k^{th}$ analysis phase for the EnKF algorithm. To refine its estimation $\boldsymbol{u}_k^a$ as done with the EnKF in the analysis step, it is necessary to incorporate another RFR model, denoted by the operator $\mathcal{F}_{\textrm{s.e}}^r$ for a single decision tree $r$, which is able to predict the correction given to the velocity $\Delta \hat{\boldsymbol{u}}_k^{+/\ast}$, so that $\hat{\boldsymbol{u}}_k^{a^{+/\ast}} = \boldsymbol{u}_{k}^{f^{+/\ast}} + \Delta \hat{\boldsymbol{u}}_k^{+/\ast}$. Then, the dimensional $\hat{\boldsymbol{u}}_k^a$ is estimated.

\begin{equation}
    \Delta \hat{\boldsymbol{u}}_k^{+/\ast} = \frac{1}{R}\sum_{r = 1}^R \mathcal{F}_{\textrm{s.e}}^r \left(\boldsymbol{u}_{k}^{f^{+/\ast}} \right)
    \label{eqn:prediction_Vel_RFR}
\end{equation}

\bibliographystyle{elsarticle-num} 
\bibliography{references}





\end{document}